\documentclass{article}
\usepackage{amssymb}
\usepackage[dvipdfmx]{graphicx}
\usepackage{bmpsize}
\usepackage{amsmath}
\usepackage{xpiano}
\usepackage{latexsym}
\usepackage{lscape}
\usepackage{fullpage}
\usepackage{float}
\usepackage{hyperref}
\usepackage{color}
\usepackage{amsthm}
\usepackage[minimal]{leadsheets}
\usepackage{yfonts}
\usepackage{epigraph}
\usepackage{afterpage}
\usepackage{parskip}
\useleadsheetslibraries{chords}

\newcommand{\fancy}{\textbf}

\begin{document}
\begin{titlepage}
    \begin{center}
        \vspace*{1cm}
 
        \textbf{\huge{Harmony and Duality}}
 
        \vspace{0.5cm}
         \large{An introduction to music theory}
             
        \vspace{1.5cm}
        \includegraphics[width=0.75\textwidth]{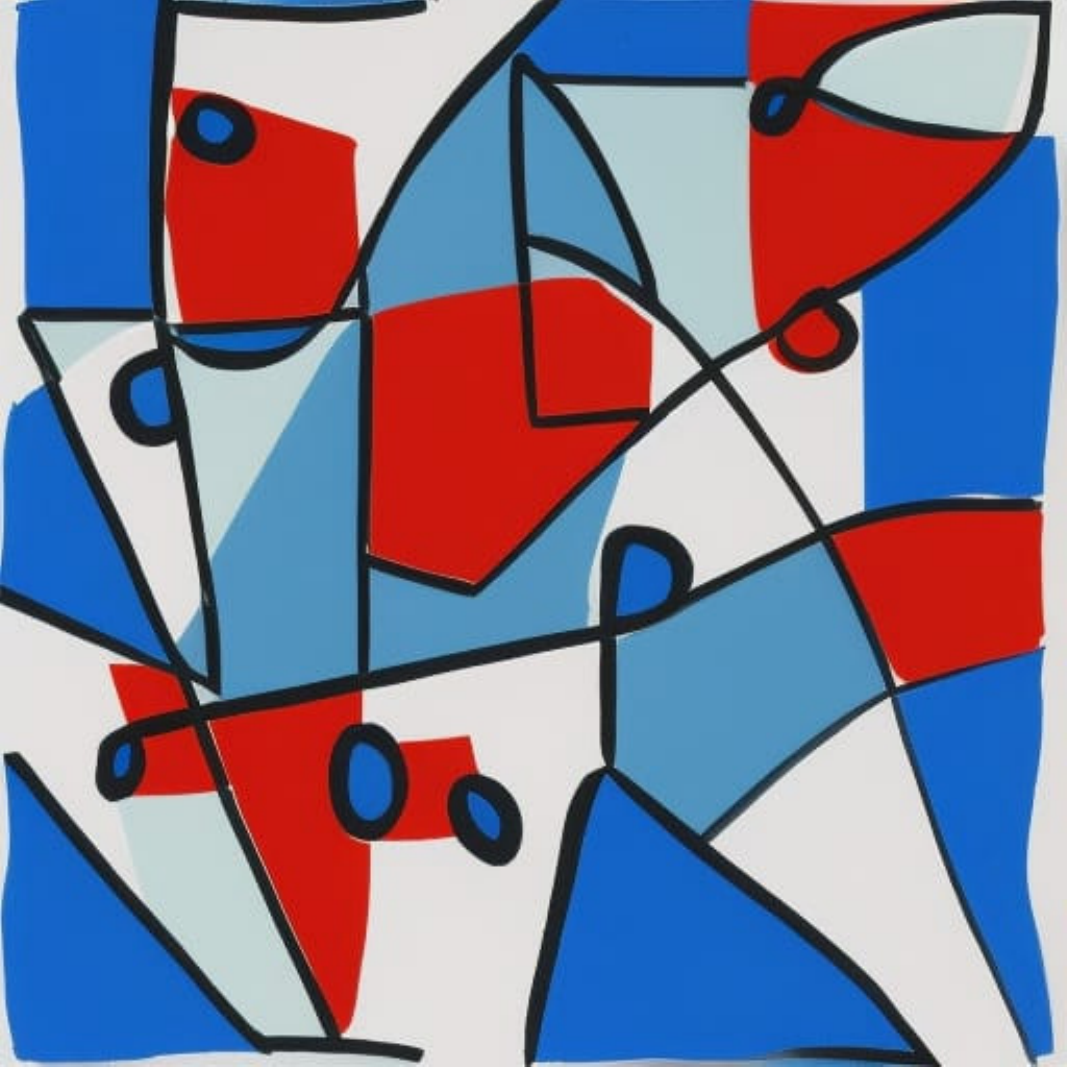}\\
        \vspace{3cm}
        \textbf{\large{Maksim Lipyanskiy}}
 
        \vspace{.5cm}
     
             
        \vspace{0.8cm}

    \end{center}
\end{titlepage}

\null\thispagestyle{empty}\newpage

\setcounter{tocdepth}{1}
\tableofcontents
\newpage

\section{Introduction}
\epigraph{``Writing about music is like dancing about architecture''}{---anonymous}
We develop aspects of music theory---scales, chord formation, and improvisation---from a combinatorial perspective.  The goal is to derive the basic harmonic structure from a few assumptions, rather than presenting long lists of chords and scales to memorize without an underlying principle.  While learning to read lead sheets and studying jazz improvisation were the initial motivation, this approach is equally applicable to arrangement and composition in a much broader context.

A longer-term goal is to provide an alternative approach for learning to play the piano.  Traditionally, one starts by learning to read notation and then progresses linearly through pieces of increasing difficulty.  We propose instead to practice basic harmonic structures and start improvising from the beginning.  Much like a beginning art student who does not spend all their time copying the masters (though that too can be useful), we advocate for a more personalized approach where the student is the creator from day one.  This is complementary to the traditional method: it builds a set of skills quite different from sight-reading or finger dexterity.  As we will demonstrate, armed with a few basic rules, even a beginner can start creating their own arrangements.  For a concrete illustration, see Section~\ref{sec:leadag}.

Here are the key concepts:
\begin{itemize}
    \item \textit{Constraints}: Harmony involves several voices moving simultaneously, and a scale dictates which tones are available.  A natural constraint is to avoid voice ``collisions'': for example, we can require that no two notes in a scale are a semitone apart.  A more refined constraint forbids three notes separated only by semitones.
    \item \textit{Completeness}: We require that our scales are \textit{complete}---they are the maximal sets of tones satisfying a given constraint.  That is, we cannot add further tones without violating the constraint.  Remarkably, completeness applied to the simple constraints above characterizes the scales commonly used in music.
    \item \textit{Duality}: There is a surprising correspondence between scales subject to the two-note constraint and those subject to the three-note constraint.  This duality provides a way to understand one family of scales in terms of the other.
\end{itemize}
Unpacking these concepts will lead to a full classification of harmony and chords as well as a principled approach to core topics in music theory.

\textbf{Outline of Contents:} The first part (Sections~\ref{sec:piano}--\ref{sec:white}) is introductory and focuses on major and pentatonic scales.  The next part (Sections~\ref{sec:harm}--\ref{sec:dual}) generalizes these ideas to a full discussion of harmony and packings.  Sections~\ref{sec:modes}--\ref{sec:voicings} classify voicings for the various harmonies.  Sections~\ref{sec:lead} and~\ref{sec:leadag} connect our approach to standard lead sheet notation.  Several appendices review basic concepts and cover tangential topics.

\textbf{Prerequisites:}  We aim to make this work accessible to beginners familiar with the initial chapters of \textit{Alfred's Basic Adult All-in-One Course} (Palmer, Manus, and Lethco).  We assume the reader can read sheet music and has some basic exposure to the piano; familiarity with chord notation is also helpful.  Some basics are summarized in the appendices, but the main text focuses on material not readily available elsewhere.

\textbf{Math and Music:} This work does not assume a strong background in mathematics, though we borrow mathematical language where it is the clearest way to express an idea.  The aim is concrete musical applications, not to make music theory a branch of mathematics.  That said, a willing music student may find here an accessible introduction to concepts such as classification, completeness, reducibility, and duality.  The connection between mathematics and music has a long history going back at least to Pythagoras; for a recent treatment, see Benson~\cite{benson2006}.  We believe the present work connects the two subjects in a novel way.

\textbf{What this work is not about:}  We aim to be universal, avoiding context-specific notions like ``happy'' or ``sad'' chords and not tying our foundation to any particular style.  Each generation of musicians reevaluates existing conventions and develops new rules; we focus on the parts of harmony that can be formalized and leave the art to the practitioner.  By analogy with writing, we build the student's vocabulary and grammar, not teach them to write in a particular style.

\textbf{Future Work:}  Ideally, one would develop a full course based on this approach with many more worked examples and exercises.  We hope to return to this in the future.  For now, concrete practice suggestions appear in Section~\ref{sec:exblack} and Section~\ref{sec:vchord}---enough to keep a beginner busy for months.

\textbf{Acknowledgement:} The combinatorial approach to harmony originates with the work of the pianist, composer, and author's teacher, Gustavo Casenave, who introduced the notion of a semitone cell and used it to classify harmony and modes.  We take his foundational insight in a different direction: focusing on harmony with zero cells, introducing the dual notion of ``blocks'' and ``packings,'' and developing a theory of chords and voicings from these ingredients.  A key motif is to associate to each harmony a simpler ``shadow'' object---its packing---which, as we will show, helps us understand the harmony itself.  To our knowledge, this relationship between harmony and packings is new.  For more related work, see Appendix~\ref{sec:related}.

We would like to thank Ilya Elson, Sam Kaufman, Mikhail Lipyanskiy, Otar Sepper, and Alex Sotirov for their help with earlier drafts of this paper.  
\newpage
\section{Overture: The Piano Layout and the Circle of Fifths}
\label{sec:piano}
We begin by justifying the piano key layout.  While typically taken for granted, the layout can be motivated from first principles, and doing so introduces the main constructions used throughout this work.

\subsection{The Piano Layout}

\begin{figure}[H]
    \begin{center}
    \includegraphics[width=6.4in]{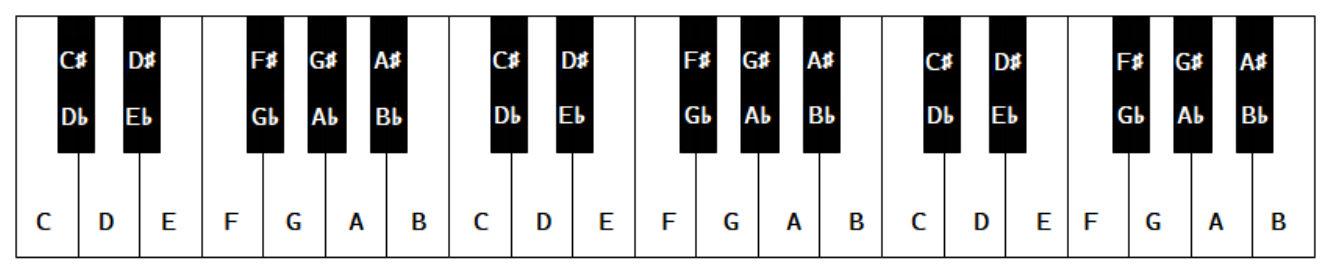}
    \caption{Piano keys.}
    \label{fig:pianol}
    \end{center}
\end{figure}

Looking at the piano as if for the first time, we notice a few things right away (see Figure~\ref{fig:pianol}):
\begin{enumerate}
    \item The pattern of keys is \textbf{periodic}: it repeats every 12 keys.
    \item The \textbf{period is 12}: within each octave there are exactly 12 distinct tones.
    \item The keys are divided into a \textbf{distinct pattern of black and white keys}, singling out a particular 7-note scale (the C major scale) as the white keys.
\end{enumerate}
Each of these features has a musical justification, which we now explain.

\subsection{Periodicity}

Any key on the piano corresponds to a vibrating string of some length.  If we halve this length, or equivalently double the frequency, we obtain a key that is an octave higher.  Tones that are an octave apart are so consonant that we tend not to distinguish between them harmonically.  This is the reason behind the periodicity of the piano layout.  Effectively, it reduces the harmonic possibilities to the range inside a single octave.

\subsection{Twelve Tones}

Why 12 tones in an octave?  Start with a single tone at frequency $f$.  Doubling the frequency gives higher octaves ($2f$, $4f$, \dots), which are harmonically equivalent.  To get a genuinely new note, we triple the frequency to $3f$.  Since $3f$ lies above the octave, we halve it to $\frac{3f}{2}$, landing inside the octave.  This interval---frequency ratio 3:2---is the \textit{perfect fifth}, the most consonant interval within an octave.  For example, starting from F, the perfect fifth gives us C.

There are two ways to continue subdividing.  One is to take higher \textit{harmonics} (higher multiples of $f$), which yields consonant intervals but ties every note to the starting frequency, producing an asymmetric layout (see Appendix~\ref{sec:appE}).  The alternative, known as Pythagorean tuning, is to keep applying the perfect fifth: from F we get C, from C we get G, and so on.  This treats every note symmetrically.  Iterating gives:
\begin{equation}
    \text{F, C, G, D, A, E, B, F{\sharp}, C{\sharp}, G{\sharp}, D{\sharp}, A{\sharp}, F }\dots
\end{equation}
We cycle through all 12 notes before returning to F.  This ordering is the \textit{circle of fifths}, and it explains why there are 12 tones in the scale.

There is a slight wrinkle: tuning by exact 3:2 ratios does not land exactly back on F after 12 steps.  The modern piano uses \textit{equal temperament}, replacing the perfect fifth with the slightly smaller ratio $2^{7/12}\approx 1.4983$, chosen so that 12 steps return \textit{exactly} to the starting frequency.  The near-coincidence of 12 perfect fifths with 7 octaves is why 12 tones work so well.  See Appendices C and D for details.

\subsection{Black and White Keys}

The last point to explain is why the 12 notes are divided into black and white keys.  This division singles out the C major scale (all white keys) and is not merely a visual convenience---it lies at the foundation of Western music.  Standard notation for any instrument takes a major scale as its basis; the remaining notes are ``accidental.''

\begin{figure}[H]
    \begin{center} 
    \begin{tabular}{cc}
    \includegraphics[width=2.0in]{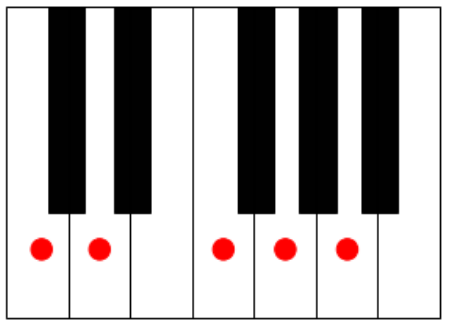} &  \includegraphics[width=2.0in]{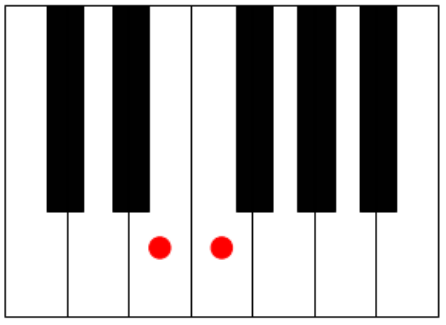} \\
    (a) F pentatonic scale & (b) semitone block \\    
     \end{tabular}
\caption{First stopping point.}
\label{fig:scales}
\end{center}
\end{figure}

We now motivate this division, introducing concepts central to our treatment of harmony.  In practice, we rarely use all 12 notes at once; the subject of harmony is about choosing interesting subsets.  A useful heuristic: stop adding notes before too many cluster together.  Following the circle of fifths, the first natural stopping point comes after five notes, giving the F pentatonic scale: 
\begin{equation}
    \text{F, C, G, D, A}
    \label{eq:penta}
\end{equation}
The next note, E, is the first that lies a semitone from an existing note (E and F).  Moreover, every subsequent note (E through A${\sharp}$) is a semitone away from some note already in the pentatonic scale.  See Figure~\ref{fig:scales}(a).

\begin{figure}[H]
\begin{center} 
\begin{tabular}{cc}
\includegraphics[width=2.0in]{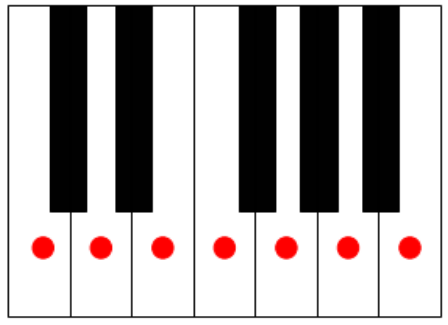} &  \includegraphics[width=2.0in]{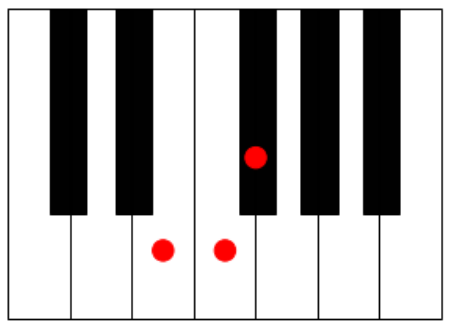} \\
(a) C major scale & (b) semitone cell  \\
\end{tabular}
\caption{Second stopping point.}
\label{fig:cell}
\end{center}
\end{figure}

We call two notes a semitone apart a \textbf{semitone block} (or simply a \textbf{block}).  The pentatonic scale contains no blocks, and no note can be added without creating one (check this!).  It is the largest block-free set of tones.  We believe the prevalence of pentatonic scales in world music is related to this characterization.
Another harmonically significant stopping point occurs at
\begin{equation}
   \text{F, C, G, D, A, E, B}
    \label{eq:major}
\end{equation} 
when we have exhausted all the white keys.  Here a different form of dissonance appears: adding F${\sharp}$ creates for the first time three consecutive semitones (E, F, F${\sharp}$).  We call such a triple a \textbf{dissonant cell} (see Figure~\ref{fig:cell}(b)).  Just as with blocks, every remaining note $\{$F${\sharp}$, C${\sharp}$, G${\sharp}$, D${\sharp}$, A${\sharp}$$\}$ would create a cell if added to the white keys (check this).  Thus the C major scale is the largest cell-free set of tones.  This observation, due to Gustavo Casenave, helps explain the prevalence of major scales in Western music.
To summarize: we have two notions of dissonance---the 2-note block and the 3-note cell---and two corresponding families of maximal scales.  Block-free maximal sets will be called \textbf{packings} (the pentatonic scale is an example); cell-free maximal sets will be called \textbf{harmonies} (the major scale is an example).  Defining and exploring these concepts in general is a main goal of the sections that follow, and combining them will be the foundation of our theory of chords and voicings.

\subsection{A Glimpse of Duality}

The black keys---the complement of the white keys---form a pentatonic scale!  This is surprising: the white keys were defined by avoiding cells, a criterion that has nothing to do with blocks.  Yet their complement turns out to be a packing.  By symmetry, this holds for every major scale and its pentatonic complement.  As we will show, it reflects a deeper duality between harmonies and packings, or equivalently, between the two notions of dissonance.
\begin{figure}[H]
\begin{center}
\includegraphics[width=2.0in]{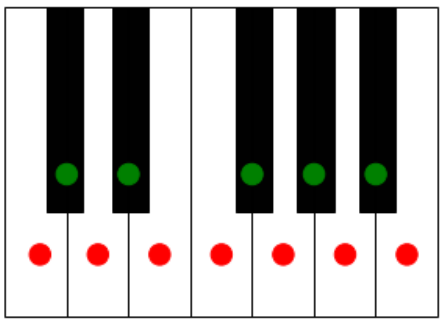}
\caption{C major scale (red) and its pentatonic complement (green).}
\label{fig:major}
\end{center}
\end{figure}
\textbf{Remark:} We started the circle of fifths on F rather than C in order to highlight the C major scale.  Starting on C would highlight G major instead.  This suggests that the most natural starting point for C major is F, not C.  More generally, a major scale has seven starting points called \textit{modes}; starting on F gives the Lydian mode, the brightest of all modes and the basis of George Russell's Lydian perspective on tonal organization~\cite{russell1953}.  We return to modes in Section~\ref{sec:modes}.
\newpage
\section{Black Keys}
\label{sec:black}
\subsection{Getting Started}
We begin in the simplest setting: the pentatonic scale.  Typically, students start with the white keys (C major) since they are visually easiest to identify and music notation is designed around them.  Instead, we focus on the black keys, which form a simpler harmonic structure and, in our view, provide a better entry point for harmony, improvisation, and composition.  Anyone who has played around with a piano may have noticed that playing black keys randomly already sounds melodic---this is because the pentatonic scale avoids dissonant semitone blocks.

The only drawback is notation: we need six flats to write the F${\sharp}$ pentatonic scale.  We feel this is a small price to pay for the conceptual and technical benefits.  Moreover, as we will see in the next section, the black keys are not merely a toy example---they form the building blocks of major harmony.

\begin{figure}[H]
    \begin{center}
    \includegraphics[width=6.5in]{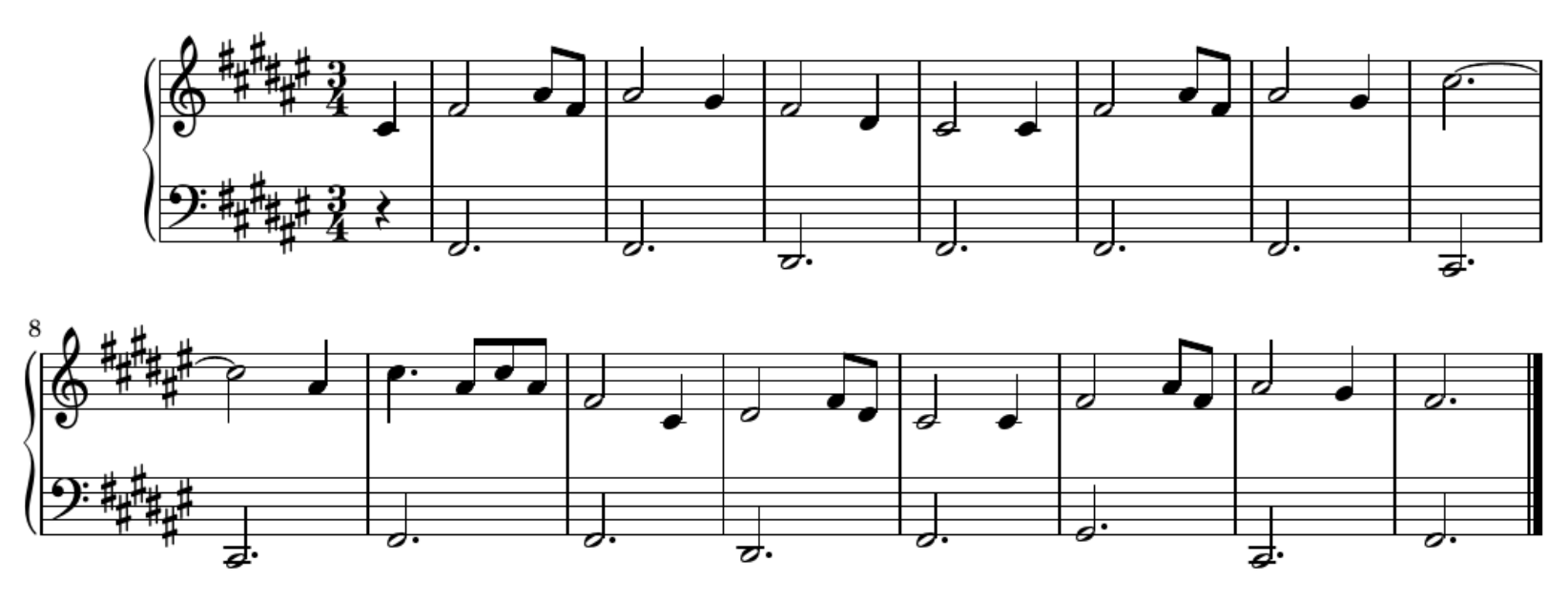}
    \caption{\textit{Amazing Grace} in F${\sharp}$.}
    \label{fig:ag0}
    \end{center}
\end{figure}

The pentatonic scale is rich enough to form basic melodies.  \textit{Amazing Grace} by John Newton (18th century) is a great example.\footnote{See \url{https://en.wikipedia.org/wiki/Amazing_Grace} for more information.} Figure~\ref{fig:ag0} shows a simple arrangement.  We use this piece as the main example throughout the text.
\subsection{Chords}
Let's form chords using the pentatonic scale.  Restricting to five notes means few clusters, so almost any combination sounds reasonable---giving the performer freedom to experiment.  Still, we want a systematic approach with an eye toward generalization.

\begin{figure}[H]
    \begin{center} 
    \begin{tabular}{cc}
    \includegraphics[width=2.0in]{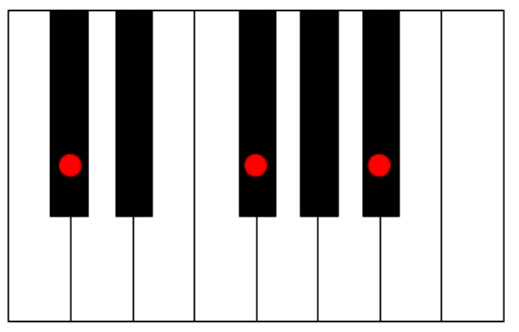} &  \includegraphics[width=2.0in]{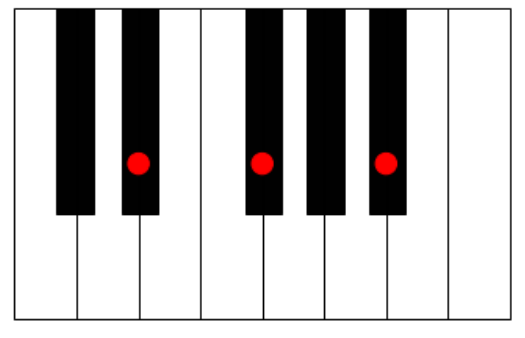} \\
    (a) F$\sharp$ major triad (inverted) & (b) D$\sharp$ minor triad \\
    \end{tabular}
    \caption{Triad options.}
    \label{fig:bkfig1}
    \end{center}
\end{figure}

\begin{figure}[H]
    \begin{center} 
    \begin{tabular}{cc}
    \includegraphics[width=2.0in]{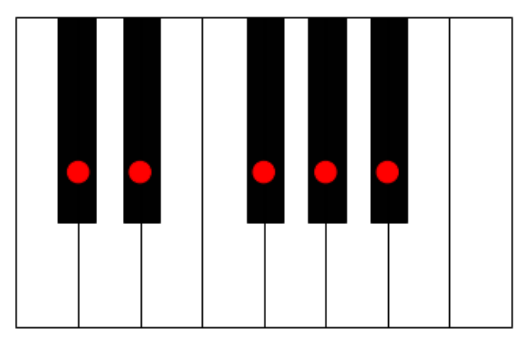} &  \includegraphics[width=3.3in]{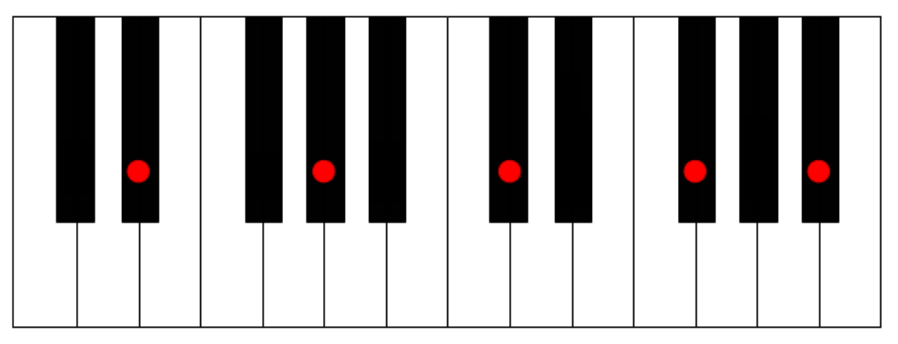} \\
    (a) clustered 5-chord & (b) spread out 5-chord \\
    \end{tabular}
    \caption{A few 5-chord inversions.}
    \label{fig:bkfig2}
    \end{center}
\end{figure}

\begin{figure}[H]
    \begin{center} 
    \begin{tabular}{ccc}
    \includegraphics[width=1.9in]{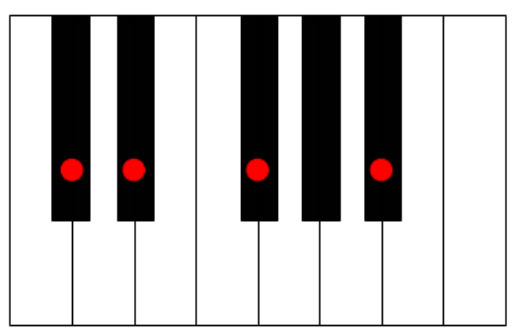} &  \includegraphics[width=1.9in]{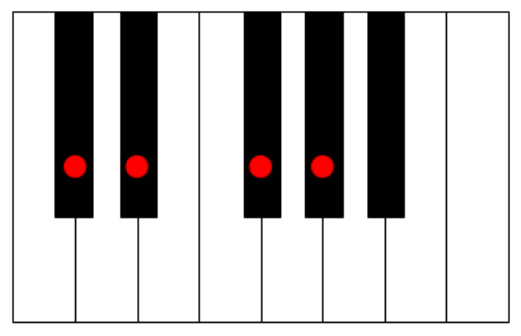}  &  \includegraphics[width=1.9in]{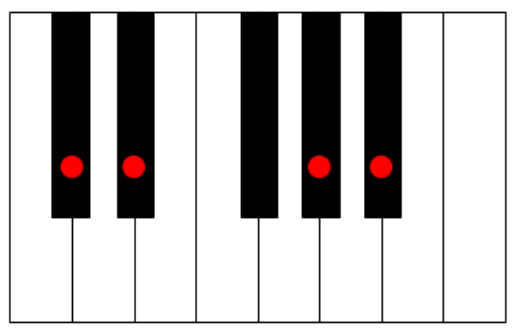} \\
    (a)  & (b) & (c) \\
    \end{tabular}
    \caption{4-chord variants.}
    \label{fig:bkfig3}
    \end{center}
\end{figure}

\begin{figure}[H]
    \begin{center} 
    \begin{tabular}{cc}
        \includegraphics[width=1.73in]{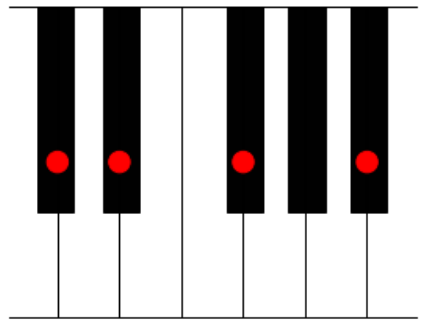} & \includegraphics[width=1.9in]{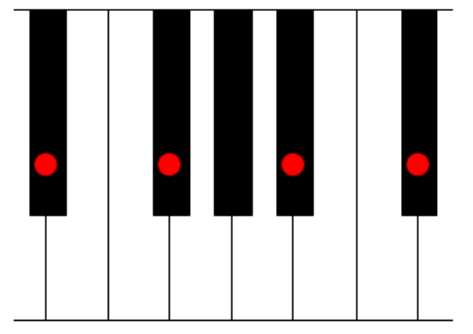}  \\  \vspace{.4cm}
        (a)  & (b) \\ 
        \includegraphics[width=1.73in]{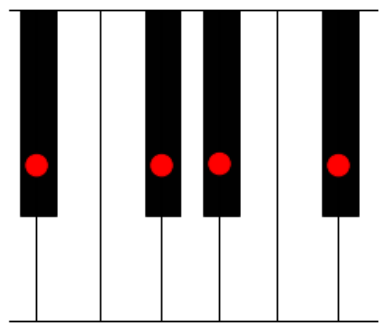}  &  \includegraphics[width=1.9in]{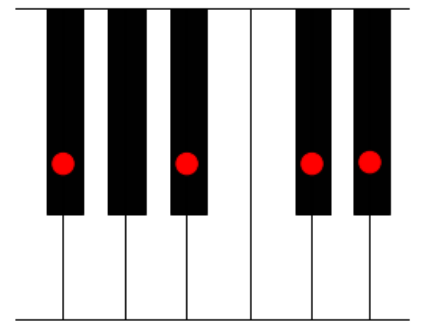} \\
    (c)  & (d) \\
    \end{tabular}
    \caption{Basic chord inversions.}
    \label{fig:invchord}
    \end{center}
\end{figure}

On one extreme, we can avoid notes that are at most a tone apart.  The maximal chords satisfying this are the two triads in Figure~\ref{fig:bkfig1}: an F$\sharp$ major triad (a) and a D$\sharp$ minor triad (b).  Triads are the basis of classical harmony.
On the other extreme, we may use all five notes, creating a 5-chord (pentachord).  Clustered together as in Figure~\ref{fig:bkfig2}(a), this can sound muddy; spread across both hands as in (b), it sounds much better.  Harmonically these are equivalent---we are always free to move any note by octaves.  The 5-chord is used in practice, notably in the jazz piano playing of Bill Evans and McCoy Tyner.
We propose a middle ground: form chords that avoid three notes in a row at most a tone apart.  Figure~\ref{fig:bkfig3} shows the three maximal chords of this type in the pentatonic scale.  These 4-chord shapes are the basic building blocks of our approach.  Other chords are derived from them:
\begin{itemize}
    \item Invert the chord by moving a tone up/down by octaves.  One can ``spread'' the notes of the chord to be played with both hands. See the examples in Figure~\ref{fig:invchord}.
    \item Omit any number of notes from the chord.  For example, we can view the triad chords as ``derived'' from the 4-chords by omitting notes.
    \item Add the missing note to get the ``composite'' 5-chord.  Once again, we view the 5-chord as derived from the fundamental 4-chord construction.  
\end{itemize}
\textbf{Remark:}  The 4-chords we obtained in Figure~\ref{fig:bkfig3} are known as the F$\sharp$maj6 (a), D$\sharp$min11 (b), and A$\sharp$min11 (c).  These standard naming conventions are meant to reflect that we add tones to basic minor/major triads.  For instance, F$\sharp$maj6 is obtained by adding the 6th degree of the F$\sharp$ major scale to the F$\sharp$ major triad.  Concretely, this adds the D$\sharp$ to (F$\sharp$, A$\sharp$, C$\sharp$).  For D$\sharp$min11, we add the 4th degree to the standard 7th chord tones.  This adds the G$\sharp$ to (D$\sharp$, F$\sharp$, C$\sharp$). We will review basic chord notation in Appendix~\ref{sec:appA}.

\begin{figure}[H]
    \begin{center}
    \includegraphics[width=6.5in]{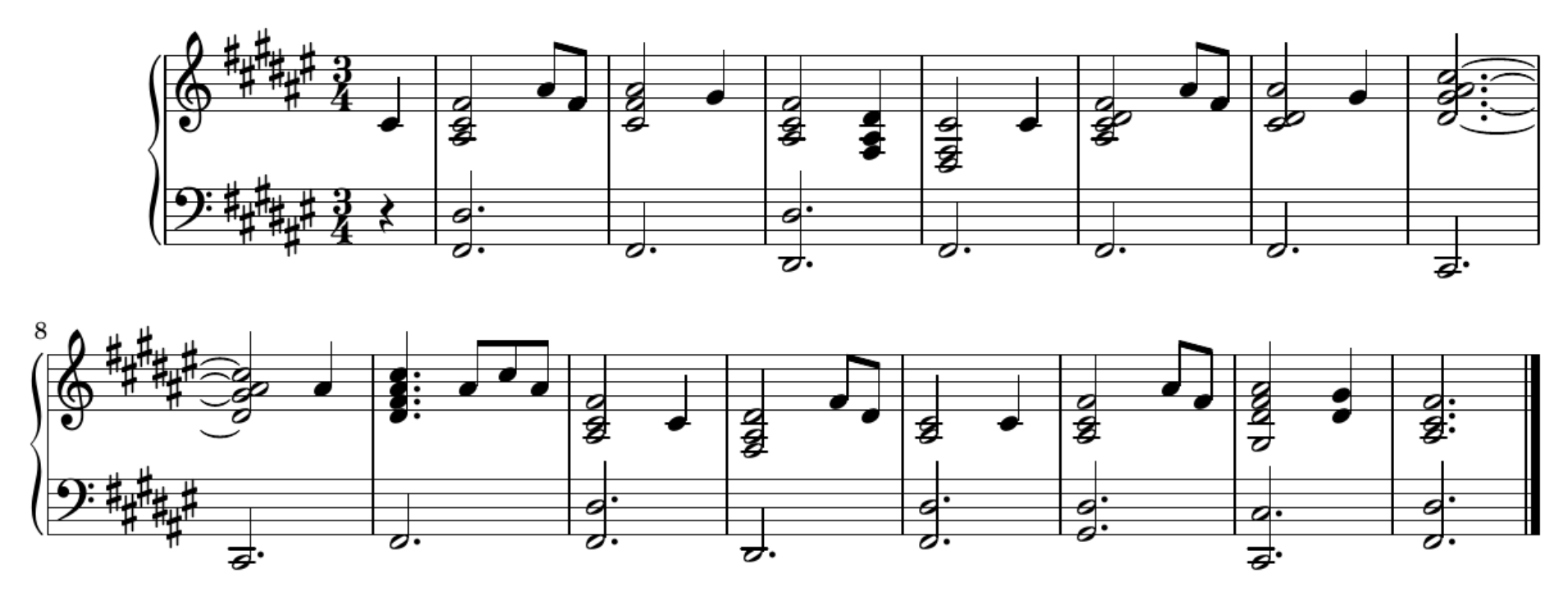}
    \caption{\textit{Amazing Grace} with chords in F${\sharp}$.}
    \label{fig:ag1}
    \end{center}
\end{figure}

\textbf{Remark:} We will generally not distinguish between chords that are related by an inversion.  For instance, if we start with D$\sharp$, then the inversion of F$\sharp$maj6 is also D$\sharp$min7.  It may seem odd that inversions have such different names and we will clarify this when we discuss modes in Section~\ref{sec:modes}. 
Let's apply this to the tune.  The goal is to replace the single melodic line with chords that have the melody as their top note.  The process combines \textit{structure} (pick a 4-chord shape) with \textit{randomization} (omit tones or add the missing note for a 5-chord).  Figure~\ref{fig:ag1} illustrates the result.  One useful technique is ``drop 2'': remove the second-highest note from the right hand and move it to the left, giving the melody more separation (see the F$\sharp$ in bar 2).  In bar 3 we use only two notes of a 4-chord; in bar 15 we use the full 5-chord.
Even restricted to a pentatonic scale, there is considerable freedom to create harmony.  Having concrete chord shapes in mind is liberating for the beginner: one can experiment from the start without worrying about dissonance, yet the number of alternatives is already large enough to sustain exploration.
\subsection{Basic Improvisation}
\begin{figure}[H]
    \begin{center}
    \includegraphics[width=6.5in]{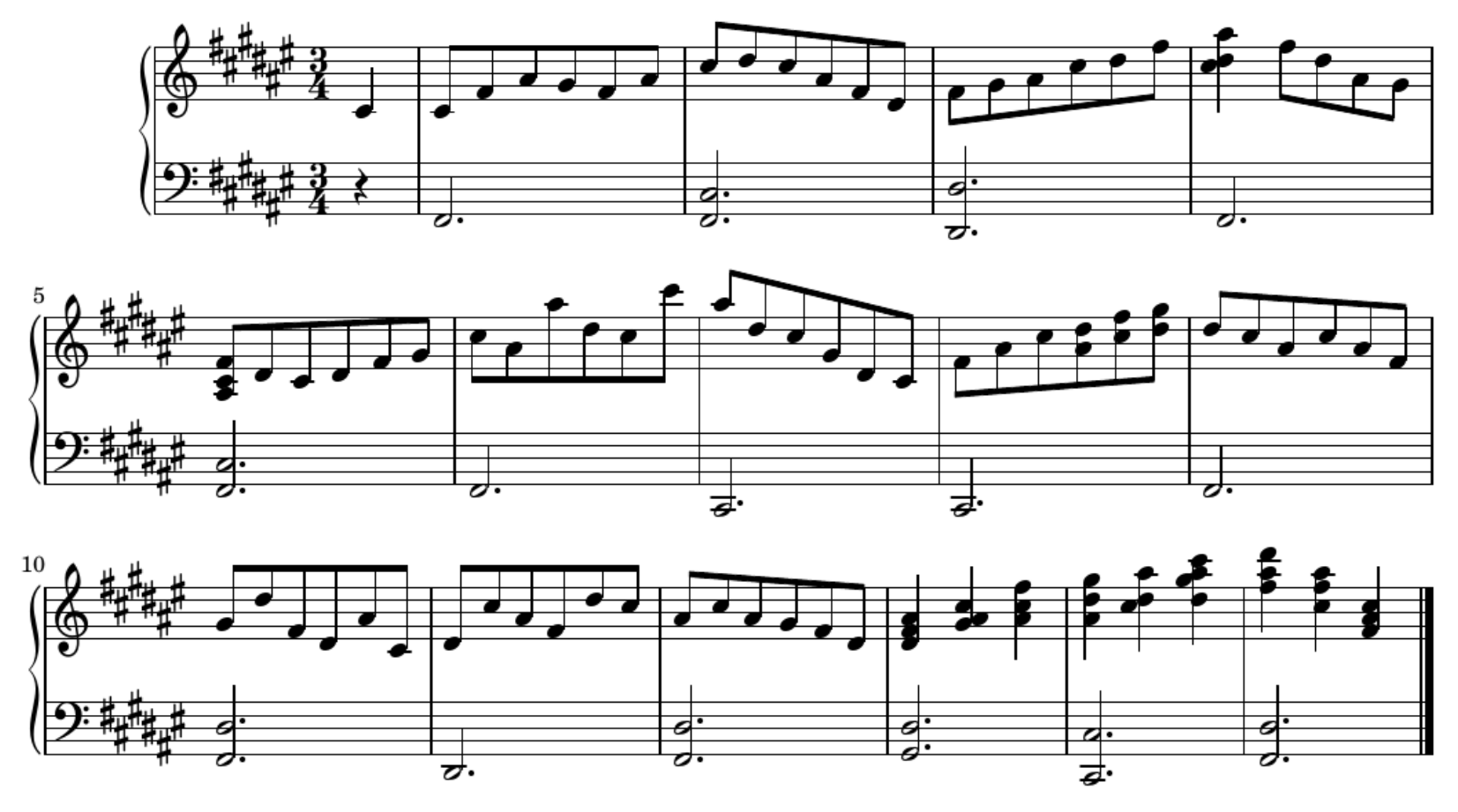}
    \caption{\textit{Amazing Grace} with basic improvisation in F${\sharp}$.}
    \label{fig:ag2}
    \end{center}
\end{figure}

Improvisation can be intimidating for beginners, largely because of the overwhelming number of choices.  Restricting to a pentatonic scale helps: virtually any melodic line sounds reasonable.  Conceptually, we think of improvisation as either adding small variants to the melody (ending on the same tones) or filling gaps in the original line.  Figure~\ref{fig:ag2} gives a simple illustration.  Of course, improvisation is as much art as science---we are limited only by imagination.
\subsection{Playing Outside}
\label{sec:outside}
So far we have worked exclusively with the black keys, illustrating that even this restricted setting has enough complexity for interesting melodic development and chord formation.  Ultimately, however, we want access to the entire keyboard.

\begin{figure}[H]
    \begin{center} 
    \begin{tabular}{cc}
    \includegraphics[width=2.0in]{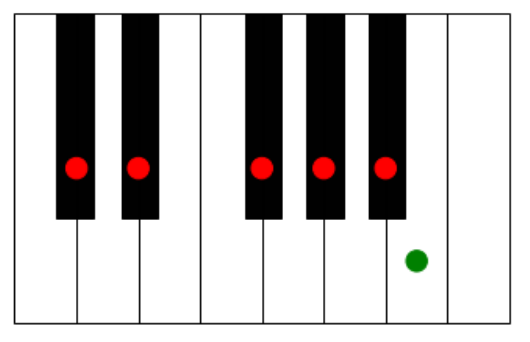} &  \includegraphics[width=2.0in]{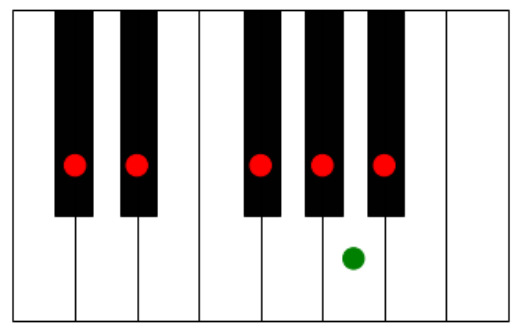} \\
    (a) B with one resolution & (b) ``blue'' note A with two resolutions  \\
    \end{tabular}
    \caption{Adding white notes.}
    \label{fig:whitenotes}
    \end{center}
\end{figure}

\begin{figure}[H]
    \begin{center}
    \includegraphics[width=6.5in]{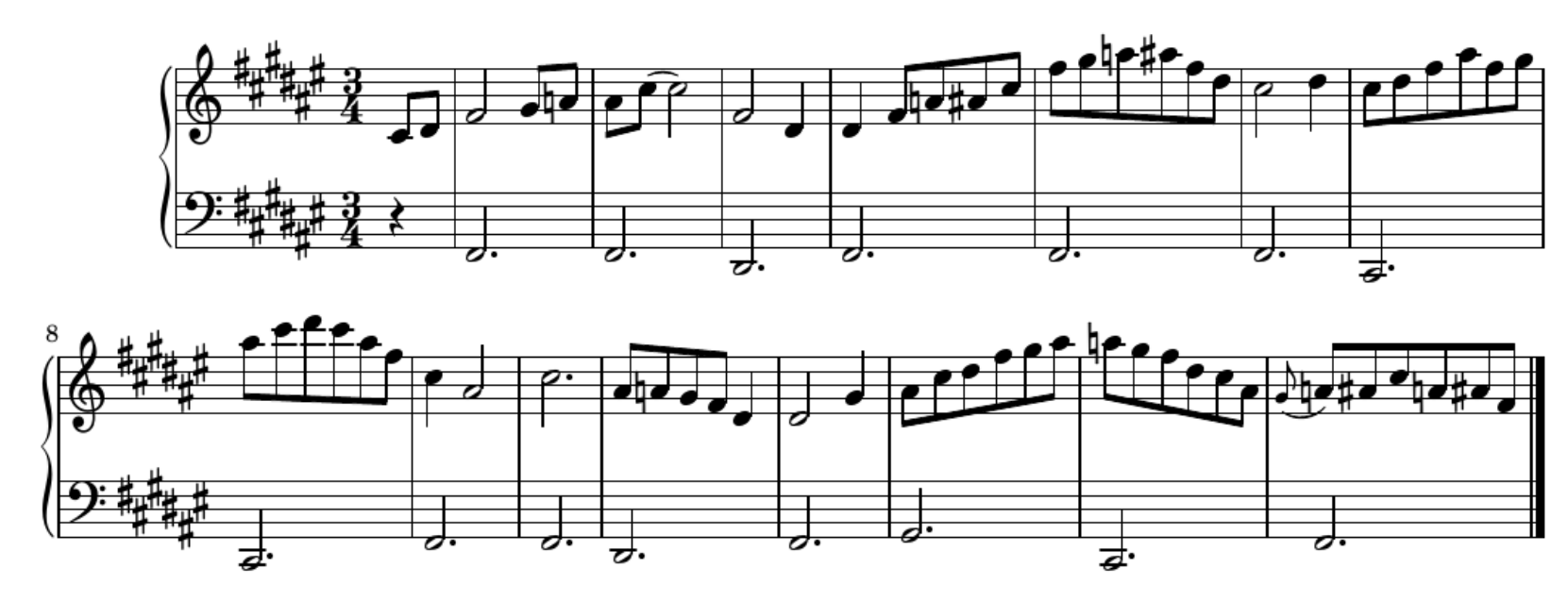}
    \caption{\textit{Amazing Grace} with a ``blue'' note.}
    \label{fig:ag5}
    \end{center}
\end{figure}

\begin{figure}[H]
    \begin{center}
    \includegraphics[width=6.5in]{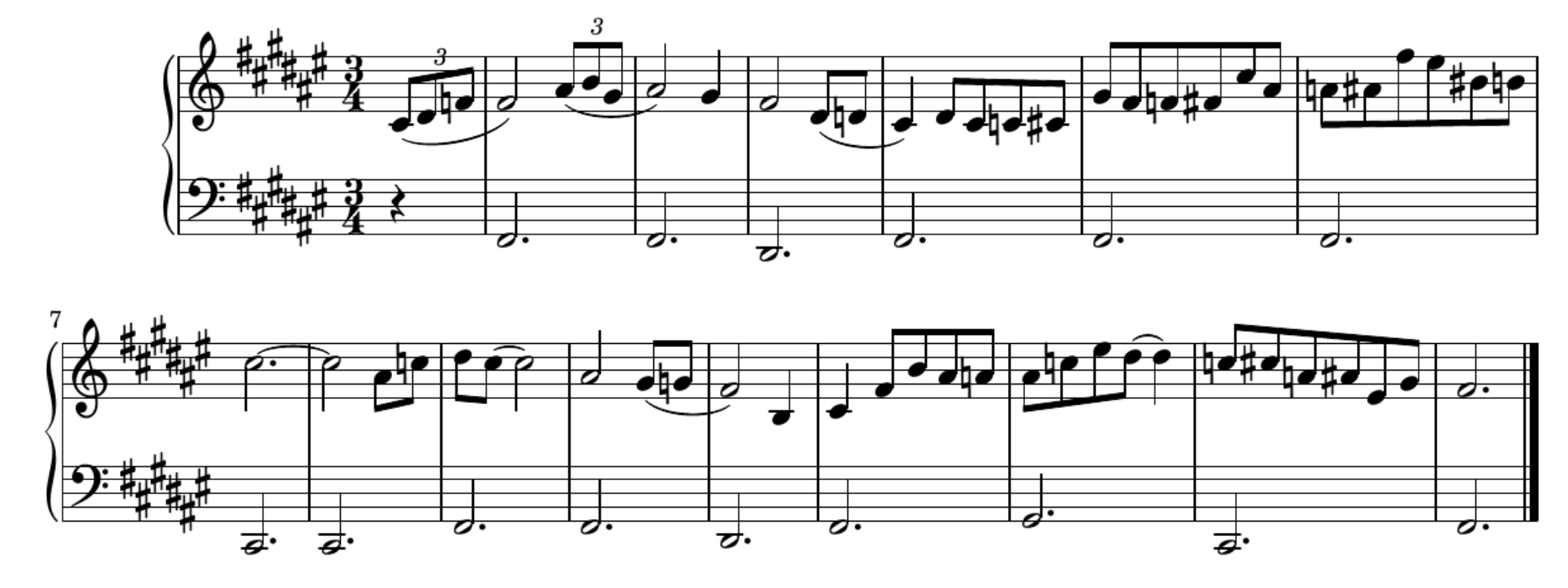}
    \caption{\textit{Amazing Grace} with many ``outside'' tones.}
    \label{fig:ag4}
    \end{center}
\end{figure}

Recall that the pentatonic scale contains no semitone blocks and is maximal: every note outside the scale is a semitone from some note inside it.  For the black keys, this means each white key sounds ``dissonant'' and has one or two resolutions to a neighboring black key.  Figure~\ref{fig:whitenotes} illustrates: in (a), B resolves to A$\sharp$; in (b), A can resolve to either G$\sharp$ or A$\sharp$.  Adding A produces the well-known ``blues'' scale, with A as an external ``blue'' note.  From the pentatonic perspective, such outside notes add tension and complexity but should be used sparingly---seasoning rather than main course.
We illustrate with two examples.  In Figure~\ref{fig:ag5}, we improvise using just the single blue note A.  The resulting blues scale is very common in jazz and is often the main ingredient in professional-sounding improvisational lines.  See Oscar Peterson's ``Georgia On My Mind'' for a great example.
In our second example, we try to incorporate all extraneous white notes into our pentatonic scale.  See Figure~\ref{fig:ag4}. Once again, these are used as a sort of embellishment to add complexity to the melody without deviating too much from the original idea.  
\textbf{Remark:} Notice that some of the white keys have a single black note that is a semitone away (like B), while others have two such notes (like A).  When we discuss harmony, we will show that the notes with a single resolution can be included in a harmony such as a major scale while the others cannot be included into a harmonic structure.  Using the language from Section~\ref{sec:piano}, notes with two resolutions would form a semitone cell if added to the scale.  See Section~\ref{sec:proj1}.

For simplicity, we have focused on right-hand improvisation.  The same ideas apply to the left hand, with one caveat: complex chords in the lower range sound muddy, so we tend to use fewer simultaneous voices there.

\subsection{Harmony, Melody and Microtones}
An important idea implicit in the examples above is that outside notes are \textit{melodic} rather than \textit{harmonic}: they enhance the melodic line without changing the underlying chord structure.  Taken to its logical extreme, there is no reason to stay within the 12-tone system for such embellishments.  Many instruments (including the voice) can interpolate between tones---``bending'' notes to create microtonal melodic elements.  For instance, the true ``blue note'' is arguably a slightly flattened A that falls outside equal temperament.  Indeed, from the pentatonic perspective, the white keys are themselves already microtonal, lying between the black key tones.

Our perspective is that these microtonal phenomena are fundamentally melodic.  A serious study of microtonal \textit{harmony} (scales, modes, and chords) lies beyond the scope of this work.
\subsection{Exercises}
\label{sec:exblack}
Our work is mostly focused on providing a theoretical framework for harmony with an eye towards composition and improvisation.  Although much of this is accessible to the beginner, we have not devoted enough effort to building a method book full of examples and exercises.  In this section we mention a few basic exercises that further explore pentatonic scales:
\begin{itemize}
    \item Practice chord shapes on the black keys.  Get comfortable finding the three basic 4-chords as well as their inversions.  
    \item Try basic improvisation on the black keys.  You can start by modifying the tune as we did above or simply get comfortable moving in 8th notes in one hand while playing a bassline or chords in the other.  
    \item Incorporate some white notes into your improvisation and repeat the previous exercise.
    \item Once comfortable with the black keys, do this for the other 11 pentatonic scales.  It's a good idea to at first pick one scale for each practice session until you are comfortable recognizing that scale. 
    \item Try basic improvisation where you change your pentatonic scale every measure or two.  You can move along the circle of fifths, chromatically, or randomly.  
\end{itemize}
These exercises are direct applications of the concepts in this section, but will take months to master and challenge even an advanced pianist new to improvisation.  Similar exercises apply to other packings (dominant pentatonic, diminished 7th chord, etc.), which provide the basic building blocks of harmony.
\section{White Keys}
\label{sec:white}
\subsection{The Major Harmony/Pentatonic Connection}
\begin{figure}[H]
    \begin{center}
    \includegraphics[width=2.2in]{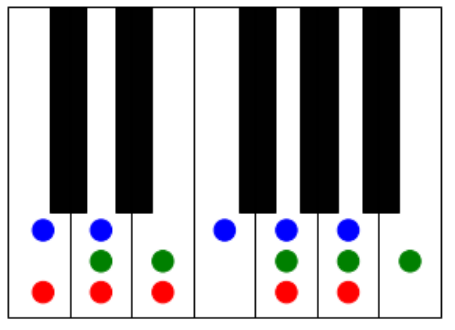}      
    \caption{Pentatonics in major harmony.}
    \label{fig:pentas}
    \end{center}
\end{figure}

Now that we have explored pentatonic scales, we turn to major harmony.  The simplest major scale is C major (the white keys).  On the surface, our focus on black keys seems unrelated---but major harmony is in fact built from pentatonic scales.  As Figure~\ref{fig:pentas} shows, three pentatonic scales (F, C, and G) cover the C major scale.  The pentatonics are elementary building blocks of major harmony.
Everything from the previous section now carries over: we have three pentatonics to choose from when constructing chords or improvising, and we can switch freely between them.  The pentatonic decomposition is an organizational tool that guides thinking about major harmony without being explicitly stated.

\begin{figure}[H]
    \begin{center}
    \begin{tabular}{ccc}  
        \includegraphics[width=1.9in]{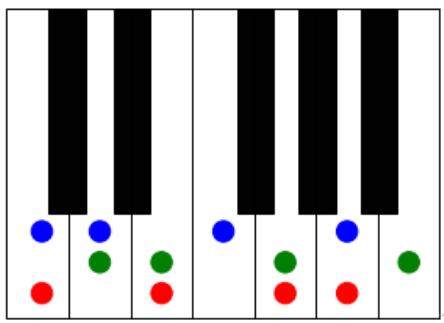} & 
        \includegraphics[width=1.9in]{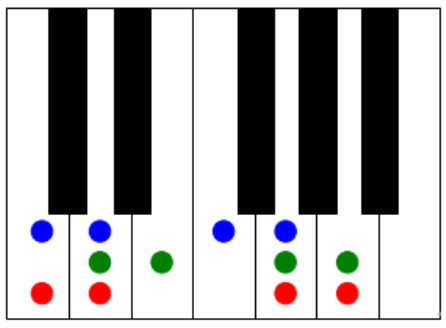} & 
        \includegraphics[width=1.9in]{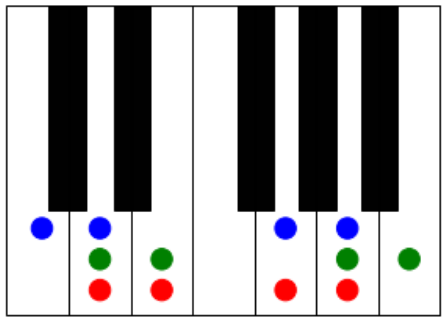} \\  
        (a) Fmaj6 (blue) & (b) Dmin11 (blue) & (c) Amin11 (blue) \\         
        Gmaj6 (green), Cmaj6 (red) & Emin11 (green), Amin11 (red) & Bmin11 (green), Emin11 (red) \\
    \end{tabular}
    \caption{4-chords from pentatonics.}
    \label{fig:pentachords}
    \end{center}
\end{figure}

Let's build 4-chords in major harmony from pentatonics.  Each pentatonic yields three 4-chord shapes (obtained by omitting one note), so three pentatonics contribute nine shapes.  However, some chords arise from more than one pentatonic.  Figure~\ref{fig:pentachords} plots the distinct shapes, color-coded by pentatonic source.
We see that there are in fact only seven distinct chords: 
\begin{center}
    \{Fmaj6, Gmaj6, Cmaj6, Amin11, Dmin11, Emin11, Bmin11\}
\end{center}
These pentatonics have distinct musical characters relative to a choice of bass note.  For example, with F in the bass (the brightest starting point for C major), the F pentatonic sounds the most consonant, the C pentatonic has a recognizable ``jazzy'' quality, and the G pentatonic sounds more ``ethereal.''  Words are inadequate here---we encourage the reader to try this.  The chords inherit these characters, most strongly for the maj6 chords, while the min11 chord is shared across scales.  We return to the relation between chords and bass notes in Section~\ref{sec:modes}.

In summary, the pentatonic decomposition reveals hidden structure in major harmony.  Three pentatonic scales embed into major harmony, giving rise to seven distinct 4-chords: three maj6 chords (at C, F, G) and four min11 chords (at A, B, D, E).
\subsection{Example: \textit{Amazing Grace}}
\label{sec:exwhite}
\begin{figure}[H]
    \begin{center}
    \includegraphics[width=6.5in]{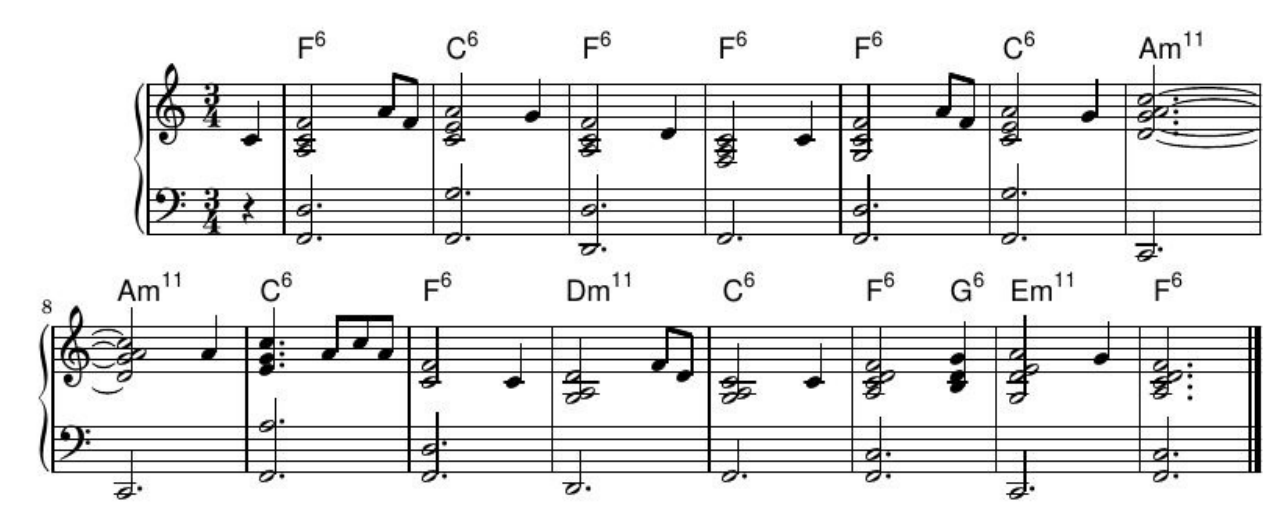}
    \caption{\textit{Amazing Grace} chord arrangement.}
    \label{fig:ag6}
    \end{center}
\end{figure}

We now harmonize \textit{Amazing Grace} again, switching from the F$\sharp$ to the F pentatonic embedded in C major.  With all white notes available, we have seven 4-chords instead of three, allowing richer harmony.  Figure~\ref{fig:ag6} shows the result, with chords annotated.  The pentatonic scales are not explicit in the score, but keeping them in mind is a helpful conceptual tool when arranging or improvising.
\subsection{Major Scale as an Extension of a Pentatonic} 
\label{sec:proj1}
We have seen that major harmony decomposes into pentatonic pieces.  Here is an alternative viewpoint: major harmony as an \textit{extension} of a pentatonic scale.  Consider the F pentatonic inside C major.  The missing tones are B and E, and each has a unique pentatonic neighbor a semitone away: B resolves to C, E resolves to F.  Thus every tone in C major has a unique ``projection'' onto the F pentatonic (tones already in the pentatonic project to themselves).  This lets us reduce any melodic line in C major to a simpler pentatonic shadow.  Crucially, tones outside C major lack a unique projection---F$\sharp$, for example, is a semitone from both F and G.  So major harmony is precisely the maximal extension of a pentatonic that preserves unique projections.

This completes our introduction via pentatonics and major harmony.  In the next section we step back to formalize these ideas and produce further examples.  While the discussion becomes more theoretical, the same practical ideas apply in the broader context.
\newpage
\section{Harmony}
\label{sec:harm}
\subsection{Harmony Basics}
Our definition of harmony builds on the notion of a scale. Since we do not distinguish notes an octave apart, any subset of the twelve tones is a \textbf{scale}.  A natural way to organize scales is by the number of dissonant semitone cells they contain: a \textbf{semitone cell} consists of three consecutive semitones (see Figure~\ref{fig:cell}(b)).

\begin{figure}[H]
    \begin{center}
    \begin{tabular}{ccc}
        \includegraphics[width=2.0in]{./figs/majorkeys-eps-converted-to.pdf} & 
        \includegraphics[width=2.0in]{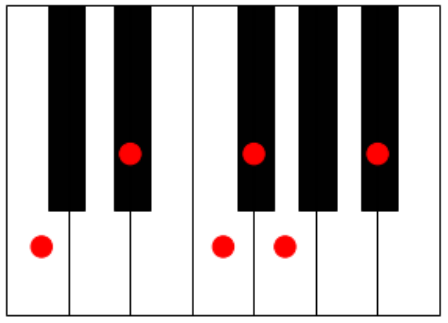} & 
        \includegraphics[width=2.0in]{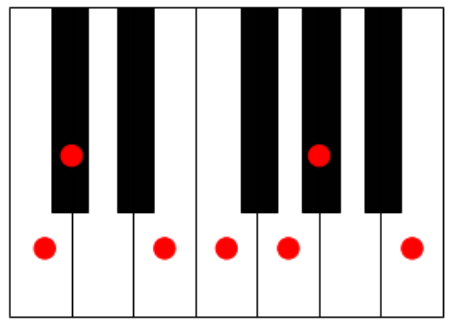} \\ 
        (a) C major & (b) C minor blues & (c) C major double harmonic\\
    \end{tabular}
    \caption{Scale examples.}
    \label{fig:scaleex}
    \end{center}
\end{figure}

Figure~\ref{fig:scaleex} gives examples of scales: (a) is the C major scale that has no cells, (b) is a C minor blues scale that has one cell $\{$F, F$\sharp$, G$\}$, and (c) is a double harmonic C major scale that has one cell $\{$B, C, D$\flat$$\}$.  An extreme example is a \textit{chromatic} scale that encompasses all 12 tones and thus has 12 cells.
While any set of tones can be a scale, we want to restrict attention to scales suitable for multiple voices moving simultaneously---in particular, scales that avoid cells.  Since we can always add tones to a scale without introducing new cells, we are led to:

\textbf{Definition.} A scale is \textbf{semitone cell complete} if the addition of any tone not in the scale increases the number of dissonant semitone cells.

In the examples above, C major and double harmonic are complete; the blues scale is not (we can add D without creating a new cell).  We now define the central concept:

\textbf{Definition.} A \textbf{harmony} is a scale that is complete and has no cells.

Of the examples above, only C major is a harmony: the blues scale is incomplete, and the double harmonic scale contains a cell.  Since any scale can be enlarged until it is complete, we restrict attention to complete scales when classifying.

\textbf{Remark:}  ``Major harmony'' refers not to a single scale but to an equivalence class: the 12 major scales, related by shifting all notes by a fixed amount (e.g., C major shifted by a fifth gives G major).  We call such an equivalence class an \textbf{abstract harmony} and any particular scale a representation of it.  This distinction should be clear from context.

\subsection{Classification of Harmony}
\begin{figure}[H]
    \begin{center}
    \begin{tabular}{ccc}   
        \includegraphics[width=1.8in]{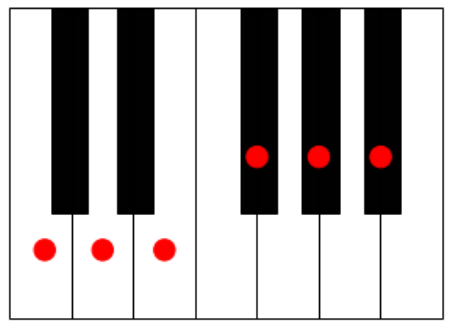} & 
        \includegraphics[width=1.8in]{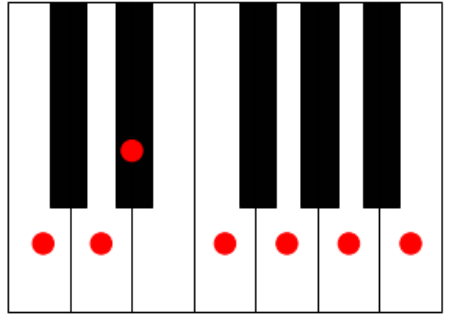} & 
        \includegraphics[width=1.8in]{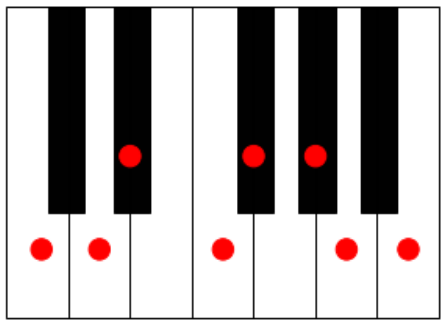} \\  \vspace{.4cm}
        (a) C Whole Tone (\textbf{WTONE})& (b) C Melodic Minor (\textbf{MEL})& (c) C Diminished (\textbf{DIM})\\    
        \includegraphics[width=1.8in]{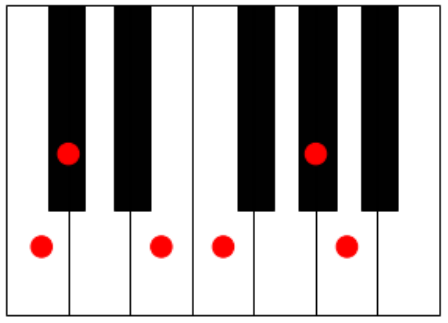} & 
        \includegraphics[width=1.8in]{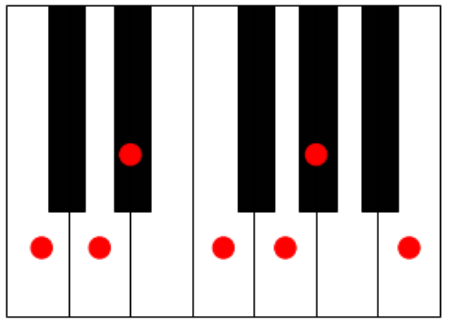} & 
        \includegraphics[width=1.8in]{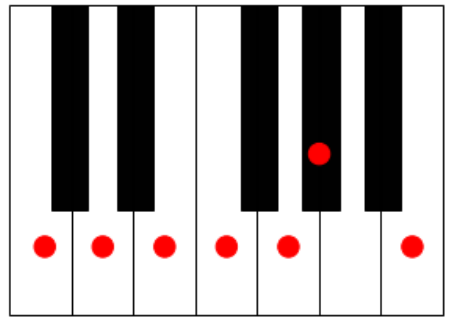} \\ 
        (d) C Symmetric Augmented (\textbf{AUG})& (e) C Harmonic Minor (\textbf{HMIN})& (f) C Harmonic Major (\textbf{HMAJ}) \\
    \end{tabular}
    \caption{Harmony examples.}
    \label{fig:harmexampl}
    \end{center}
\end{figure}

We now classify all harmonies, following Gustavo Casenave.  The classification can be verified by brute force (with computer assistance if desired):

\textbf{Classification of Harmony:}  There are a total of seven (abstract) harmonies.  They are:
\begin{itemize}
    \item Whole Tone (\fancy{WTONE}): The most symmetric harmony: six notes, all a whole tone apart.  There are only two whole tone scales, and together they exhaust the 12 tones.  Debussy made extensive use of this harmony (notably in ``Prelude to the Afternoon of a Faun'').
    \item Major Harmony (\fancy{MAJ}):  The most common harmony in music, with 12 representations (one per tone).  Western notation is built around it: notes outside the chosen major scale are ``accidental.''  Adding any accidental introduces a cell---this observation is the heart of our approach.
    \item Melodic Minor (\fancy{MEL}):  Probably the most common scale after major harmony.  It has 12 representations and is related to major harmony by flatting the major third to a minor third.  
    \item Diminished Harmony (\fancy{DIM}): An 8-note scale obtained by alternating semitone and whole-tone steps.  This yields only 3 distinct representations.  Equivalently, any two of the three diminished 7th chords---$(B, D, F, A{\flat})$, $(G, B{\flat}, D{\flat}, E)$, and $(C, E{\flat}, G{\flat}, A)$---combine to form a diminished scale.  Note that no diminished chord fits inside major harmony; diminished harmony exists to accommodate them.
    \item Symmetric Augmented (\fancy{AUG}): Analogous to diminished but built on augmented triads.  There are 4 augmented chords: $(C, E, G{\sharp})$, $(B, D{\sharp}, G)$, $(A{\sharp}, D, F{\sharp})$, and $(A, C{\sharp}, F)$.  Combining two that are a semitone apart gives an augmented scale (equivalently, alternate semitone and minor-third steps).  There are 4 representations.  Used less commonly---one example is Liszt's ``Faust Symphony.''
    \item Harmonic Minor (\fancy{HMIN}): Obtained from melodic minor by flatting the 6th note.  There are 12 representations of this harmony. 
    \item Harmonic Major (\fancy{HMAJ}): Obtained from major harmony by flatting the 6th note.  There are 12 representations of this harmony.  This harmony is used quite rarely in composition.
\end{itemize}
Figure~\ref{fig:harmexampl} shows representatives of each of these harmonies.
 Much of this work is devoted to studying these seven harmonies: their chord formations (voicings), melodic and improvisational possibilities, and their implicit role in composition.

\begin{figure}[H]
    \begin{center}
    \begin{tabular}{ccc}   
        \includegraphics[width=0.7in]{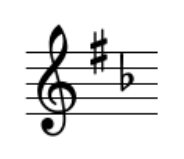} & 
        \includegraphics[width=0.7in]{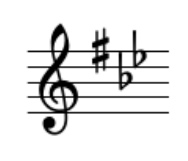} & 
        \includegraphics[width=0.7in]{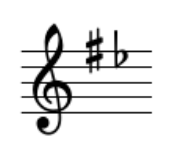} \\  
        (a) G Melodic Minor & (b)  G Harmonic Minor & (c)  G Harmonic Major \\           
    \end{tabular}
    \caption{Staff notation for non-major harmony.}
    \label{fig:notation}
    \end{center}
\end{figure}

 \textbf{Harmony and Notation:} There is no standard notation for non-major harmony.  Some scales (e.g., melodic minor) can be notated by combining sharps and flats on the staff (see Figure~\ref{fig:notation}); others, like whole tone, do not fit easily.  Jazz lead sheets address this by using complex chord symbols that implicitly indicate the underlying harmony.
We conclude this section with a simple invariant for distinguishing harmonies: the maximal number of consecutive notes a harmony shares with a whole tone scale.  Table~\ref{tbl:relwt} displays this overlap.  \textbf{MEL} has the most in common with \fancy{WTONE}, while \fancy{AUG} has the least; the two harmonic scales both share three notes.  From a practical standpoint, this begins to address the question ``how can we identify the underlying harmony?''  For instance, a passage containing four sequential whole tones can only belong to \fancy{WTONE} or \fancy{MEL}---the simplest example of recognizing harmonic structure by ear.
\begin{table}[H]
    \begin{center}
    \begin{tabular}{l|l|l|l|l|l|l|l}
    harmony & \fancy{WTONE} & \fancy{MEL} & \fancy{MAJ} & \fancy{HMIN} & \fancy{HMAJ} & \fancy{DIM} & \fancy{AUG} \\
    \hline
    overlap with \fancy{WTONE}  & 6  & 5  & 4  & 3  & 3  & 2  & 1 \\
    \end{tabular}
    \end{center}
    \caption{Harmony by overlap with the whole tone scale.}
    \label{tbl:relwt}
\end{table}
\section{Packings}
\label{sec:packings}
\subsection{Initial Classification}
We now turn from cells to blocks, developing a parallel classification.  Recall that a dissonant semitone block consists of two notes a semitone apart.  We reserve the term ``scale'' for sets of notes organized by cells and use \textbf{packing} for sets organized by blocks (though both are simply subsets of the 12 tones).

\textbf{Definition.} A packing is (semitone block) \textbf{complete} if adding any tone not in the packing increases the number of dissonant (semitone) blocks.

\begin{figure}[H]
    \begin{center}
    \begin{tabular}{cc}   
        \includegraphics[width=2.0in]{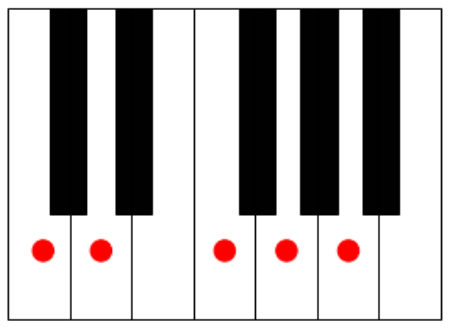} & 
        \includegraphics[width=2.0in]{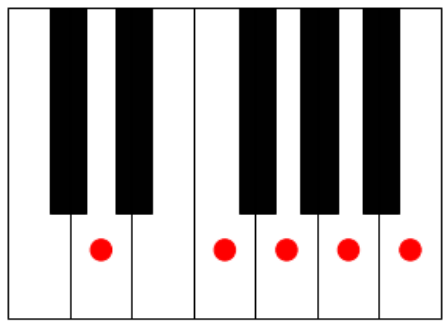} \\  \vspace{.4cm}
        (a) F pentatonic  (\textbf{penta}) & (b) G dominant pentatonic  (\textbf{dpenta}) \\    
        \includegraphics[width=2.0in]{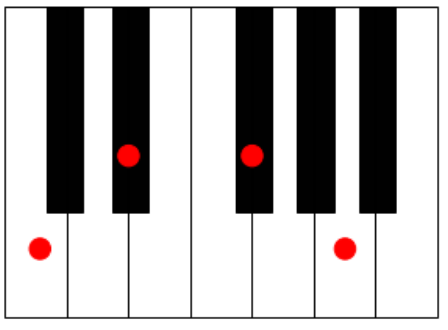} & 
        \includegraphics[width=2.0in]{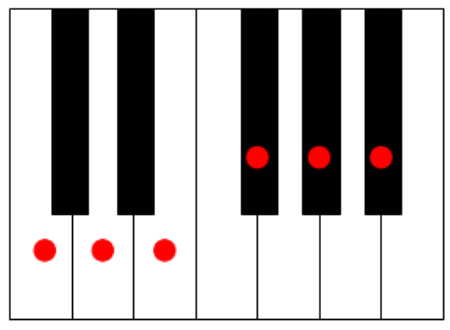} \\ 
        (c) C diminished  (\textbf{dim})& (d) C whole tone  (\textbf{wtone}) \\
    \end{tabular}
    \caption{Complete packings with no dissonant blocks.}
    \label{fig:pack0}
    \end{center}
\end{figure}

The name ``packing'' reflects the goal: pack as many notes as possible without adjacent semitones forming blocks.  In analogy with harmony:

\textbf{Classification of complete packings with no blocks:}  Up to shifts, there are a total of four complete packings that have no blocks.  They are:
\begin{itemize}
\item Pentatonic packing (\textbf{penta}): The pentatonic scale discussed in Section~\ref{sec:black}.  See Figure~\ref{fig:pack0}(a).
\item Dominant pentatonic packing (\textbf{dpenta}):  Obtained from \textbf{penta} by raising the 5th degree, introducing a tritone---particularly useful for dominant chords.  See Figure~\ref{fig:pack0}(b).
\item Diminished packing (\textbf{dim}):  Simply the diminished 7th chord (only 3 distinct chords up to shifts).  See Figure~\ref{fig:pack0}(c).
\item Whole tone packing (\textbf{wtone}):  The whole tone scale, now viewed as a packing---a rare case where a packing and a harmony coincide.  See Figure~\ref{fig:pack0}(d).
\end{itemize}
\subsection{Irreducible Packings}
The classification of block-free packings is straightforward.  We now consider complete packings that contain some blocks.  Many useful chords (e.g., the major 7th) contain dissonant blocks, so allowing some dissonance is necessary for a general theory of chord formation.  There are 132 complete packings in total---far too many.  We identify a much smaller generating set.

Given a complete packing $P$, we can add missing tones to obtain larger complete packings.  A packing obtained this way is \textbf{reducible}---it is derived from a smaller one.  The fundamental building blocks are those that cannot be so derived:

\textbf{Definition:}  A complete packing is \textbf{irreducible} if no subset of it is complete.

Irreducible packings generate all others, so we focus on them.  Note that all block-free packings from the previous classification are automatically irreducible: if a complete block-free packing $P$ contained a smaller complete subset $P'$, then adding back the missing tones to recover $P$ would introduce a block for each, contradicting the assumption that $P$ is block-free.

\begin{figure}[H]
    \begin{center}
    \begin{tabular}{ccc}   
        \includegraphics[width=1.8in]{./figs/aug-eps-converted-to.pdf} & 
        \includegraphics[width=1.8in]{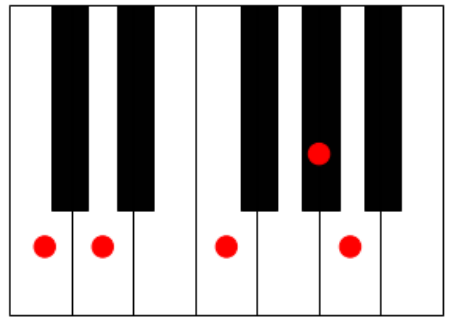} & 
        \includegraphics[width=1.8in]{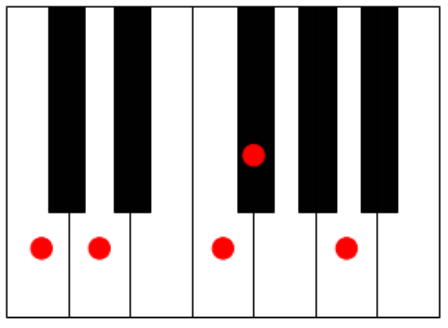} \\ 
        (a) C augmented (\textbf{aug}) & (b) F harmonic major  (\textbf{hmaj}) & (c) D harmonic minor  (\textbf{hmin}) \\           
    \end{tabular}
    \caption{Irreducible packings with dissonant blocks.}
    \label{fig:pack1}
    \end{center}
\end{figure}

We now classify all irreducible packings:

\textbf{Classification of Irreducible Packings:}  There are a total of seven irreducible packings, up to shifts.  They are given by the 4 block-free packings in Figure~\ref{fig:pack0} and the 3 packings with blocks in Figure~\ref{fig:pack1}.
The harmonic major (minor) packing is obtained from a pentatonic by raising (lowering) the 2nd degree.  The augmented packing has 3 blocks; the harmonic major/minor have one each.  Note that the augmented packing coincides with the augmented harmony \textbf{AUG}---so together with \textbf{wtone}, there are two cases where a harmony and an irreducible packing are the same object.  As with the black keys of Section~\ref{sec:black}, all irreducible packings can serve as the basis for improvisation and chord formation.  We now turn to the fundamental relation between packings and harmony.

\textbf{Remark:}  While most irreducible packings we described are well-known, \textbf{hmaj}/\textbf{hmin} seem more obscure and it was not clear whether they are commonly used in practice.  However, we recently found this blues practice session that demonstrates the utility of these two scales over the standard blues progression.  See \url{https://www.youtube.com/watch?v=mviSr-aggEg} for the video and attached PDF.

\textbf{Remark:}  We have introduced the concept of irreducibility, as applied to packings.  One can formulate a similar notion for scales and classify irreducible scales that may have dissonant cells.  We explore this in Appendix~\ref{sec:appG}.
\section{The Harmony-Packing Duality}
\label{sec:dual}
\subsection{Statement of the Correspondence}
As suggested in the previous sections, there is a correspondence between harmony and packings.  We now make this explicit:

\textbf{Harmony-Packing Duality (or Block-Cell Duality):}  There is a one-to-one correspondence between the seven harmonies and the seven irreducible packings:
\begin{center}
\textbf{MAJ} $\Leftrightarrow$  \textbf{penta}  \\
\textbf{MEL} $\Leftrightarrow$  \textbf{dpenta} \\ 
\textbf{WTONE} $\Leftrightarrow$  \textbf{wtone}  \\
\textbf{DIM} $\Leftrightarrow$  \textbf{dim} \\
\textbf{AUG} $\Leftrightarrow$  \textbf{aug}  \\
\textbf{HMIN} $\Leftrightarrow$  \textbf{hmin}  \\
\textbf{HMAJ} $\Leftrightarrow$  \textbf{hmaj}  \\
\end{center}
Moreover, given a harmony or a packing, the corresponding dual object is obtained by taking the complementary tones.

This generalizes the black-key/white-key duality from Section~\ref{sec:piano}: the major scale (defined by avoiding cells) has as its complement a pentatonic scale (defined by avoiding blocks).  The duality result says this complementarity holds for all seven harmonies and their packings.  Two cases are exceptional: \textbf{AUG}/\textbf{aug} and \textbf{WTONE}/\textbf{wtone} are complementary to themselves.  Such objects are called \textbf{self-dual}.
\subsection{Proof of Duality}
Given the classifications, verifying duality is a matter of checking complements in all seven cases.  However, this does not explain \textit{why} two seemingly different constraints are related.  Here we give a simple argument that establishes the correspondence \textit{a priori}, without reference to the explicit classifications.  The proof requires nothing beyond basic set theory.

To prove the correspondence, we will prove an auxiliary statement that will allow us to explicitly see the relation between blocks and cells:

\textbf{Auxiliary Result:} By taking complements, there is a one-to-one correspondence between scales with cells and incomplete packings.

To prove this, let $S$ be a scale with complementary packing $S'$.  Suppose $S$ has a cell, say $\{$E, F, F$\sharp$$\}$ (see Figure~\ref{fig:cell}).  None of these tones belong to $S'$, so we can add F to $S'$ without creating a block (its neighbors E and F$\sharp$ are absent).  Hence $S'$ is incomplete.

Conversely, suppose $S'$ is incomplete: some tone, say F, can be added without creating blocks.  Then neither E nor F$\sharp$ belongs to $S'$ (either would form a block with F), so all three belong to $S$---that is, $S$ contains the cell $\{$E, F, F$\sharp$$\}$.

We now prove the duality.  By contrapositive, the auxiliary result says: scales with no cells correspond to complete packings.  We show that if $S$ is a harmony, its complement $S'$ is irreducible.  Since $S$ has no cells, $S'$ is complete.  Take any proper subset $T' \subset S'$; its complement $T \supset S$ is strictly larger than $S$.  Since $S$ is a harmony (maximal cell-free), $T$ must contain a cell.  By the auxiliary result, $T'$ is not complete.  Since no proper subset of $S'$ is complete, $S'$ is irreducible.

Conversely, given an irreducible packing $S'$, we show its complement $S$ is a harmony.  Since $S'$ is complete, the auxiliary result gives that $S$ has no cells.  It remains to show $S$ is complete.  Adding any tone to $S$ produces a larger scale $T$ whose complement $T' \subset S'$ is a proper subset.  Since $S'$ is irreducible, $T'$ is not complete, so by the auxiliary result $T$ has a cell.  Thus $S$ is complete, hence a harmony.

\textbf{Remark:}  The key insight is that harmonies are \textit{maximal} cell-free scales while irreducible packings are \textit{minimal} complete packings.  Complementation reverses inclusions, swapping maximal and minimal---the rest follows.

\subsection{Relating Harmony and Packings}
What is the practical use of duality?  Recall that we started with the black-key/white-key example: the complement of a harmony (the packing) turned out to be key to understanding the harmony itself, since pentatonic scales embed inside major harmony and provide the building blocks for chords and improvisation.  The same idea applies more generally.

\begin{figure}[H]
    \begin{center}
    \begin{tabular}{ccc}   
        \includegraphics[width=1.8in]{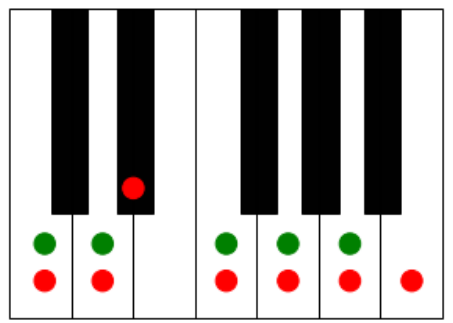} & 
        \includegraphics[width=1.8in]{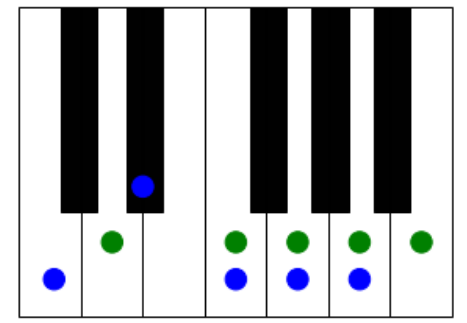} & 
        \includegraphics[width=1.8in]{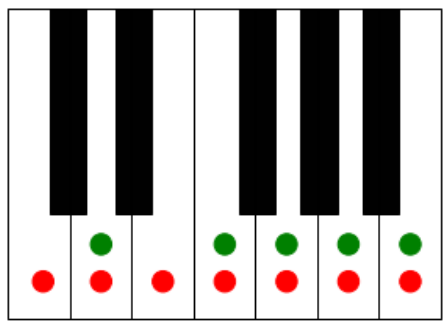} \\
        (a) \textbf{penta} (green) & (b) \textbf{dpenta} (green and blue)  & (c) \textbf{dpenta} (green)  \\
        inside \textbf{MEL}        & covering \textbf{MEL}     & inside \textbf{MAJ}\\
    \end{tabular}
    \caption{Embedding \textbf{penta} and \textbf{dpenta} in harmony.}
    \label{fig:melpent}
    \end{center}
\end{figure}

Table~\ref{tbl:harmpackemb} summarizes the number of embeddings of each packing into each harmony.  For concreteness, we fix harmonies starting in C and list the starting tones of the embedded packings.  For instance, there are three embeddings of $\textbf{penta}$ into C $\textbf{MAJ}$, at tones C, F, and G (as discussed in Section~\ref{sec:white}).
\begin{table}[H]
    \begin{center}
    \begin{tabular}{l|l|l|l|l|l|l|l}
          & \fancy{wtone} &  \fancy{aug} & \fancy{penta} & \fancy{dpenta} & \fancy{dim} & \fancy{hmin} & \fancy{hmaj}  \\
    \hline
    \textbf{C WTONE} & C & & & & & &  \\
    \hline
    \textbf{C AUG} &  & C & & & & &  \\
    \hline
    \textbf{C MAJ} &  & & C,F,G & G & & &  \\
    \hline
    \textbf{C MEL} &  & & F & F,G & & &  \\
    \hline
    \textbf{C DIM} &  & & & & C,D & D,F,A$\flat$,B & D,F,A$\flat$,B \\ 
    \hline
    \textbf{C HMIN} &  & & & & D &  & A$\flat$ \\
    \hline
    \textbf{C HMAJ} &  & & & & D & G &  \\
    \end{tabular}
    \end{center}
    \caption{Embedding packings into harmonies starting in C.}
    \label{tbl:harmpackemb}
\end{table}
The table has nonzero diagonal entries: each packing embeds into its dual harmony.  For instance, \textbf{penta} embeds 3 times into \textbf{MAJ} and once into \textbf{MEL} (see Figure~\ref{fig:pentas} and Figure~\ref{fig:melpent}(a)), and into no other harmony---illustrating the close relationship between the pentatonic scale and major harmony.  The shared embedding of \textbf{penta} also highlights the kinship between \textbf{MAJ} and \textbf{MEL}.

\begin{figure}[H]
    \begin{center}
    \begin{tabular}{ccc}   
        \includegraphics[width=1.8in]{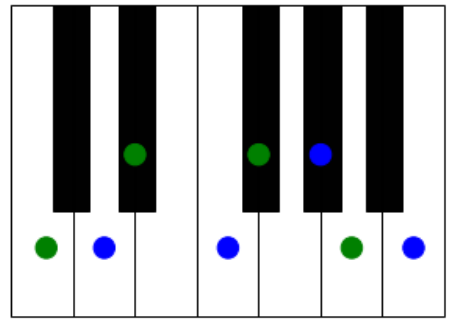} & 
        \includegraphics[width=1.8in]{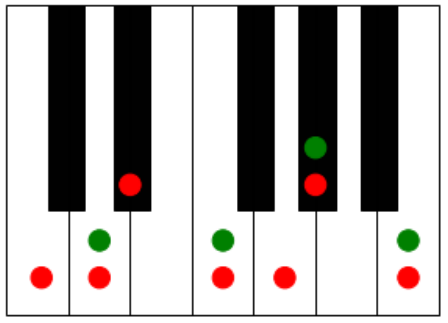} & 
        \includegraphics[width=1.8in]{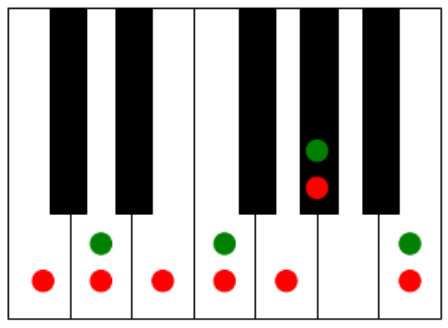} \\
        (a) \textbf{dim} (green and blue) & (b)  \textbf{dim} (green) & (c) \textbf{dim} (green) \\           
        covering C \textbf{DIM} & inside C \textbf{HMIN} & inside C \textbf{HMAJ}  \\
    \end{tabular}
    \caption{Embedding \textbf{dim} in harmony.}
    \label{fig:dimdim}
    \end{center}
\end{figure}

\begin{figure}[H]
    \begin{center}
    \begin{tabular}{cc}   
        \includegraphics[width=2.0in]{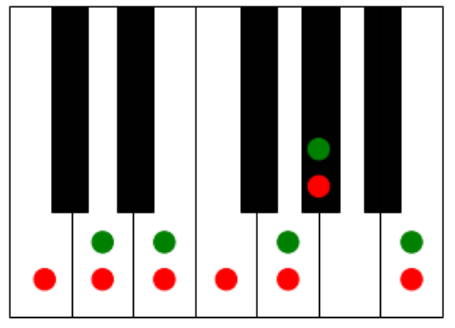} & 
        \includegraphics[width=2.0in]{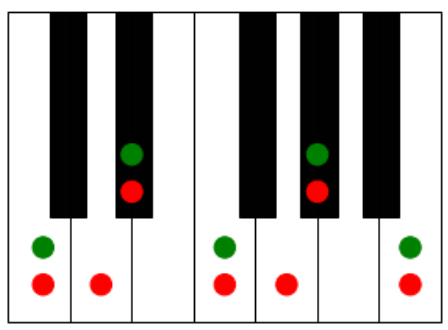} \\ 
        (a) G \textbf{hmin} (green) inside C \textbf{HMAJ} & (b) A$\flat$ \textbf{hmaj} (green) inside C \textbf{HMIN}\\           
    \end{tabular}
    \caption{Embedding \textbf{hmin}/\textbf{hmaj} in harmony.}
    \label{fig:hembed}
    \end{center}
\end{figure}

\begin{figure}[H]
    \begin{center}
    \begin{tabular}{cc}   
        \includegraphics[width=2.0in]{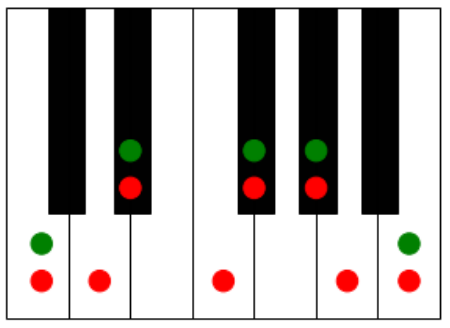} & 
        \includegraphics[width=2.0in]{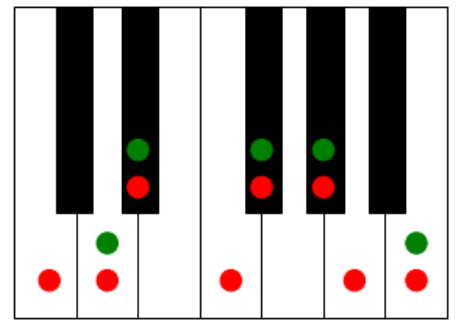} \\ 
        (a) B \textbf{hmin} (green) inside C \textbf{DIM} & (b) B \textbf{hmaj} (green) inside C \textbf{DIM}\\           
    \end{tabular}
    \caption{Embedding \textbf{hmin}/\textbf{hmaj} in harmony (cont).}
    \label{fig:hembed2}
    \end{center}
\end{figure}

Similarly, $\textbf{dpenta}$ has two embeddings into its dual $\textbf{MEL}$ that cover the entire scale (Figure~\ref{fig:melpent}(b)), and one embedding into $\textbf{MAJ}$ (Figure~\ref{fig:melpent}(c))---again confirming the close relationship between $\textbf{MEL}$ and $\textbf{MAJ}$.
The self-dual cases $\textbf{aug}$ and $\textbf{wtone}$ embed only into their own harmonies, covering them perfectly.  The diminished packing $\textbf{dim}$ has two embeddings covering $\textbf{DIM}$ (recall that $\textbf{DIM}$ is built from two adjacent diminished chords; see Figure~\ref{fig:dimdim}(a)) and a unique embedding into each of $\textbf{HMIN}$ and $\textbf{HMAJ}$ ((b) and (c)).
The case of $\textbf{hmin}$/$\textbf{hmaj}$ is exceptional because the embeddings switch:  $\textbf{hmin}$ embeds into $\textbf{HMAJ}$, while $\textbf{hmaj}$ embeds into $\textbf{HMIN}$.  See Figure~\ref{fig:hembed} for an illustration. Finally, both $\textbf{hmin}$ and $\textbf{hmaj}$  embed inside $\textbf{DIM}$.  Due to the symmetry of that harmony, we can always shift by a minor third and still land in the scale.  For this reason, there are in fact 4 distinct embeddings of $\textbf{hmin}$ and $\textbf{hmaj}$ inside $\textbf{DIM}$. See Figure~\ref{fig:hembed2} for an illustration of the packings starting at B.  There are three more such embeddings starting at D, F and A$\flat$.
In all cases except $\textbf{HMIN}$/$\textbf{HMAJ}$, the dual packing embeds enough times to cover the entire harmony.  The harmonic cases are exceptional: the embeddings \textit{switch} ($\textbf{hmin}$ embeds into $\textbf{HMAJ}$, not $\textbf{HMIN}$, and vice versa), and even with the help of $\textbf{dim}$ one tone remains uncovered.  We return to this in Section~\ref{sec:voicings}.

Overall, the table demonstrates that a harmony and its dual packing are intimately related despite being complementary.  Finding packings inside a harmony unlocks improvisational and chordal structure, just as in Sections~\ref{sec:black} and~\ref{sec:white}.  The off-diagonal entries point to more exotic embeddings that enrich the available harmonic vocabulary.
\subsection{Projecting Harmony to a Packing}
We can take this further by viewing harmony as an extension of a packing ``shadow'' contained within it.  In Section~\ref{sec:exwhite} we saw that every tone in major harmony has a unique projection to an embedded pentatonic.  This generalizes:

\textbf{Projecting a Harmony to a Packing:}  Given a harmony $H$ and a packing $P\subset H$ contained in that harmony, there is a unique projection of every tone in $H$ to $P$.  In other words, every tone in $H$ is either already in $P$ or has a \textit{unique} neighbor a semitone away that is in $P$.

To see this, take any note F in $H$ that is not in $P$.  At least one of its neighbors (E or F$\sharp$) must be in $P$; otherwise $\{$E, F, F$\sharp$$\}$ would be a cell disjoint from $P$, and we could add F to $P$ without creating blocks, contradicting completeness.  Moreover, only one neighbor can be in $P$, since both would give a cell in $H$.  So F has a unique semitone neighbor in $P$.

A consequence of this observation is that any melodic line in $H$ has a unique projection to $P$.  This provides a simplification of the melodic line by creating a ``skeletal'' version that removes some of the tension.  This can be used in helping create a harmonic profile of a piece of music. 
\subsection{Blue Notes}
Projections also shed light on ``blue'' notes.  There is considerable ambiguity around the definition of a blues scale; we offer a perspective based on our framework.  Consider the major pentatonic.  As noted in Section~\ref{sec:outside}, its complement consists of two types: tones that can extend the pentatonic into a harmony (unique projection) and tones that cannot (they form a cell, having \textit{two} resolutions to the pentatonic).  The latter are the ``blue'' notes---dissonant by nature and not part of any chord structure.

\begin{figure}[H]
    \begin{center}
    \begin{tabular}{cc}   
        \includegraphics[width=2.0in]{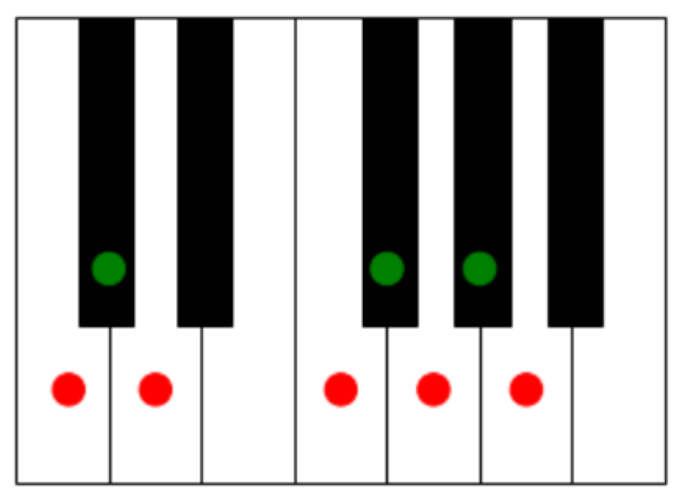} & 
        \includegraphics[width=2.0in]{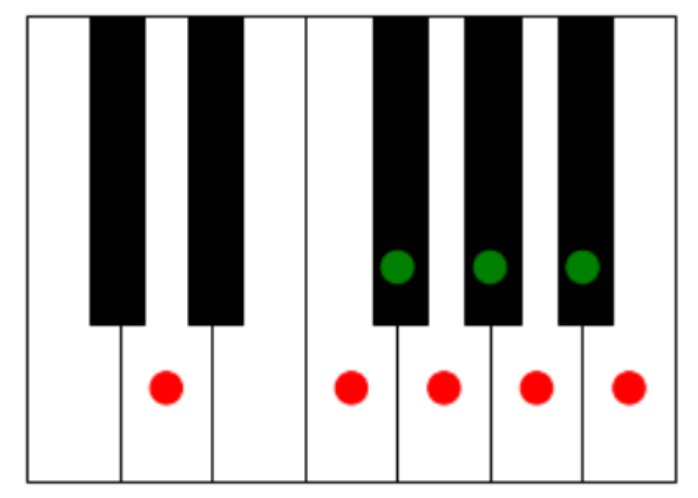} \\ 
        (a) \textbf{penta} & (b) \textbf{dpenta} \\     
        \includegraphics[width=2.0in]{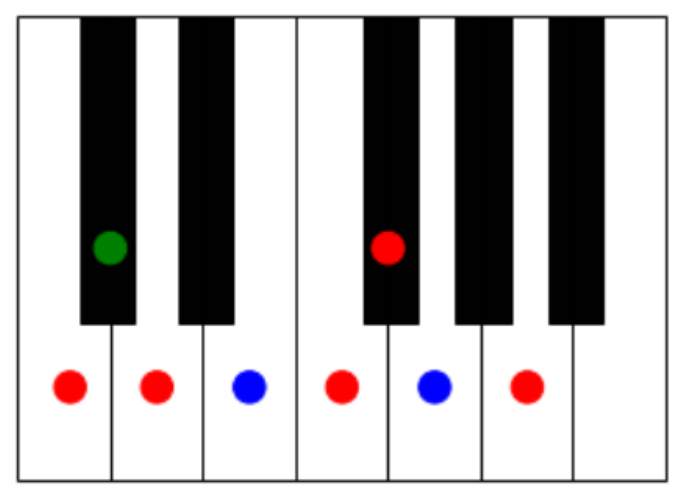} & 
        \includegraphics[width=2.0in]{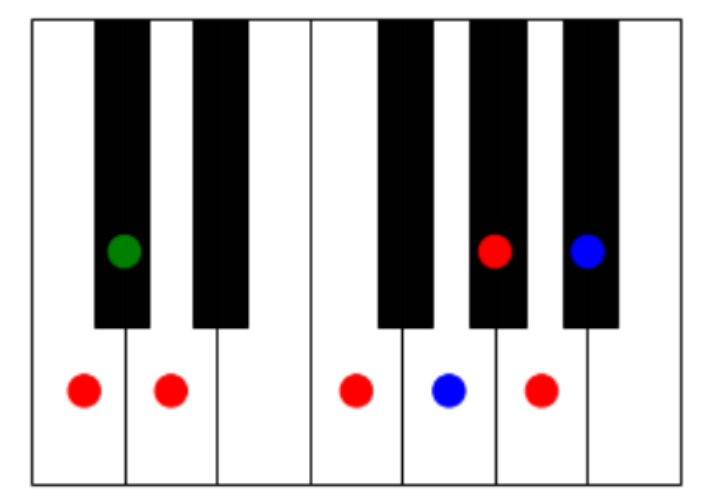} \\ 
        (c) \textbf{hmin}  & (d)  \textbf{hmaj}  \\         
    \end{tabular}
    \caption{Blue note variants.}
    \label{fig:blue}
    \end{center}
\end{figure}

For $\textbf{penta}$ and $\textbf{dpenta}$, there are exactly three such blue notes each (Figure~\ref{fig:blue}(a),(b)).  The packing $\textbf{dim}$ has no blue notes at all.  An interesting case arises with $\textbf{hmin}$ and $\textbf{hmaj}$: among the tones that cannot extend to a harmony, some still have a unique projection (blue) while others do not (green); see Figure~\ref{fig:blue}(c),(d).  For $\textbf{wtone}$, all complementary tones satisfy the blue-note property, while for $\textbf{aug}$ the complementary tones form cells but still project uniquely.
We do not insist on a single definition of ``blue'' note, but note that for the most common packings ($\textbf{penta}$ and $\textbf{dpenta}$) the two characterizations coincide.  Awareness of these notes provides a systematic way to incorporate dissonance into improvisation.
\subsection{Harmony and Piano Layouts}
\begin{figure}[H]
    \begin{center} 
    \begin{tabular}{cc}
    \includegraphics[width=3.2in]{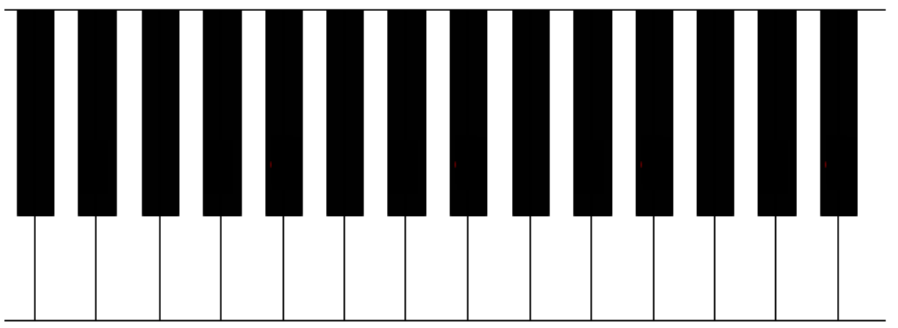} &  \includegraphics[width=3.11in]{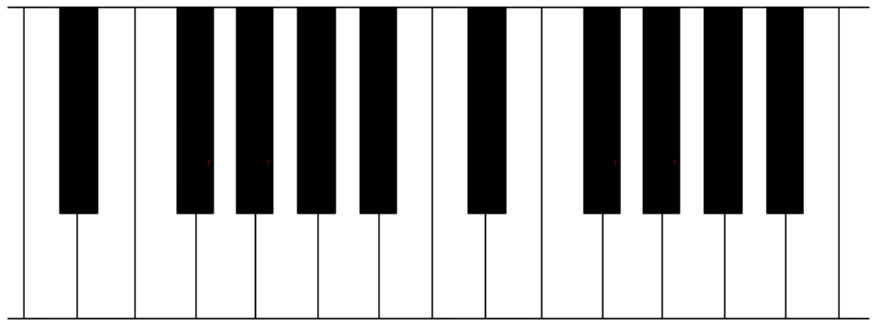} \\ \vspace{.4cm}
    (a) Whole Tone & (b) Melodic Minor \\
     & \\
    \includegraphics[width=3.2in]{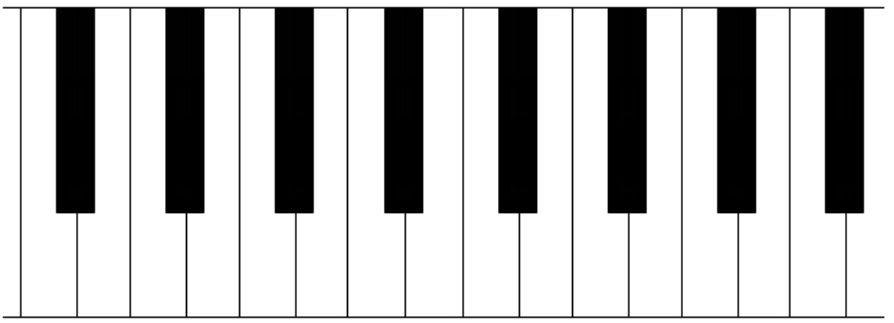} &  \includegraphics[width=3.11in]{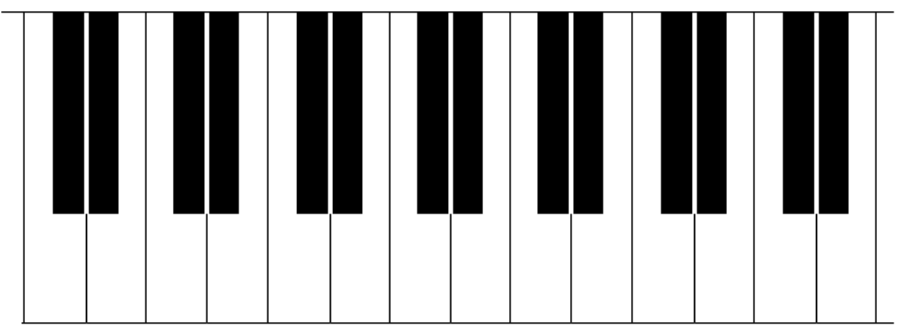} \\ \vspace{.4cm}
    (c) Diminished & (d) Augmented \\
    & \\
    \includegraphics[width=3.2in]{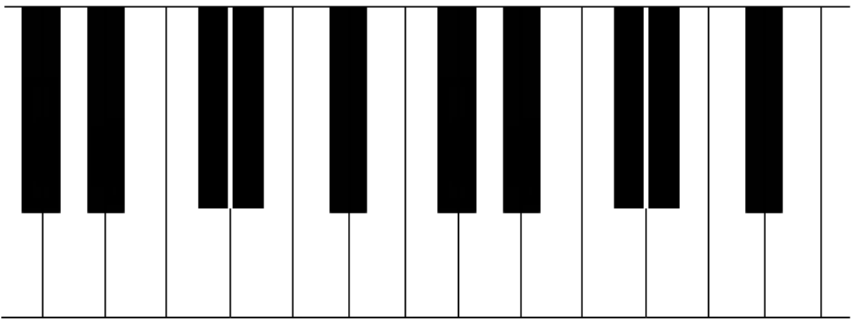} &  \includegraphics[width=3.11in]{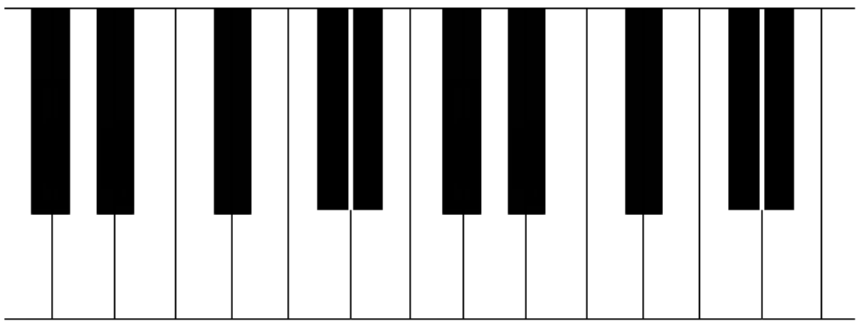} \\
    (e) Harmonic Minor & (f) Harmonic Major\\
    \end{tabular}
    \caption{Keyboard layouts for non-major harmonies.}
    \label{fig:keysnew}
    \end{center}
\end{figure}
We return to the opening question: why are the piano keys arranged that way?  The standard layout highlights major harmony (white keys) and its dual pentatonic packing (black keys).  Figure~\ref{fig:keysnew} imagines layouts for the other six harmonies, with white keys as harmony and black keys as dual packing.  We see the symmetry of whole tone and diminished harmony, and the mirror-image relationship between harmonic minor and major.  All of this structure is hidden in the standard piano layout.

\textbf{Remark (Duality in Mathematics):}  The ideas in this section are instances of a broad mathematical concept.  In the words of Sir Michael Atiyah, ``duality in mathematics is not a theorem, but a \textit{principle}''~\cite{atiyah2007}.  Duality relates two types of structures in a one-to-one fashion and appears across set theory, algebra, optimization, geometry, analysis, and beyond.  Often the dual object is more tractable, providing insight into the original---exactly the role packings play for harmony here.  For a general overview, see the Wikipedia entry on duality (\url{https://en.wikipedia.org/wiki/Duality_(mathematics)}).  
\newpage
\section{Interlude: Introduction to Modes}
\label{sec:modes}
\subsection{Modes of Major Harmony}
Let's introduce modes, a concept central to Western and Eastern musical traditions.  While the material is mostly standard, it can be confusing, and we want to clarify the aspects most relevant to our treatment of harmony.

Consider the C major scale.  On its own it sounds neither ``happy'' nor ``sad''---we need a tonal center.  The seven modes of the major scale correspond to the seven choices of tonal center within it.  In practice, the tonal center is established by fixing a bass note with fundamental frequency $f$; other tones are heard relative to this bass, with lower overtones sounding more consonant.
Modes of major harmony are closely connected to the circle of fifths:
\begin{equation}
    \text{F, C, G, D, A, E, B, F{\sharp}, C{\sharp}, G{\sharp}, D{\sharp}, A{\sharp}, F }\dots
\end{equation}
or, equivalently
\begin{equation}
    \text{F, C, G, D, A, E, B, G{\flat}, D{\flat}, A{\flat}, E{\flat}, B{\flat}, F}\dots
\end{equation}
By symmetry, any major scale consists of seven consecutive tones in this sequence.  To find all major scales containing a given bass note, say F, we start with the default $\{$F, C, G, D, A, E, B$\}$ (closest to F in the overtone series) and shift the window one step at a time:
\begin{itemize}
    \item \{F, C, G, D, A, E, B\} (F Lydian)
    \item \{B$\flat$, F, C, G, D, A, E\} (F Ionian)
    \item \{E$\flat$, B$\flat$, F, C, G, D, A\} (F Mixolydian)
    \item \{A$\flat$, E$\flat$, B$\flat$, F, C, G, D\} (F Dorian) 
    \item \{D$\flat$, A$\flat$, E$\flat$, B$\flat$, F, C, G\} (F Aeolian)
    \item \{G$\flat$, D$\flat$, A$\flat$, E$\flat$, B$\flat$, F, C\} (F Phrygian)
    \item \{B, G$\flat$, D$\flat$, A$\flat$, E$\flat$, B$\flat$, F\} (F Locrian)
\end{itemize}
The modes are labeled by their traditional Greek names, ordered from brightest to darkest.  Each step replaces one tone by another further from F in the overtone series---this is the physical reason modes darken progressively.  The standard F major scale is the Ionian mode, though as we noted earlier, Lydian is actually brighter.

\begin{figure}[H]
    \begin{center}
    \includegraphics[width=5in]{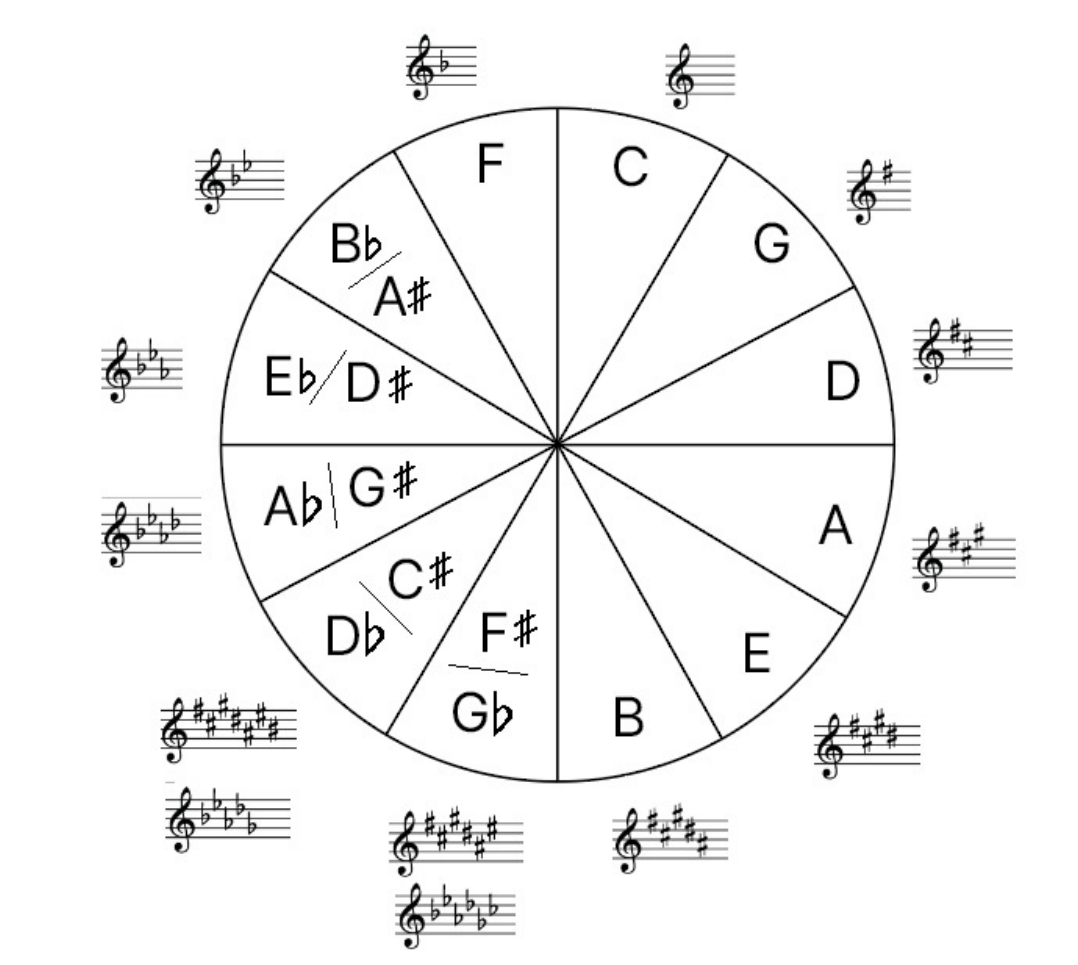}
    \caption{Circle of fifths.}
    \label{circle}
    \end{center}
\end{figure}

Presenting modes by fixing a bass and following the circle of fifths is the most practical approach: it shows that modes correspond to one-note-at-a-time alterations of the scale.  See Figure~\ref{circle} for an illustration.  We start with C major (F Lydian) and move counterclockwise to G$\flat$ major (F Locrian); beyond that, F itself drops out of the scale.
An alternative view is to fix the scale and vary the bass.  For the C major scale:
\begin{itemize}
    \item C Ionian \{C, D, E, F, G, A, B\}
    \item D Dorian \{D, E, F, G, A, B, C\}
    \item E Phrygian \{E, F, G, A, B, C, D\}
    \item F Lydian \{F, G, A, B, C, D, E\}
    \item G Mixolydian \{G, A, B, C, D, E, F\}
    \item A Aeolian \{A, B, C, D, E, F, G\}
    \item B Locrian \{B, C, D, E, F, G, A\}
\end{itemize}
Both perspectives are useful.  For example, to find A Dorian: Dorian is the second mode, so A must be the second degree of some major scale---namely G major.  We encourage the reader to play the C major scale over different bass notes to hear the effect: F in the bass gives the bright Lydian mode, while B gives the very dark Locrian.

We can also view mode progression through the pentatonic lens.  C major (F Lydian) decomposes into the F, C, and G pentatonics.  Moving to the next mode (F Ionian) swaps the G pentatonic for B$\flat$; the next (F Mixolydian) swaps C for E$\flat$.  Each mode change replaces one pentatonic with the next one along the circle of fifths in reverse.

\subsection{Modes of Other Harmonies}
Modes apply equally to other harmonies: pick a bass tone and hear the remaining notes relative to it.  For whole tone harmony (6 notes, all a whole tone apart), all six modes are equivalent up to a shift---there is only one distinct mode.  In general:
\begin{itemize}
    \item Whole Tone Harmony: one mode
    \item Augmented Harmony: two modes
    \item Diminished Harmony: two modes
    \item Major, Melodic Minor, Harmonic Major/Minor: seven modes each
\end{itemize}
We refer to modes by number rather than naming them all.  For example, the 4th mode of $\textbf{MEL}$ starting from C melodic minor begins on F.  This is the Mixolydian scale with a raised 11th (also called the \textit{Lydian Dominant} scale).  To verify: F is the 5th degree of B$\flat$ major, so F Mixolydian is the B$\flat$ major scale, which differs from C melodic minor only by B$\flat$ versus B---exactly the sharp 11th.

Another important example is the 7th mode of $\textbf{MEL}$.  For C melodic minor, this starts on B.  It contains the major third (B to E$\flat$) and minor 7th (B to A), so it supports a dominant chord on B---but all other tones are altered from B Mixolydian.  This is the \textit{altered scale}, equivalent to the Lydian Dominant a tritone away (starting on F).

Modes play a key role in lead sheet notation: the chord symbol's letter denotes the bass note of the mode.  For example, G$^7$ denotes G Mixolydian, Cmin$^7$ denotes C Dorian, F$^{7(\sharp 11)}$ denotes Lydian Dominant, and B$^{\text{7alt}}$ denotes the altered scale.  We return to this in Section~\ref{sec:lead}.

We can also order modes by brightness for other harmonies.  For $\textbf{DIM}$, fix a diminished chord, say $(C, E\flat, G\flat, A)$.  Adding the ``upper'' chord $(G, E, D\flat, B\flat)$ or the ``lower'' chord $(D, B, A\flat, F)$ yields the two modes.  The upper chord is closer to C on the circle of fifths, so its mode is brighter.  Similar reasoning applies to the two modes of $\textbf{AUG}$.

\textbf{Remark:} In Figure~\ref{fig:harmexampl}(c),(d) we labeled $\textbf{DIM}$ and $\textbf{AUG}$ starting on C using the \textit{darker} of the two modes.  The brighter mode corresponds to the standard jazz chord (dom7$\flat$9 for $\textbf{DIM}$, maj7 for $\textbf{AUG}$), leaving the darker mode associated with the dim7 and aug chords respectively.  We discuss this further in Section~\ref{sec:lead}.
\newpage
\section{Combinatorial Approach to Chords and Voicings}
\label{sec:voicings}
\subsection{Overview of Classification}
We now classify chords using an approach similar to our classifications of packings and harmony.  The irreducibility criterion here is less canonical, so we state the goals explicitly:
\begin{itemize}
    \item The list should be small: every other chord is obtained by removing notes (incomplete) or adding notes (reducible).  A short list lets the musician produce many arrangements, as in Sections~\ref{sec:black} and~\ref{sec:white}.
    \item Irreducible chords should include most commonly used chords.
    \item The criterion should relate to a simple property of tones, not an arbitrary combinatorial rule.
\end{itemize}
These goals do not determine a unique criterion, but different reasonable choices yield similar chord lists.

Our criterion involves both blocks and cells.  Recall from Section~\ref{sec:black} that we avoided three notes each at most a tone apart---an analogue of a semitone cell at a larger scale:

\textbf{Definition:}  A \textbf{dissonant tone cell} consists of three distinct notes in a row, which are at most a tone apart, up to shifts by an octave.

\begin{figure}[H]
    \begin{center} 
    \includegraphics[width=5.1in]{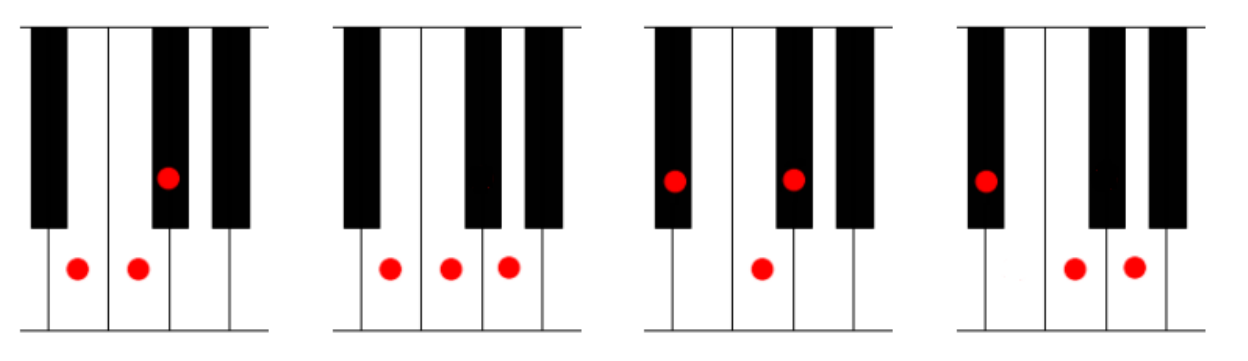} 
    \caption{Possible dissonant tone cell configurations with middle note F.}
    \label{fig:cell2}
    \end{center}
\end{figure}

See Figure~\ref{fig:cell2} for possible tone cell configurations with middle note F.
Our completeness criteria for chords will combine semitone blocks and tone cells:

\textbf{Definition:}  A chord is \textbf{complete} if adding further notes will increase either the number of semitone blocks or the number of tone cells (or both).

\textbf{Definition:}  A chord is \textbf{irreducible} if it is complete, but no subset of the chord is complete.

There are exactly eleven irreducible chords.  We go through each below, discussing their embeddings into harmony and packings.  Most are well known and we use standard notation.

\textbf{Remark (Triads):}  We are focused mainly on 4-note chords, which are more common in jazz harmony than classical harmony which is focused on triads (or 3-note chords).  We will explain how triads fit into our framework in Appendix~\ref{sec:appF}.

\textbf{Remark (Inversions):}  We consider chords related by inversion to be equivalent (e.g., Amin7 and Cmaj6 differ by moving the root up an octave).  The ambiguity is resolved by choosing a bass note, which determines the chord quality.  See Section~\ref{sec:lead}.

\textbf{Remark (Guide to Reader):}  The rest of this section is a reference listing all irreducible voicings and their embeddings.  We recommend reading it slowly, starting with the most common voicings.
\subsection{min7 (maj6)}
\label{sec:maj6}
\begin{figure}[H]
    \begin{center} 
    \begin{tabular}{cc}
    \includegraphics[width=2.0in]{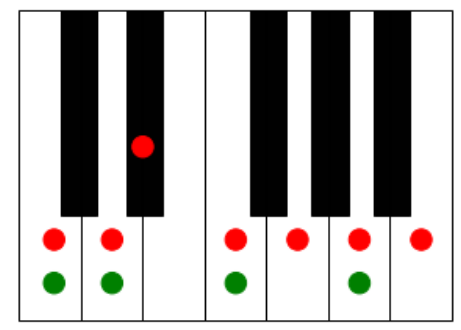}  &  \includegraphics[width=2.0in]{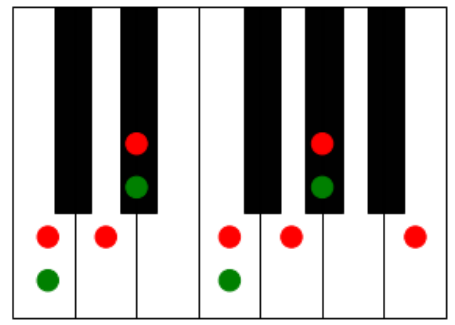}\\ \vspace{.4cm}
    (a) Dmin7 in C \textbf{MEL} & (b) Fmin7 in C \textbf{HMIN} \\
    \includegraphics[width=2.0in]{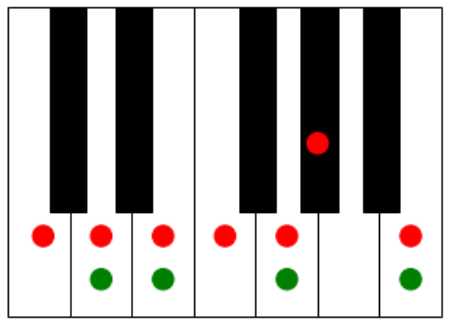}  &  \includegraphics[width=2.0in]{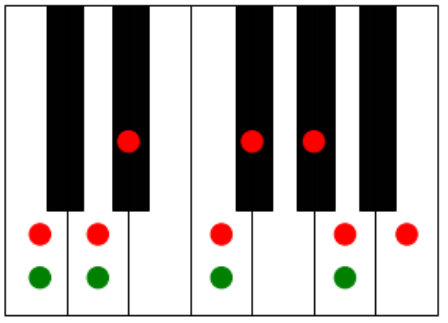} \\  
    (c) Emin7 in C \textbf{HMAJ} & (d) Dmin7 in C \textbf{DIM} \\
                                  & \{F,A$\flat$,B\}min7 not shown\\
    \end{tabular}
    \caption{Possible embeddings of min7.}
    \label{fig:maj6}
    \end{center}
\end{figure}

We encountered this chord with the black keys in Section~\ref{sec:black}; its inversion gives the major 6th chord.  It occurs in the packings \textbf{penta}, \textbf{hmin}, and \textbf{hmaj}, yielding three embeddings into \textbf{MAJ} (Section~\ref{sec:white}), one into \textbf{MEL}, and one each into \textbf{HMIN} and \textbf{HMAJ}.  See Figure~\ref{fig:maj6}(a)--(c), where the chord is red and the scale green.

There are also four embeddings into \textbf{DIM} (Figure~\ref{fig:maj6}(d)), corresponding to the four embeddings of \textbf{hmin} into \textbf{DIM}.  By the symmetry of the diminished scale, C \textbf{DIM} contains Dmin7, Fmin7, A$\flat$min7, and Bmin7 (these bass notes form a diminished chord).  Since Dmin7 is also Fmaj6, the scale contains both the major and minor 4-chords on the same bass---a rare occurrence.
This chord is most commonly used to voice the Lydian mode.  With C in the bass, we use Amin7/C or Emin7/C as in Section~\ref{sec:white}.  The latter is the standard ``rootless'' jazz voicing: since C is already in the bass, we replace it with the 9th (D), giving the tones (E, G, B, D)---the 3rd, 5th, 7th, and 9th of C Lydian.  Of course, min7 can voice any mode of any harmony where it embeds, yielding many exotic possibilities.

\textbf{Remark:}  A Cmaj$^7$ chord symbol in sheet music suggests the C Lydian mode.  The standard voicing is Emin7/C rather than the literal Cmaj7 chord---we use a minor 7th to voice a major 7th symbol.  This can confuse beginners, but it is standard practice: with C already in the bass, we gain the extra 9th tone (D).  See Section~\ref{sec:lead} for further discussion.
\subsection{maj7}
The major 7th chord is a major triad plus a major 7th.  It is commonly used to voice the Dorian mode (which, confusingly, is associated with the min7 chord symbol).  For D Dorian (second mode of C major), the voicing is Fmaj7/D.

\begin{figure}[H]
    \begin{center} 
    \begin{tabular}{cc} 
    \includegraphics[width=2.0in]{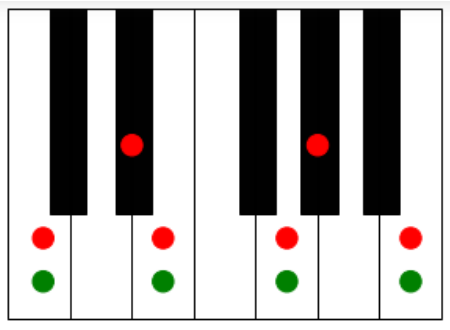}  &  \includegraphics[width=2.0in]{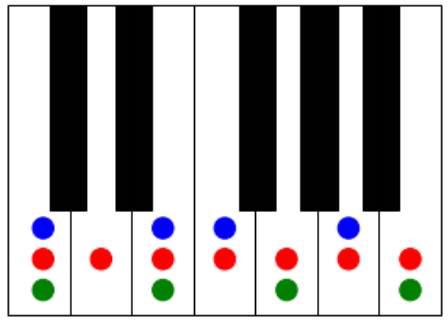}  \\ \vspace{.4cm}
    (a) Cmaj7 in B $\textbf{aug}$/$\textbf{AUG}$ & (b) Cmaj7 (green) and Fmaj7 (blue) in C \textbf{MAJ} \\ 
    \includegraphics[width=2.0in]{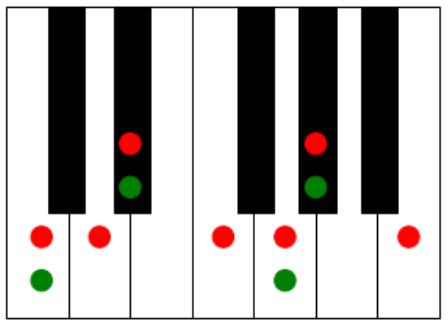}  &  \includegraphics[width=2.0in]{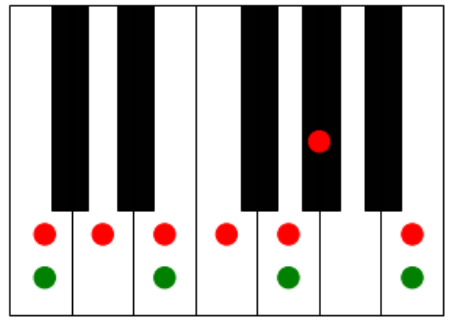}  \\
    (c) A$\flat$maj7 in C \textbf{HMIN} & (d) Cmaj7 in C \textbf{HMAJ}  \\ 
    \end{tabular}
    \caption{Embedding maj7 into harmony.}
    \label{fig:maj7}
    \end{center}
\end{figure}

Despite its pervasive use, maj7 occurs in only one packing: \textbf{aug} (Figure~\ref{fig:maj7}(a)).  It embeds twice in \textbf{MAJ} and once each in \textbf{HMIN} and \textbf{HMAJ} (Figure~\ref{fig:maj7}(b)--(d)).  Since \textbf{AUG} is self-dual, there are also three embeddings there.  This is an interesting case of a chord ``borrowed'' from one packing (\textbf{aug}) for use in a quite different harmony (\textbf{MAJ}).
Another common use of this chord is to give a ``suspended'' voicing for the Mixolydian mode.  For example, we can use Fmaj7 over G to get a suspended sound for G Mixolydian. We will discuss additional voicings of this mode below.  
\subsection{dom7}
\begin{figure}[H]
    \begin{center} 
    \begin{tabular}{cc}
    \includegraphics[width=2.0in]{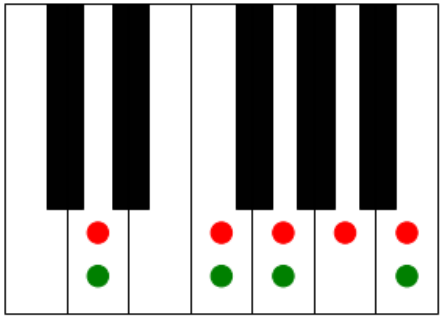}  &  \includegraphics[width=2.0in]{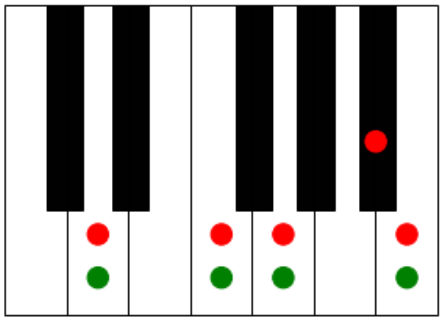}  \\
    (a) Gdom7 in G \textbf{dpenta} & (b) Gdom7 in B$\flat$ \textbf{hmin} \\ 
    \end{tabular}
    \caption{Packing embeddings of dom7.}
    \label{fig:dom7a}
    \end{center}
\end{figure}

\begin{figure}[H]
    \begin{center} 
    \begin{tabular}{cc}
    \includegraphics[width=2.0in]{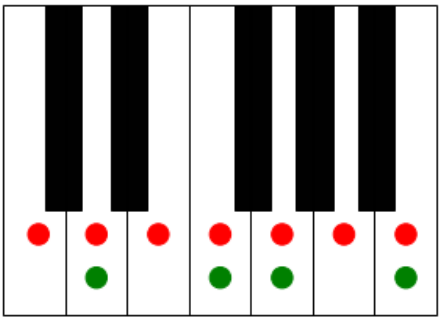}  &  \includegraphics[width=2.0in]{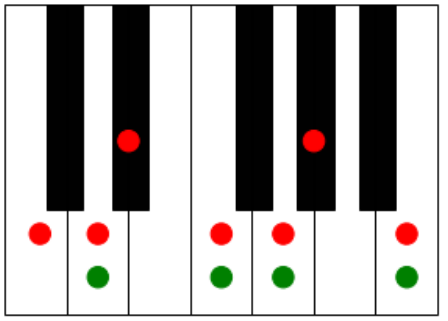}  \\
    (a) Gdom7 in C \textbf{MAJ} & (b) Gdom7 in C \textbf{HMIN} \\ 
    \end{tabular}
    \caption{Embedding dom7 into harmony.}
    \label{fig:dom7b}
    \end{center}
\end{figure}

The standard dominant chord: a major triad plus a minor 7th, containing the tritone interval that creates tension before resolution.  It occurs in the packings \textbf{dpenta} and \textbf{hmin} (Figure~\ref{fig:dom7a}).
In major harmony, dom7 has one embedding and is the classical voicing of the Mixolydian mode (e.g., Gdom7 = G, B, D, F in C major; see Figure~\ref{fig:dom7b}(a)).  Since this chord already contains the bass note G, it is a less complex voicing than, say, Emin7/C for Lydian; jazz often uses richer alternatives.  Dom7 also embeds twice in \textbf{MEL} and \textbf{HMAJ}, once in \textbf{HMIN}, and four times in \textbf{DIM}.  See the figures below.
\begin{figure}[H]
    \begin{center} 
    \begin{tabular}{ccc}
    \includegraphics[width=1.8in]{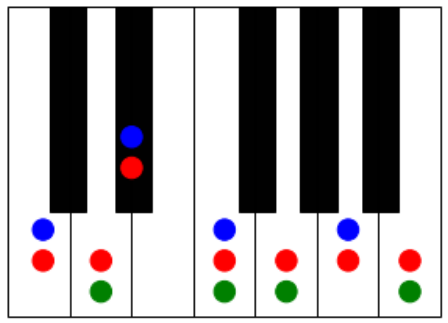}  &  \includegraphics[width=1.8in]{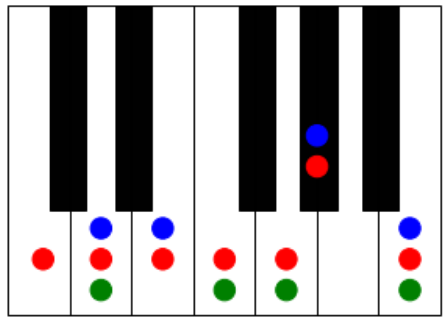}  &  \includegraphics[width=1.8in]{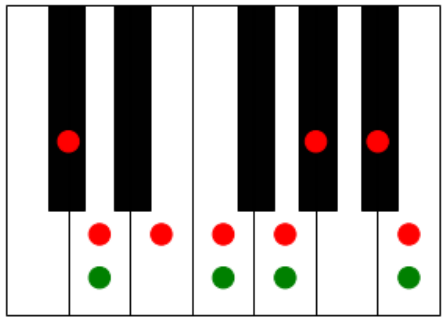}  \\
    (a) Gdom7 (green) and Fdom7 (blue)  & (b) Gdom7 (green) and Edom7 (blue) & (c) Gdom7 in D \textbf{DIM} \\ 
    in C \textbf{MEL} & in C \textbf{HMAJ} & \{B$\flat$,D$\flat$,E\}dom7 not shown\\
    \end{tabular}
    \caption{Embedding dom7 into harmony (cont.)}
    \label{fig:dom7c}
    \end{center}
\end{figure}
\subsection{dim7}
\begin{figure}[H]
    \begin{center} 
    \begin{tabular}{ccc}
        \includegraphics[width=1.8in]{./figs/dimdim-eps-converted-to.pdf}  & \includegraphics[width=1.8in]{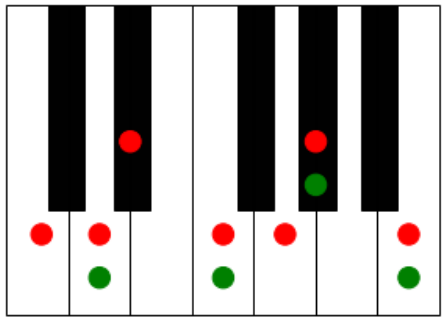}  &  \includegraphics[width=1.8in]{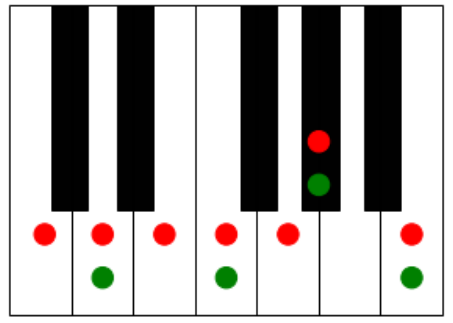}  \\
    (a) Cdim7 (green) and Ddim7 (blue)  & (b) Ddim7 in C \textbf{HMIN} & (c) Ddim7 in C \textbf{HMAJ}  \\
    in C \textbf{DIM} & & \\    
    \end{tabular}
    \caption{Embedding dim7 into harmony.}
    \label{fig:dimcho}
    \end{center}
\end{figure}

The diminished 7th is both a chord and the diminished packing \textbf{dim}.  Diminished harmony is built from two dim7 chords a semitone apart (Figure~\ref{fig:dimcho}(a)); there are also unique embeddings into \textbf{HMIN} and \textbf{HMAJ} ((b) and (c)).  The dim7 chord voices both modes of the diminished scale.  By convention, the ``upper'' green chord of Figure~\ref{fig:dimcho}(a) is used: e.g., Cdim7 over both C and F.
\subsection{min7$\flat$5}
\begin{figure}[H]
    \begin{center} 
    \begin{tabular}{cc}
    \includegraphics[width=2.0in]{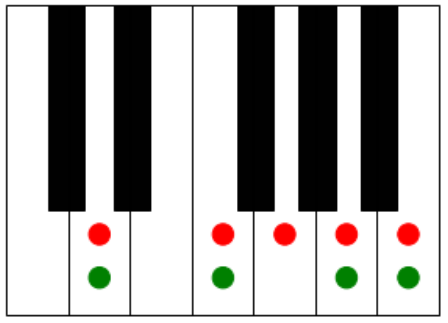}  &  \includegraphics[width=2.0in]{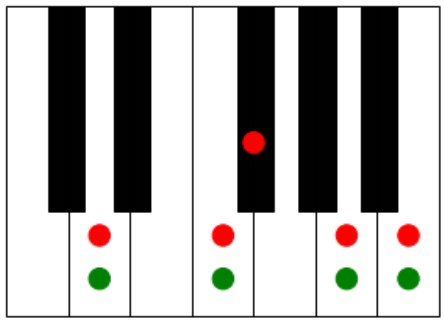}  \\
    (a) Bmin7$\flat$5 in G \textbf{dpenta} & (b) Bmin7$\flat$5 in D \textbf{hmaj}  \\ 
    \end{tabular}
    \caption{Embedding min$\flat$5 into packings.}
    \label{fig:min7b5a}
    \end{center}
\end{figure}

\begin{figure}[H]
    \begin{center} 
    \begin{tabular}{cc}
    \includegraphics[width=2.0in]{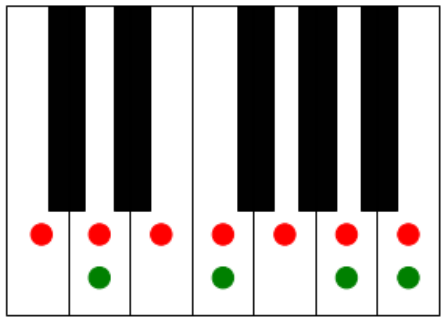}  &  \includegraphics[width=2.0in]{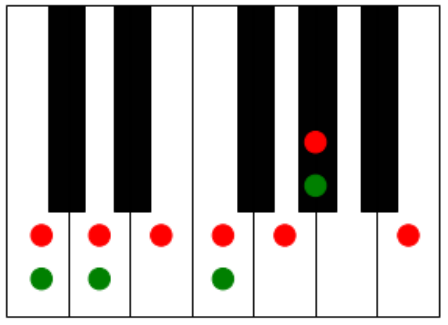} \\
    (a) Bmin7$\flat$5 in C \textbf{MAJ} & (b) Dmin7$\flat$5 in C \textbf{HMAJ} \\ 
    \end{tabular}
    \caption{Embedding min$\flat$5 into harmony.}
    \label{fig:min7b5b}
    \end{center}
\end{figure}

\begin{figure}[H]
    \begin{center} 
    \begin{tabular}{ccc}
    \includegraphics[width=1.8in]{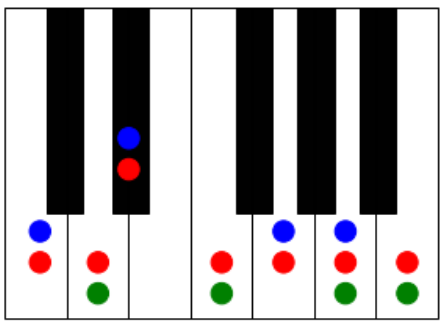}  &  \includegraphics[width=1.8in]{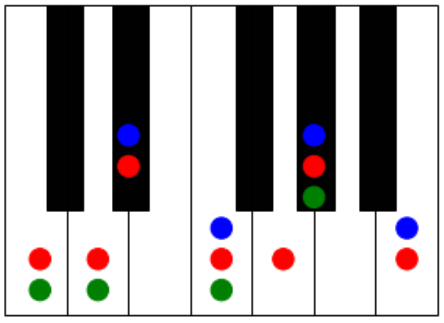} &  \includegraphics[width=1.8in]{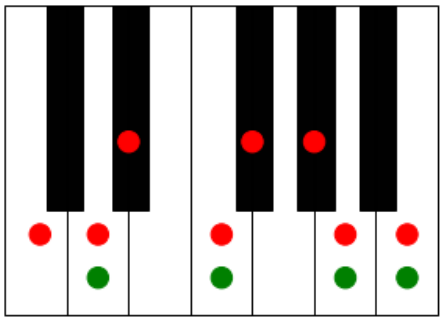}  \\
    (a) Bmin7$\flat$5 (green)   & (b) Dmin7$\flat$5 (green) and  & (c) Bmin7$\flat$5 in C \textbf{DIM}\\ 
    and Amin7$\flat$5 (blue) in C \textbf{MEL} & Fmin7$\flat$5 (blue) in C \textbf{HMIN}  & (\{D,F,A$\flat$\}min7$\flat$5 not shown) \\
    \end{tabular}
    \caption{Embedding min$\flat$5 into harmony (cont).}
    \label{fig:min7b5c}
    \end{center}
\end{figure}

Obtained from min7 by flatting the 5th (also called the half-diminished chord, denoted {\O}).  It occurs in the packings \textbf{dpenta} and \textbf{hmaj} (Figure~\ref{fig:min7b5a}).
As far as embeddings into harmony, there are unique embeddings into \textbf{MAJ} and \textbf{HMAJ}, as well as two embeddings into \textbf{MEL} and \textbf{HMIN}.  Finally, there are four embeddings into \textbf{DIM}.  See Figure~\ref{fig:min7b5b} and Figure~\ref{fig:min7b5c}.
Commonly used to voice the Locrian mode of \textbf{MAJ} and the 2nd mode of \textbf{HMIN}.  For example, B Locrian (C major scale) uses Bmin7$\flat$5/B.
\subsection{maj7$\flat$5}
\begin{figure}[H]
    \begin{center} 
    \begin{tabular}{cc}
    \includegraphics[width=2.0in]{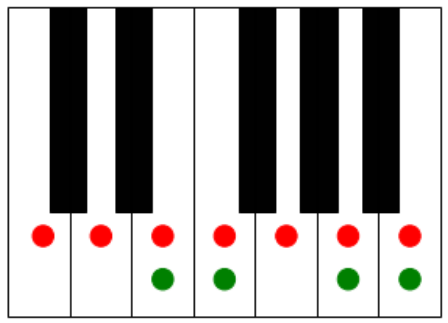}  &  \includegraphics[width=2.0in]{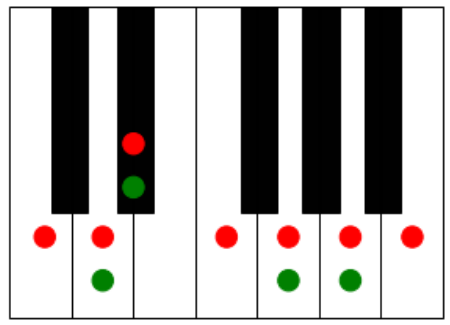}  \\ \vspace{.4cm}
    (a) Fmaj7$\flat$5 in C \textbf{MAJ} & (b) E$\flat$maj7$\flat$5 in C \textbf{MEL}  \\    
    \includegraphics[width=2.0in]{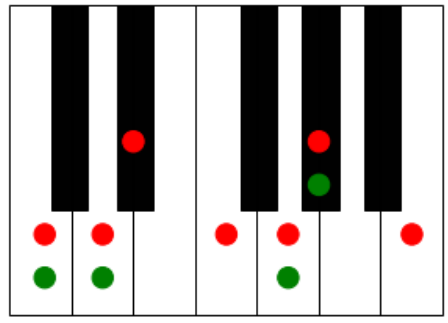}  &  \includegraphics[width=2.0in]{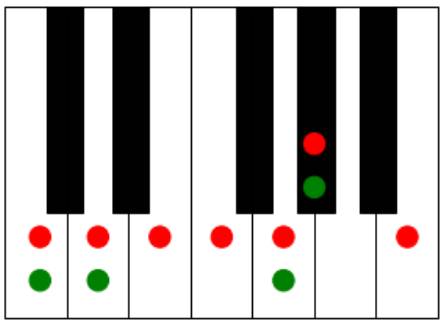}  \\
    (c) A$\flat$maj7$\flat$5 in C \textbf{HMIN} & (d) A$\flat$maj7$\flat$5 in C \textbf{HMAJ}  \\   
    \end{tabular}
    \caption{Embedding maj7$\flat$5 into harmony.}
    \label{fig:maj7b5}
    \end{center}
\end{figure}

Obtained from maj7 by flatting the 5th---our first chord not derived from any packing.  It contains a block and a tritone, making it fairly dissonant.  It has a unique embedding in each of \textbf{MAJ}, \textbf{MEL}, \textbf{HMIN}, and \textbf{HMAJ}, and no others (Figure~\ref{fig:maj7b5}).  Commonly used in jazz for a richer Mixolydian voicing: e.g., Fmaj7$\flat$5/G for G Mixolydian.
\subsection{aug}
\begin{figure}[H]
    \begin{center} 
    \begin{tabular}{cc}
    \includegraphics[width=2.0in]{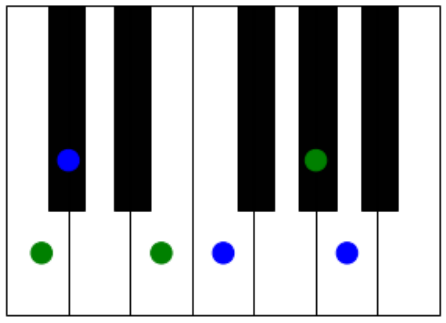}  &  \includegraphics[width=2.0in]{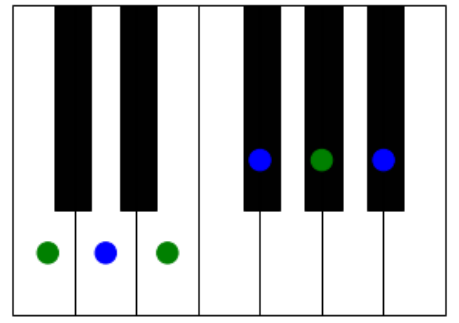}\\
    (a) Caug (green) and Faug (blue)& (b) Caug (green) and Daug (blue)  \\  
    covering C \textbf{AUG}   & covering C \textbf{WTONE} \\ 
    \end{tabular}
    \caption{Packings/harmony that is covered by aug.}
    \label{fig:chordaug}
    \end{center}
\end{figure}

\begin{figure}[H]
    \begin{center} 
    \begin{tabular}{ccc}
    \includegraphics[width=1.8in]{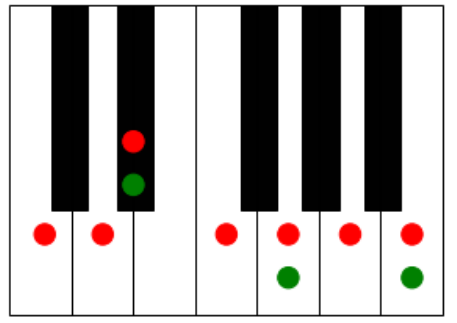}  &  \includegraphics[width=1.8in]{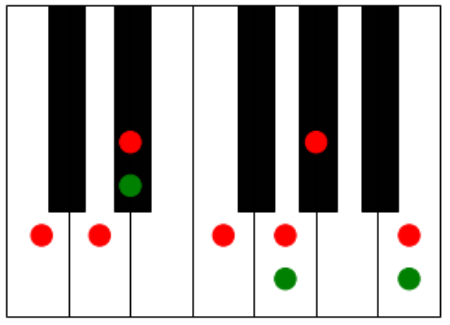} &  \includegraphics[width=1.8in]{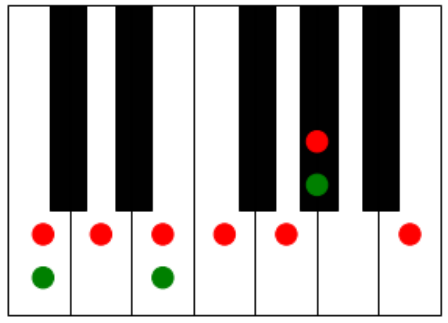} \\
    (a) Gaug in C \textbf{MEL} & (b) Gaug in C \textbf{HMIN} & (c) Caug in C \textbf{HMAJ}\\ 
    \end{tabular}
    \caption{Other harmony embeddings of aug.}
    \label{fig:chordaug2}
    \end{center}
\end{figure}

The augmented triad---the only 3-note irreducible chord---is built by stacking two major thirds.  It occurs twice in both \textbf{AUG} and \textbf{WTONE} (Figure~\ref{fig:chordaug}), and once each in \textbf{MEL}, \textbf{HMIN}, and \textbf{HMAJ} (Figure~\ref{fig:chordaug2}).  For C melodic minor, Gaug/C gives the Cmin(maj7) chord.
\subsection{min11}
\begin{figure}[H]
    \begin{center} 
    \begin{tabular}{ccc}
    \includegraphics[width=1.8in]{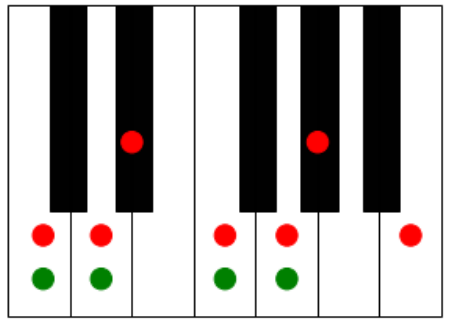}  &  \includegraphics[width=1.8in]{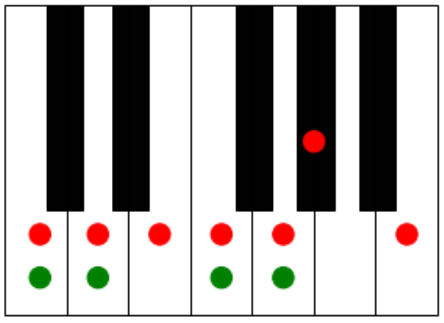}&  \includegraphics[width=1.8in]{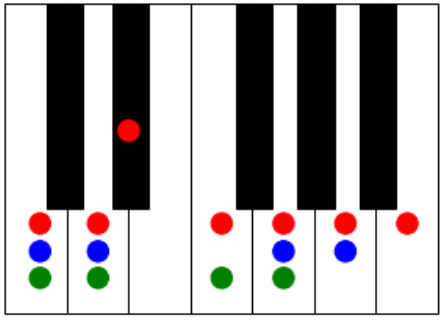}\\
    (a) Dmin11 in C \textbf{HMIN} & (b) Dmin11 in C \textbf{HMAJ} & (c) Dmin11 (green) and Amin11 (blue)  \\
     & & in C \textbf{MEL} \\
    \end{tabular}
    \caption{Embeddings of min11.}
    \label{fig:min11}
    \end{center}
\end{figure}

Encountered in Section~\ref{sec:black}, min11 comes from the pentatonic packing \textbf{penta}.  It has four embeddings into \textbf{MAJ} (Section~\ref{sec:white}), two into \textbf{MEL}, and one each into \textbf{HMIN} and \textbf{HMAJ} (Figure~\ref{fig:min11}).  Like min7, it is most commonly used to voice the Lydian mode.
\subsection{dom$\sharp$11}
\begin{figure}[H]
    \begin{center} 
    \begin{tabular}{ccc}
    \includegraphics[width=1.8in]{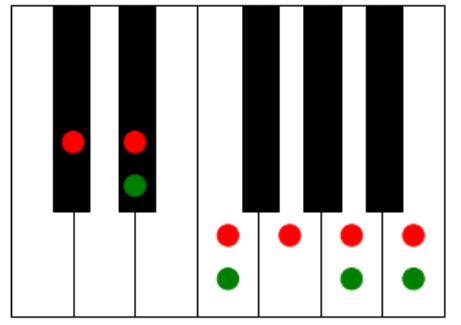}  &  \includegraphics[width=1.8in]{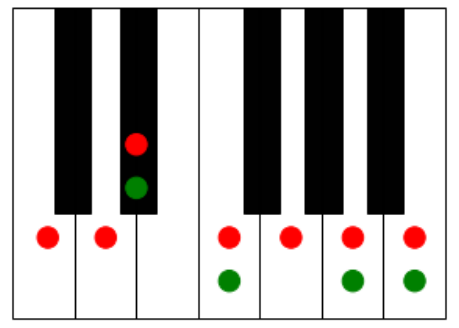}  &  \includegraphics[width=1.8in]{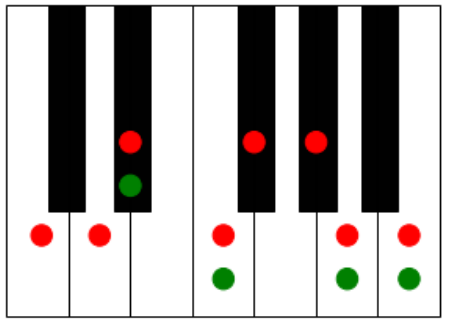}  \\
    (a) F$\sharp$11 in F \textbf{WTONE} & (b) F$\sharp$11 in C \textbf{MEL} & (c) F$\sharp$11 in C \textbf{DIM} \\ 
    (\{G,A\}$\sharp$11 not shown)                  &                                     & (D$\sharp$11 not shown)                \\
    \end{tabular}
    \caption{Embedding dom$\sharp$11 into harmony.}
    \label{fig:dom7b5}
    \end{center}
\end{figure}

A dominant 7th with a sharp 11th replacing the 5th---two tritones a whole tone apart.  It occurs only in the whole tone packing (three embeddings) and embeds once in \textbf{MEL} and twice in \textbf{DIM} (Figure~\ref{fig:dom7b5}).

This chord voices the 4th mode of \textbf{MEL} (e.g., Fdom$\sharp$11/F for C \textbf{MEL}) and also the 5th mode (Mixolydian $\flat$13): Fdom$\sharp$11/G.
\subsection{dom11}
\begin{figure}[H]
    \begin{center} 
    \begin{tabular}{cc}
    \includegraphics[width=2.0in]{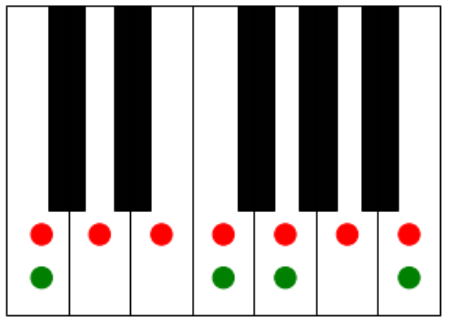}  &  \includegraphics[width=2.0in]{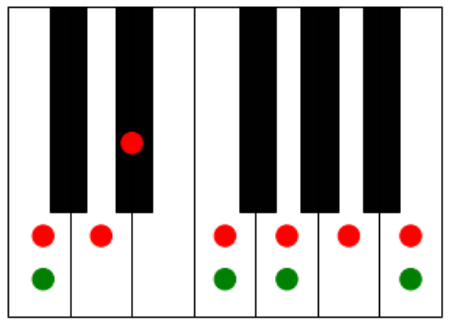}  \\ \vspace{.4cm}
    (a) Gdom11 in C \textbf{MAJ} & (b) Gdom11 in C \textbf{MEL}  \\ 
    \includegraphics[width=2.0in]{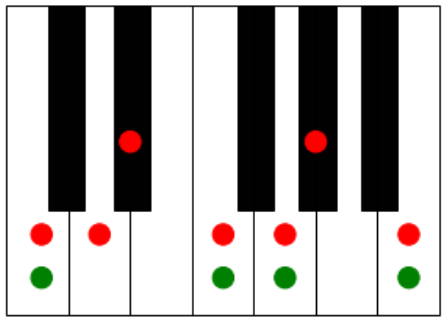}  &  \includegraphics[width=2.0in]{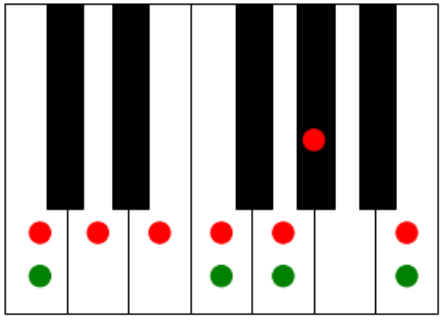}  \\
    (c) Gdom11 in C \textbf{HMIN} & (d) Gdom11 in C \textbf{HMAJ} \\ 
    \end{tabular}
    \caption{Embedding dom11 into harmony.}
    \label{fig:dom11}
    \end{center}
\end{figure}

Another chord not derived from any packing: dom7 with the 5th replaced by a 4th.  It has a unique embedding in each of \textbf{MAJ}, \textbf{MEL}, \textbf{HMIN}, and \textbf{HMAJ} (Figure~\ref{fig:dom11}).
A possible use is to voice the Mixolydian mode of  \textbf{MAJ}.  For instance, for G Mixolydian,  if our top melody note is a G we may use the inversion (F, B, C, G) as a more complex variant of the vanilla (F, B, D, G) voicing of that mode.  The chord Gdom11 can be abbreviated to G11.  
\subsection{dem}
\begin{figure}[H]
    \begin{center} 
    \begin{tabular}{cc}
    \includegraphics[width=2.0in]{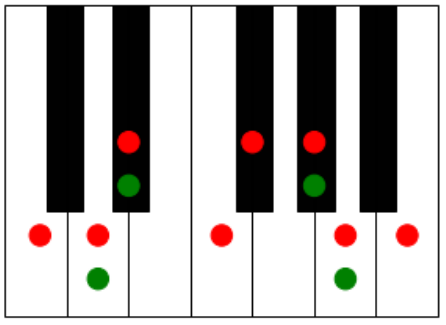}  &  \includegraphics[width=2.0in]{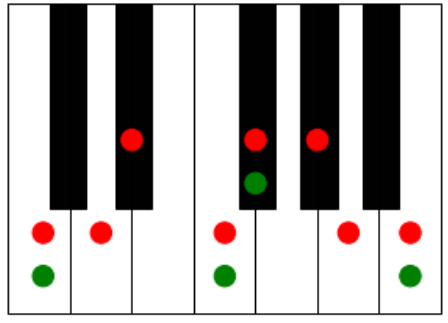}  \\
    (a) Adem (or E$\flat$dem)& (b) Cdem  (or G$\flat$dem)\\    
    \end{tabular}
    \caption{Embedding dem into C \textbf{DIM}.}
    \label{fig:triton}
    \end{center}
\end{figure}

Two tritone intervals a semitone apart.  With two semitone blocks, this is quite dissonant and rarely used.  It occurs in no packing, has two embeddings in \textbf{DIM} and no others---making it a unique fingerprint of diminished harmony (Figure~\ref{fig:triton}).
The author is not aware of a standard name for this chord and, since we already have ``dim'' and ``dom'' chords, we decided to use ``dem'' for the name.  In the 19th century, the tritone was associated with the devil for its harsh sound and since this chord shape contains two tritones we may think of ``dem'' as short for demon.

\begin{figure}[H]
    \begin{center}     
    \includegraphics[width=6.3in]{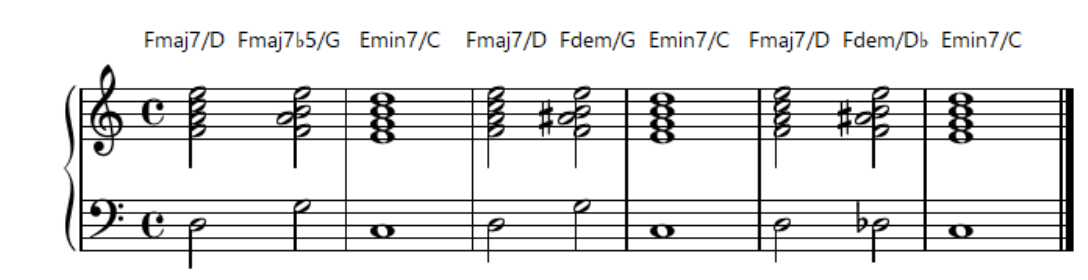}      
    \caption{251 progressions in C.}
    \label{fig:dem251}
    \end{center}
\end{figure}

One use is to voice the dominant chord inside diminished harmony.  Figure~\ref{fig:dem251} shows a jazz ``251'' progression with a standard dominant chord, then a variant using dem.\footnote{See Appendix~\ref{sec:appH} for more on the 251 progression and substitutions.}  This voicing singles out the diminished scale as its only supporting harmony.  The figure also shows the ``tritone'' substitution (replacing G by D$\flat$).  This substitution works because G and D$\flat$ dominant chords share the same tritone (with 3rd and 7th swapped).  The dem voicing is symmetric with respect to this substitution: the $\sharp$9th and 13th for G become the 13th and $\sharp$9th for D$\flat$.  Symbolically, $$\text{Fdem = Bdem}$$ since we identify chord shapes up to inversion.

This concludes our discussion of irreducible chord shapes.  Most come from packings; the exceptions are dom11, maj7$\flat$5, and dem.  Understanding these shapes and their embeddings lies at the heart of our approach to harmony.
\subsection{Reducible Voicings and Completeness Results}
\label{sec:complete}
\begin{figure}[H]
    \begin{center} 
    \begin{tabular}{cc}
    \includegraphics[width=2.0in]{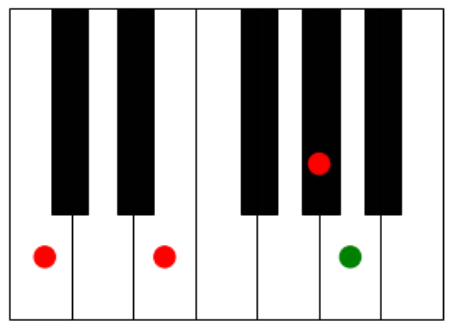}  &  \includegraphics[width=2.0in]{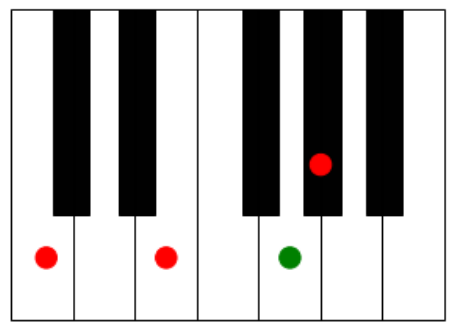}\\ \vspace{.4cm}
    (a) Caug$^{(\text{add 13})}$ & (b) Caug$^{(\text{add 5})}$\\
    \includegraphics[width=2.0in]{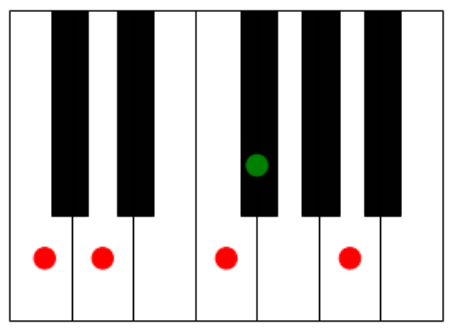}  &  \includegraphics[width=2.0in]{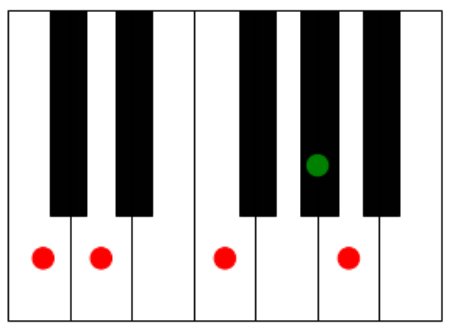} \\  
    (c) Dmin7$^{(\text{add $\flat$11})}$ (\textbf{hmin}) & (d) Dmin7$^{(\text{add $\sharp$11})}$  (\textbf{hmaj})\\
    \end{tabular}
    \caption{Reducible chords with one semitone block and zero tone cells.}
    \label{fig:comp}
    \end{center}
\end{figure}

We conclude by mentioning common reducible voicings.  A complete chord is either irreducible (already on our list) or obtained by adding notes to an irreducible chord.  There are 265 complete chords in total, but most are too dissonant for standard use.  The simplest nontrivial family consists of chords with zero tone cells and at most one block:

\textbf{Characterization of chords with one block:}  Consider the set of chords that are (chord) complete, contain at most one semitone block and zero tone cells.  These chords are either irreducible as above or one of the four reducible chords in Figure~\ref{fig:comp}.
 Only four reducible chords qualify.  The first two ((a),(b)) add one note to the augmented chord (highlighted green); these voice modes of \textbf{MEL}.  The last two ((c),(d)) are the \textbf{hmin}/\textbf{hmaj} packings.

\begin{figure}[H]
    \begin{center} 
    \begin{tabular}{cc}
    \includegraphics[width=2.0in]{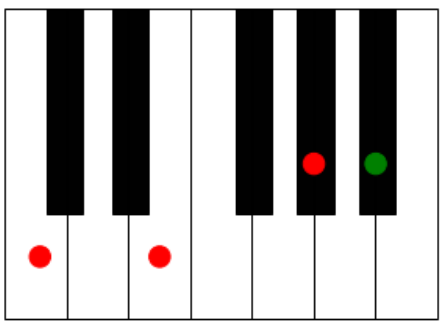}  &  \includegraphics[width=2.0in]{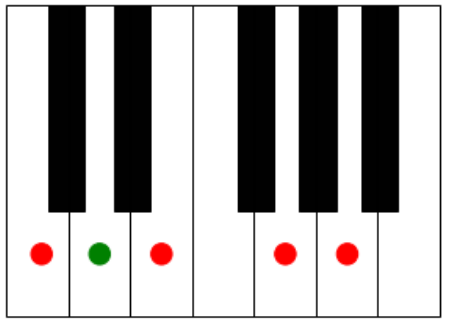}\\
    (a) Caug$^{(\text{add min7})}$ (Cdom$\flat$13 or Cdom7$\sharp5$) & (b) Cmaj6$^{(\text{add 9})}$ (Cmaj$^{6/9}$)\\
    \end{tabular}
    \caption{Reducible chords with zero semitone blocks and one tone cell.}
    \label{fig:comp2}
    \end{center}
\end{figure}

The complementary case: complete chords with no blocks and at most one tone cell.  Since these have no blocks, they embed in block-free packings (\textbf{penta}, \textbf{dpenta}, \textbf{wtone}, \textbf{dim}):

\textbf{Characterization of chords with one cell:}  Consider the set of chords that are (chord) complete, contain at most one tone cell and zero (semitone) blocks.  These chords are either irreducible as above or one of the two reducible chords in Figure~\ref{fig:comp2}.
We include standard lead sheet notation alongside our irreducible-chord notation.  The chord in (a) voices the 5th mode of \textbf{MEL} (Mixolydian $\flat$13).  The chord in (b) creates pentatonic 5-note voicings---we saw an example in Section~\ref{sec:black} (penultimate bar of Figure~\ref{fig:ag1}).  Since it contains an entire pentatonic, it has three embeddings in $\textbf{MAJ}$ and one in $\textbf{MEL}$; as noted earlier, it should be spread across both hands.

We end this section by returning to melody harmonization from Section~\ref{sec:black}: given a melodic line, form a chord with the melody as its top note.  The irreducible chords suffice:

\textbf{Harmonizing tones:}  For any harmony and any tone inside that harmony, there is an irreducible chord in that harmony that contains the tone as its top note.

\begin{figure}[H]
    \begin{center} 
    \begin{tabular}{cc}
    \includegraphics[width=2.0in]{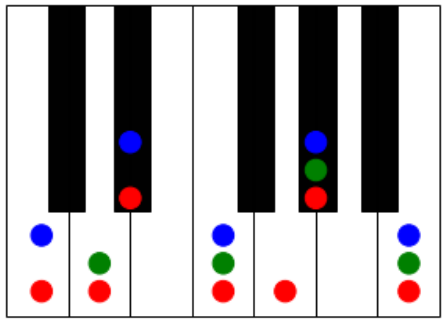}  &  \includegraphics[width=2.0in]{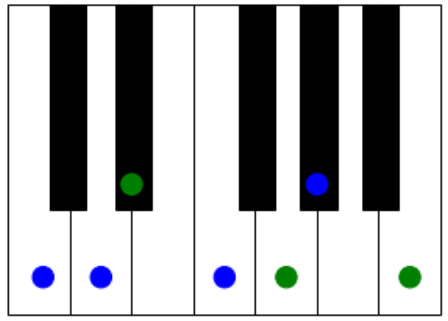}\\
    (a) A$\flat$ \textbf{hmaj} (blue) and D \textbf{dim} (green) & (b) E$\flat$aug (green) and Dmin7$\flat$5 (blue)  \\
    in C \textbf{HMIN}  & covering  C \textbf{HMIN} \\
    \end{tabular}
    \caption{Covering \textbf{HMIN} by irreducible chords.}
    \label{fig:coverhmin}
    \end{center}
\end{figure}

This can be verified by checking all seven harmonies.  For most, the dual packings cover the harmony, so their chords suffice.  For example, \textbf{MEL} is covered by two copies of \textbf{dpenta}, giving dom7 and min7$\flat$5 for every tone.  The exceptions are \textbf{HMIN} and \textbf{HMAJ}, where the embedded packings leave one tone uncovered.  For \textbf{HMIN}, supplementing with min7$\flat$5 and aug fills the gap (Figure~\ref{fig:coverhmin}).
We will refine this discussion further when we will go over concrete voicings for the most common modes in Section~\ref{sec:vchord}.  This will provide examples of explicit chords that harmonize every tone in a given scale.
\subsection{Summary of Harmony-Chord Embeddings}
For convenience, we present this information from the complementary perspective: starting with a harmony, we list all chord embeddings.  Table~\ref{tbl:harmchord} summarizes all embeddings for harmonies starting in C.  For example, C \textbf{WTONE} supports the chords Caug, Daug, Cdom$\sharp$11, Ddom$\sharp$11, and Edom$\sharp$11.  While the table does not indicate when to use each chord, it maps out the harmonic options available for composition and arrangement.  We discuss common usage in Section~\ref{sec:lead}.
\begin{table}[H]
    \begin{center}
    \begin{tabular}{l|l|l|l|l|l|l|l|l|l|l|l}
      C harmony      & min7 & maj7 & dom7 & dim7 & min7$\flat$5 &  maj7$\flat$5 & aug & min11 & dom$\sharp$11 & dom11 & dem \\
    \hline
    \textbf{WTONE} &      &      &       &     &              &               & C,D &       &   C,D        &       &  \\
                     &      &      &       &     &              &               &     &       &   E           &       &  \\
    \hline
    \textbf{MAJ}   & A,D & C,F  & G   &      &  B           & F             &     & D,E &             & G      &  \\      
                     & E &        &     &      &              &               &     & A,B &             &       &  \\    
    \hline
    \textbf{MEL}  &   D   &      &  G,F   &     & B,A          & E$\flat$             &  G  &  D,A   &  F           &  G    &  \\   
    \hline
    \textbf{HMIN}  &  F$\flat$    & A$\flat$     &  G     &  D   & D,F   &  A$\flat$  &  G   & D  &        & G      &  \\    
    \hline
    \textbf{HMAJ} & E & C & G,E & D & D &  A$\flat$ & C & D &  & G &  \\
    \hline
    \textbf{DIM} & D,F        &  & D,F        & C,D & D,F       &   &   &  & D,F &  & C,A \\   
                   & B,A$\flat$ &  & B,A$\flat$ &     & B,A$\flat$ &   &   &  &     &  &    \\   
        \hline
    \textbf{AUG} &  & D$\flat$,F & & & &   & C,D$\flat$ & &  & & \\    
                   &  & A & & & &   &  & &  & & \\    

    \end{tabular}
    \end{center}
    \caption{Embedding irreducible chords into harmony starting at C.}
    \label{tbl:harmchord}
\end{table}
While it is useful to know all possibilities, in practice one wants a short list of chords per mode.  We provide this in Section~\ref{sec:vchord}.
\section{Lead Sheets and Harmony}
\label{sec:lead}
\subsection{Relating Chord Notation and Modes}
In this section, we bridge our framework to the conventional representation of harmonic ideas: lead sheet notation.

Lead sheets provide a concise way to denote harmonic structures in practice.\footnote{See Appendix~\ref{sec:appA} for a review of standard chord notation.}  We view harmony as a three-stage process:
\begin{center}
    {\Large
    HARMONY $\Rightarrow$ MODE $\Rightarrow$ VOICING \par}
\end{center}
A lead sheet chord symbol gives partial information about all three stages.  While there are standard defaults, the notation gives the performer flexibility to create their own arrangement.  Here we interpret lead sheets through the lens of harmony and voicings.

Consider Fmaj$^7$ (also written F$^{\Delta 7}$).  The letter F gives the bass note.  The ``maj7'' tells us the chord includes a major 3rd and 7th.  Any harmony containing these notes is in principle valid, so long as the melody fits.  We follow:

\textbf{Chords to Modes Correspondence Principle:}  Given the chord tones associated to a chord symbol, find the brightest mode that contains these chord tones.

For a major 7th chord, we need a mode with a major 3rd and major 7th.  F Lydian is the default: the brightest major mode containing these tones.  Alternatives exist (e.g., F Ionian, with B$\flat$ instead of B) but are less bright.  If the melody provides further information, we choose accordingly.

\begin{figure}[H]
    \begin{center}
    \begin{tabular}{ccc}  
        \includegraphics[width=0.9in]{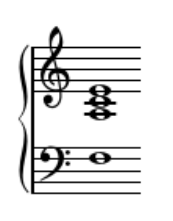} & 
        \includegraphics[width=0.9in]{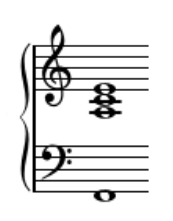} & 
        \includegraphics[width=0.9in]{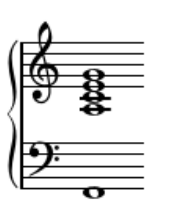} \\  
        (a) Fmaj7 & (b) Amin/F & (c) Amin7/F \\         
    \end{tabular}    
    \caption{Different voicings of Fmaj$^7$ chord.}
    \label{fig:voice}
    \end{center}
\end{figure}

Now for voicing.  Compare the three options in Figure~\ref{fig:voice}.  Option (a) is the literal Fmaj7 chord---four close tones, with the bass not low enough to stand out.  Option (b) separates the bass.  Option (c) is the standard ``rootless'' voicing: Amin7/F, which adds the 9th (G) for more color.  To request this explicitly, one writes Fmaj$^9$.  In practice, Fmaj$^7$ on a lead sheet means Amin7/F.
For Dmin$^7$: D is the bass, and we seek the brightest mode with minor 3rd and minor 7th---the Dorian mode (C major starting on D).  Separating the bass, a voicing is Fmaj/D; the standard jazz voicing highlighting the 9th is Fmaj7/D.

Our final example compares G$^7$ and G$^{7(sus4)}$.  Both map to G Mixolydian (brightest mode with minor 7th and major 3rd), but the voicings differ.  For G$^7$, a basic voicing is Bdim/G; the extended jazz voicing is Fmaj7$\sharp$11/G.  For G$^{7(sus4)}$, we avoid the major 3rd by convention, using Fmaj$^7$/G instead (which omits B).  Same harmony, different voicings to highlight different tones.

These examples illustrate both the power and subtlety of lead sheet notation.  A chord symbol does not specify a particular voicing but rather indicates which tones the harmony should include, freeing the performer to experiment.  We hope the voicing classification provides a systematic road map for creating arrangements from scratch.

\textbf{Lead Sheet vs.\ Voicing Notation:} Throughout this text, we used chord notation for concrete chords up to inversion.  Lead sheets use the same notation differently: a chord symbol suggests a harmony or voicing, not an exact chord.  To distinguish the two, we write Cmaj$^7$ for lead sheet symbols and Cmaj7 (typically over a bass) for voicings.
\subsection{Classification of Dominant Chords}
We view a chord symbol as implicitly embedded in a harmony, giving a scale for improvisation.  A given chord typically has many such embeddings, each yielding a different scale.  Let us apply this to classify all dominant chords.

A dominant chord contains a root, major third, and minor seventh.  The remaining tones can be altered.  To see what alterations are possible, we study all embeddings of the basic dominant chord into the seven harmonies.  Table~\ref{tbl:alter} lists each embedding's mode and the resulting alterations relative to the Mixolydian mode.  For example, an F dominant chord embeds as the 4th mode of C \textbf{MEL}.  Comparing to F Mixolydian (the B$\flat$ major scale), the only change is B$\flat$ $\to$ B---a sharp 11th.
\begin{table}[H]
    \begin{center}
    \begin{tabular}{l|l}
    mode & alterations\\
    \hline
    Mixolydian (5th mode of \textbf{MAJ}) & none \\
    4th mode of \textbf{MEL} & $\sharp$11 \\
    5th mode of \textbf{MEL} & $\flat$13 \\    
    7th mode of \textbf{MEL} & $\flat$9, $\sharp$9, no 5th, $\sharp$11, $\flat$13 \\    
    5th mode of \textbf{HMIN} & $\flat$9, $\flat$13 \\    
    5th mode of \textbf{HMAJ} & $\flat$9 \\    
    3rd mode of \textbf{HMAJ} & $\flat$9, $\sharp$9, no 11th, $\flat$13 \\    
    2nd mode of \textbf{DIM} & $\flat$9, $\sharp$9, $\sharp$11 \\    
    1st mode of \textbf{WTONE} & $\sharp$11, no 5th, $\flat$13 \\    
    \end{tabular}
    \end{center}
    \caption{Possible Dominant Chords.}
    \label{tbl:alter}
\end{table}
There are eight alterations of the basic dominant chord.  We generally choose voicings that reflect these alterations; some common options appear below.

\subsection{Summary of Chords/Modes Correspondence}
We conclude with a table of common chord symbols and their associated modes and voicings (Table~\ref{tbl:lead}).  Alternatives are shown in parentheses.
\begin{table}[H]
    \begin{center}
    \begin{tabular}{l|l|l}
    chord symbol & mode(s) & voicings (all /C)\\
    \hline
    Cmaj$^7$ (C$\Delta^7$,  CM$^7$)& C Lydian (or C Ionian), 4th mode of G \textbf{MAJ}& Emin7 (Cmaj6) \\ 
    Cmin$^7$ (C$-^7$,  Cm$^7$)& C Dorian, 2nd mode of B$\flat$ \textbf{MAJ} & E$\flat$maj7 \\ 
    C$^7$ (Cdom$^7$)& C Mixolydian, 5th mode of F \textbf{MAJ} & B\flat maj7$\flat$5 (Emin7$\flat$5) \\ 
    C$^{7(sus4)}$ & C Mixolydian, 5th mode of F \textbf{MAJ} & B\flat maj7\\ 
    C$^{7(\sharp 11)}$ & 4th mode of G \textbf{MEL} & Cdom$\sharp$11 \\ 
    C$^{7(\flat 13)}$ & 5th mode of F \textbf{MEL} & B$\flat$dom$\sharp$11 \\ 
    C$^{7(\flat 9, \flat 13)}$ & 5th mode of F \textbf{HMIN} & B$\flat$min7$\flat$5 \\ 
    C$^{\text{7alt}}$ (C$^{7(\sharp 9)}$) & 7th mode of D$\flat$ \textbf{MEL} & Emaj7$\flat$5 \\ 
    C$^{7(\flat 9)}$  & 2nd mode of B$\flat$ \textbf{DIM} & F$\sharp$dom7 (Gdim7)\\ 
    C$^{13(\sharp 9)}$  & 2nd mode of B$\flat$ \textbf{DIM}  & B$\flat$dem \\ 
    Cmin$^{maj 7}$  & 1st mode of C \textbf{MEL} & E$\flat$aug \\ 
    Cdim$^7$ (C$^{\circ}$)  & 1st mode of C \textbf{DIM} & Bdom7 (Cdim7)\\ 
    Cmin$^{7(\flat 5)}$ (C$^{\text{\O}}$)  & 2nd mode of B$\flat$ \textbf{HMIN} (C Locrian) & Cmin7$\flat$5 \\ 
    \end{tabular}
    \end{center}
    \caption{Chord Notation/Mode/Voicing Correspondence.}
    \label{tbl:lead}
\end{table}
\textbf{Remark:}  Chord notation can be imprecise.  A common mistake is writing $\flat 5$ for $\sharp 11$: technically the same note, but $\sharp 11$ implies the perfect 5th is still present while $\flat 5$ implies it is not.  Similarly, $\sharp 5$ is often used for $\flat 13$ even when the scale retains a perfect 5th.

\section{Lead Sheet for Amazing Grace}
\label{sec:leadag}
\begin{figure}[H]
    \begin{center}  
        \includegraphics[width=6.5in]{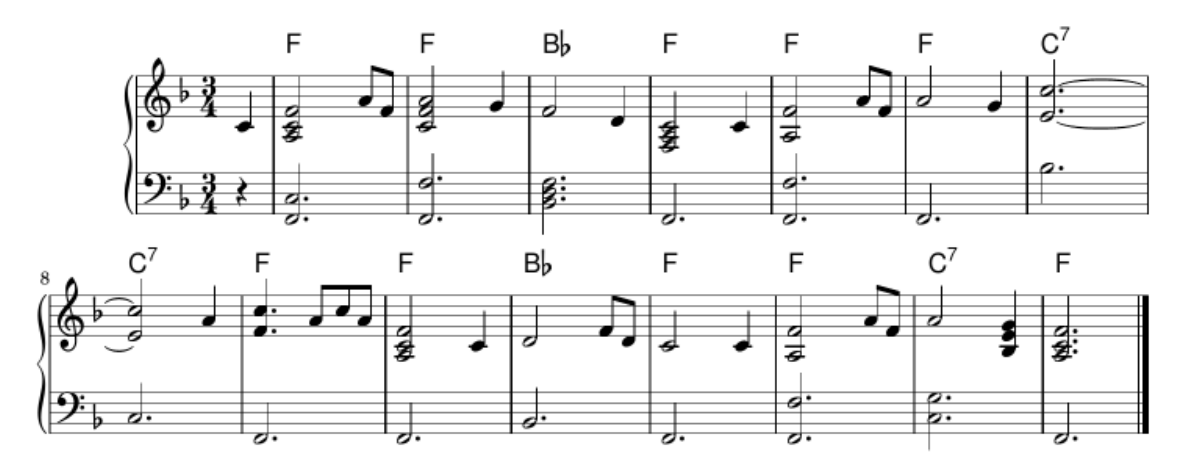}       
    \caption{Basic arrangement of \textit{Amazing Grace} with chord notation.}
    \label{fig:agl0}
    \end{center}
\end{figure}

We illustrate the lead sheet/voicing correspondence with \textit{Amazing Grace}.  Figure~\ref{fig:agl0} shows a simple arrangement with basic chord notation on top.  This lead sheet compactly represents the underlying harmony.  Such notation is rarely provided in classical sheet music, so we briefly explain how to infer it.

\begin{figure}[H]
    \begin{center}  
        \includegraphics[width=6.5in]{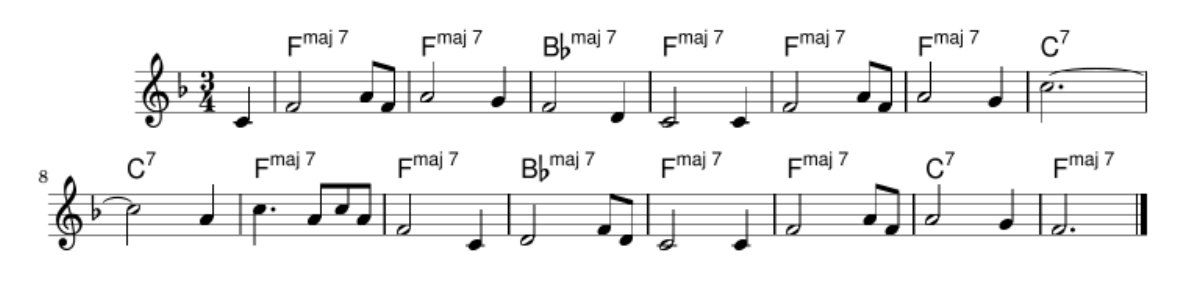}       
    \caption{Lead sheet for \textit{Amazing Grace}.}
    \label{fig:agl1}
    \end{center}
\end{figure}

\begin{figure}[H]
    \begin{center}  
        \includegraphics[width=6.5in]{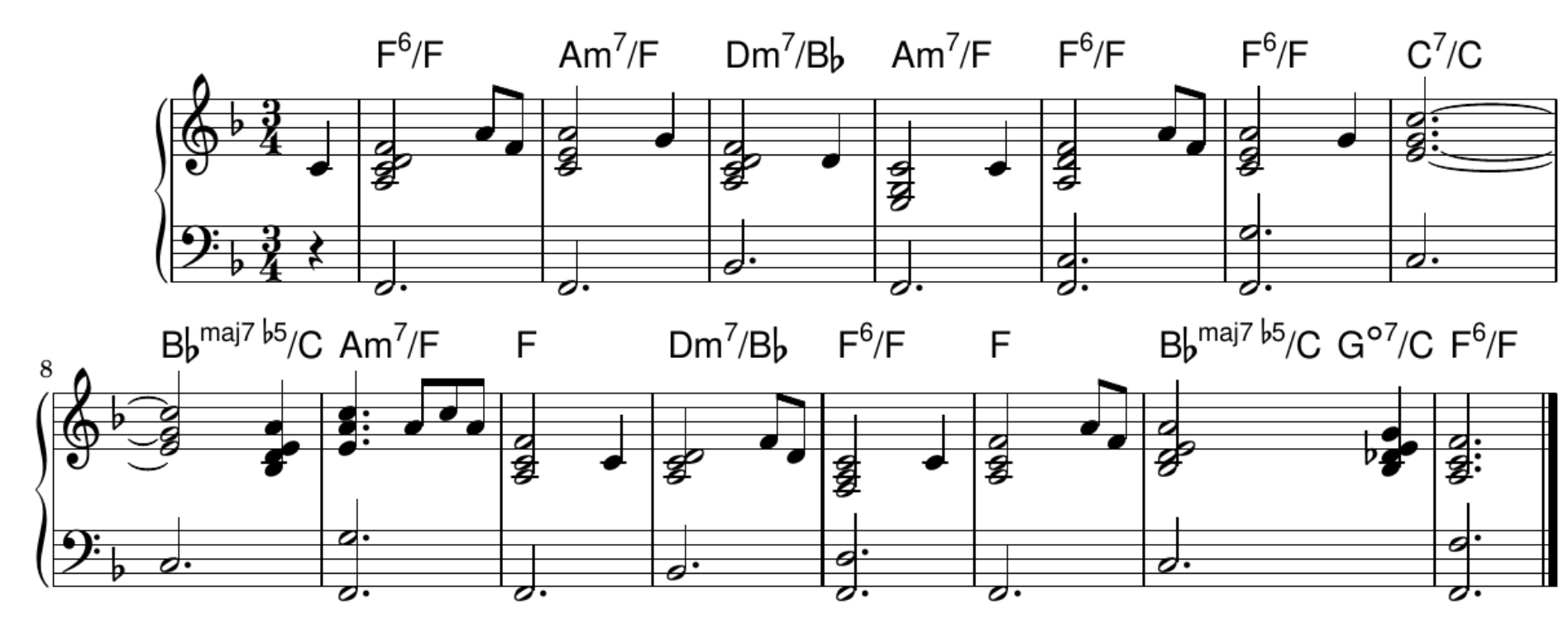}       
    \caption{Jazzy arrangement of \textit{Amazing Grace} with voicing chord notation.}
    \label{fig:agl2}
    \end{center}
\end{figure}

For each measure, we identify the chord from the notes present.  The lower note typically gives the bass of the mode, and the triads label the chords.  One exception: in the last bar of the first line, B$\flat$ in the bass with E forms the tritone of C$^7$---the true bass C is simply omitted, since the 5--1 progression (C$^7$ to F) is standard in this context.
With a basic lead sheet in hand, we add richer chord structure by replacing major (resp.\ minor) triads with major 7th (resp.\ minor 7th) chords, as in Figure~\ref{fig:agl1}.  This outlines the harmonic structure without specifying a particular arrangement.  Using the chord/harmony correspondence from the previous section, the performer can create their own arrangement, provide accompaniment, or improvise over the score.
We can go further and assign concrete voicings using Table~\ref{tbl:lead}.  Figure~\ref{fig:agl2} shows the result alongside an actual arrangement.  We use inversions of the standard voicings, possibly dropping tones and spreading chords across both hands.

Some notes on specific choices: In bar 2, we use F$^6$/F for the Fmaj$^7$ chord because the standard Amin$^{7}$/F lacks an F and our melody note is F.  Similarly, we use C$^7$ rather than B$\flat$maj$^{7(\sharp 11)}$ in the last bar of the first line because the melody requires it.  In the penultimate bar, we voice the C$^7$ chord with the diminished G$^{\text{\O}}$---a simple reharmonization replacing C$^7$ with C$^{7(\flat 9)}$ from diminished harmony.  Swapping a dominant chord for another dominant with the same tritone is a reliable reharmonization technique.
This example illustrates how to develop a lead sheet and then use our chord shapes to create an arrangement.  There is considerable flexibility in the process; we hope our framework provides a structured path through it.
\newpage
\section{Finale: Voicings the Modes}
\label{sec:vchord}
Knowing all irreducible chords and their embeddings gives a complete combinatorial picture, but it may be too much information for practical arrangement.  In Section~\ref{sec:lead} we associated a standard voicing with each mode; here we go further, associating a voicing with each \textit{tone} of a mode.  Mastering these tone-by-tone chord progressions lets the performer voice melodic lines on the spot.

\begin{figure}[H]
    \begin{center} 
    \includegraphics[width=6.7in]{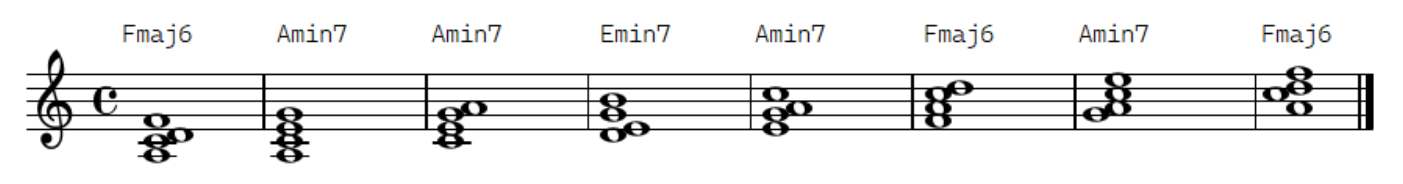} \\
    \caption{Voicing the tones of F Lydian mode (over F).}
    \label{fig:vclydian}
    \end{center}
\end{figure}

There is some ambiguity in these choices (Table~\ref{tbl:harmchord} lists several options per tone), but standard conventions exist.  Consider the voicing of F Lydian in Figure~\ref{fig:vclydian}, where all chords are played over F in the bass.
Each tone gets an irreducible chord with that tone on top.  The standard voicing Amin7/F (Table~\ref{tbl:lead}) and its inversions account for four of the seven tones; the remaining three use Fmaj6/F and Emin7/F.  These are not the only options, but they represent standard Lydian choices.  Learning this progression in all keys provides a systematic way to add chords to any melody.  In practice, as in Section~\ref{sec:black}, one need not play all notes in the right hand---some can move to the left hand or be dropped entirely.  The chord shape serves as a conceptual reference point.

\begin{figure}[H]
    \begin{center} 
    \includegraphics[width=6.7in]{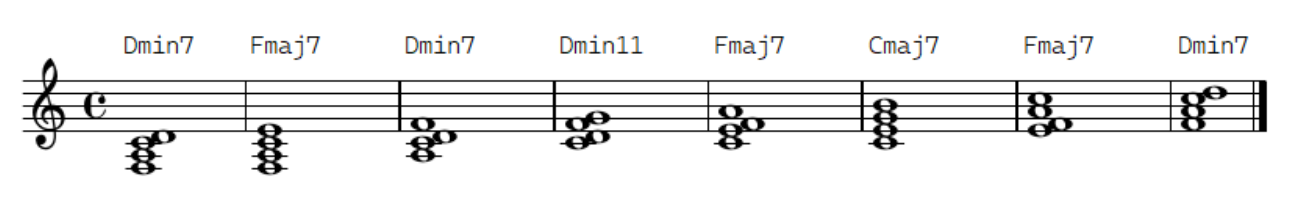} \\
    \caption{Voicing the tones of D Dorian mode (over D).}
    \label{fig:vcdorian}
    \end{center}
\end{figure}

For the D Dorian mode (same C major scale), the standard chord is Fmaj7/D.  Figure~\ref{fig:vcdorian} shows chords for every tone.  Although the scale is the same, the chords differ to emphasize the minor 3rd and 7th characteristic of Dorian.
For G Mixolydian (again the C major scale), we aim to emphasize the tritone (F and B), so standard options include Fmaj7$\flat$5/G or G7/G.  We also provide alternative, more dissonant voicings for some tones.

\begin{figure}[H]
    \begin{center} 
    \includegraphics[width=6.8in]{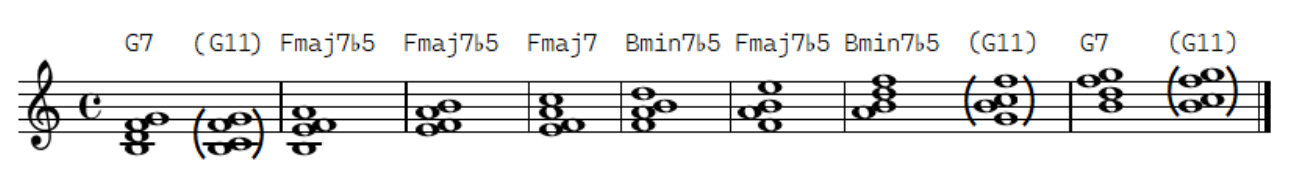} \\
    \caption{Voicing the tones of G Mixolydian mode (over G).}
    \label{fig:vcmixo}
    \end{center}
\end{figure}

The ``5-1'' progression (dominant to tonic) is one of the most common in music.  In C major, this is G7 to Cmaj.  The Mixolydian mode, associated with G7, naturally carries more dissonance before resolving; the alternative voicings can heighten this tension.  See Figure~\ref{fig:vcmixo}.
As a general principle, we choose chords so that the melody note is on top and no chord tone sits a semitone below it (which would obscure the melody).

\begin{figure}[H]
    \begin{center} 
    \includegraphics[width=6.7in]{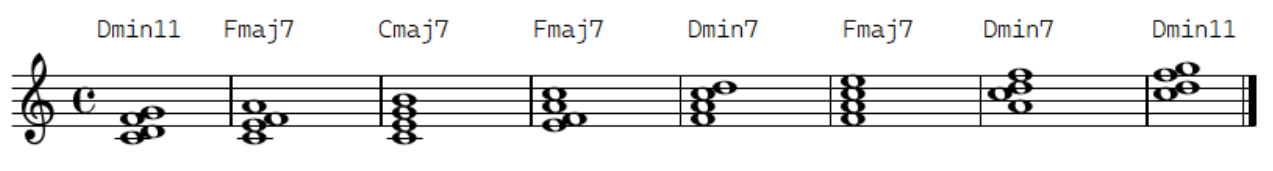} \\
    \caption{Suspended (Sus) voicing the tones of G Mixolydian mode (over G).}
    \label{fig:vcsus}
    \end{center}
\end{figure}

\begin{figure}[H]
    \begin{center} 
    \includegraphics[width=6.7in]{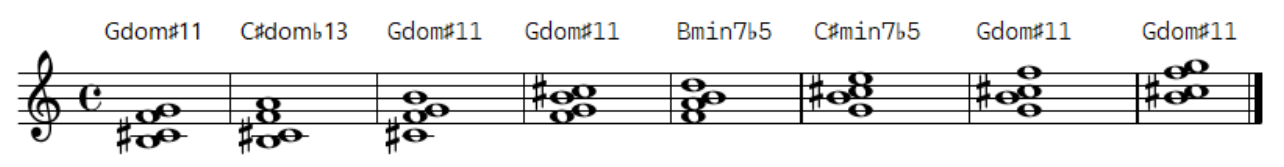} \\
    \caption{Voicing the G Lydian Dominant mode (4th mode of \textbf{MEL}) over G.}
    \label{fig:vcdomlyd}
    \end{center}
\end{figure}

\begin{figure}[H]
    \begin{center} 
    \includegraphics[width=6.2in]{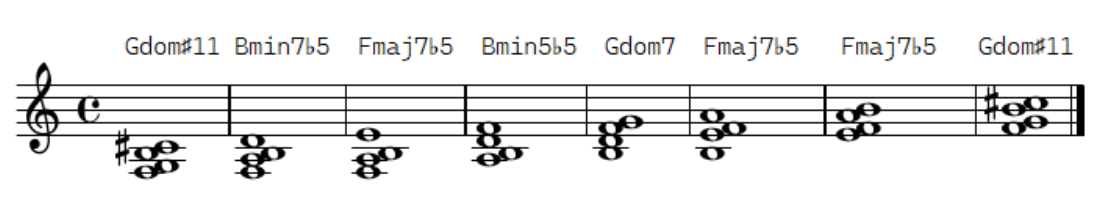} \\
    \caption{Voicing the tones of C$\sharp$ altered mode (7th mode of \textbf{MEL}) over C$\sharp$.}
    \label{fig:vcalt}
    \end{center}
\end{figure}

\begin{figure}[H]
    \begin{center} 
    \includegraphics[width=6.6in]{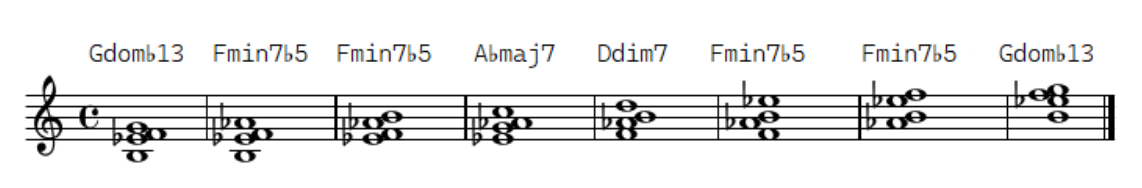} \\
    \caption{Voicing the 5th mode of \textbf{HMIN} (over G).}
    \label{fig:vchmin5}
    \end{center}
\end{figure}

\begin{figure}[H]
    \begin{center} 
    \includegraphics[width=6.6in]{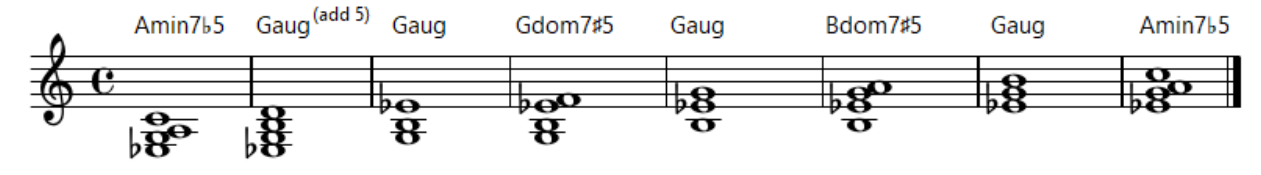} \\
    \caption{Voicing the 1st mode of \textbf{MEL} (over C).}
    \label{fig:vmel}
    \end{center}
\end{figure}

\begin{figure}[H]
    \begin{center} 
    \includegraphics[width=6.6in]{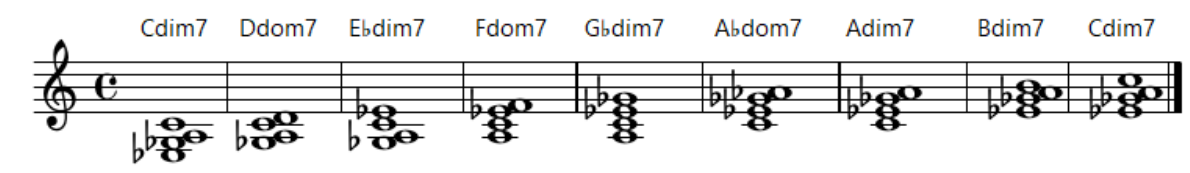} \\
    \caption{Basic voicing of $\textbf{DIM}$ (over any bass).}
    \label{fig:vcdima}
    \end{center}
\end{figure}

\begin{figure}[H]
    \begin{center} 
    \includegraphics[width=6.6in]{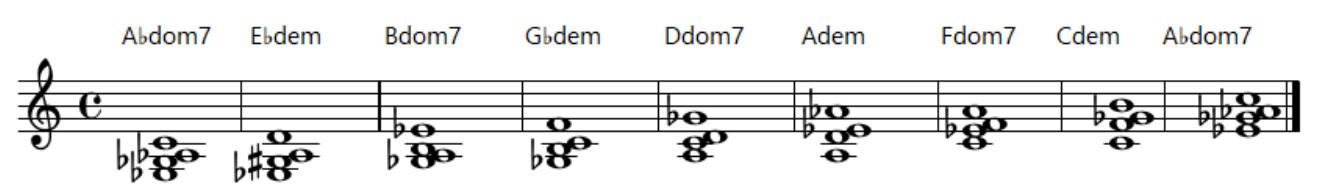} \\
    \caption{Complex voicing of $\textbf{DIM}$ (over any bass).}
    \label{fig:vdimb}
    \end{center}
\end{figure}

Even within a fixed mode, different voicings can emphasize different tones.  Figure~\ref{fig:vcsus} takes the Dorian chords from Figure~\ref{fig:vcdorian} and places them over G, giving a ``suspended'' Mixolydian voicing that avoids the tritone.  This illustrates how changing only the bass can create a new voicing of the same mode.
We now voice several other common jazz modes.  Figure~\ref{fig:vcdomlyd} shows the Lydian Dominant (4th mode of \textbf{MEL}), which differs from Mixolydian by having a $\sharp$11.  We choose chords that include this $\sharp$11 when possible.  Over the second tone (A), we use the composite chord C$\sharp$dom$\flat$13, formed from C$\sharp$aug plus a minor 7th (B).
Figure~\ref{fig:vcalt} voices the 7th mode of \textbf{MEL} (the altered scale), which alters every tone except the 3rd and minor 7th.  A quick mnemonic: the altered scale is the Lydian Dominant a tritone away (e.g., C$\sharp$ altered = G Lydian Dominant).  We use the standard G Mixolydian voicing except when the top note is the $\sharp$11, where we borrow from Lydian Dominant.
Next, we voice the 2nd and 5th modes of \textbf{HMIN}, the most commonly used harmonic minor modes.  We draw the primary chord from Table~\ref{tbl:lead} and fill in the remaining tones from Table~\ref{tbl:harmchord}, choosing more dissonant voicings for the 5th mode (associated with a dominant chord).  Note the composite chord Gdom$\flat$13 in Figure~\ref{fig:vchmin5}, formed from Gaug plus a minor 7th.
\begin{figure}[H]
    \begin{center} 
    \includegraphics[width=6.6in]{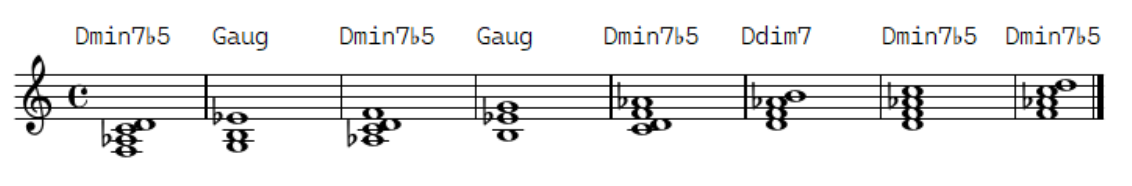} \\
    \caption{Voicing the 2nd mode of \textbf{HMIN} (over D).}
    \label{fig:vchmin2}
    \end{center}
\end{figure}
For the 1st mode of \textbf{MEL} (melodic minor), which differs from Ionian by having a minor 3rd, we highlight the aug chord---absent from major harmony---and use it heavily (Figure~\ref{fig:vmel}).  Some voicings involve composite chords such as Gdom7$\sharp$5 and Gaug$^{(\text{add 5})}$ (see Section~\ref{sec:complete}).
Finally, we turn to \textbf{DIM}.  Recall that this harmony consists of two diminished 7th chords a semitone apart: for C \textbf{DIM}, the ``upper'' Cdim7 and ``lower'' Bdim7.  Whether the mode is associated with a dominant or diminished chord, we want to highlight the upper Cdim7 (which contains the key tritones).  To voice the upper tones, we use Cdim7 directly.  For the lower tones, we move the top note of a Cdim7 inversion down a semitone: e.g., (C, E$\flat$, G$\flat$, A) becomes (C, E$\flat$, G$\flat$, A$\flat$), an inversion of A$\flat$dom7.  See Figure~\ref{fig:vcdima}.
Practicing this progression for all three diminished scales builds facility with diminished chords.  For more complex voicings (Table~\ref{tbl:harmchord}), we can replace a non-top tone in Cdim7 by moving it down a semitone (staying in the scale): e.g., altering C, E$\flat$, or G$\flat$ in (C, E$\flat$, G$\flat$, A) yields inversions of Bdom7, Ddom7, or Fdom7 respectively.  For the lower tones, dem shapes offer exotic alternatives---Adem = (A, D, E$\flat$, A$\flat$) voices the A$\flat$ tone.  Figure~\ref{fig:vdimb} shows these richer voicings.
In this section, we gave sample voicings of popular modes using the chord shapes from the text.  Practicing these chord progressions in several keys will build the ability to voice any mode on the spot and create complex arrangements.
\newpage
\appendix
\section{Basic Music Notation}
\label{sec:appA}
We summarize basic interval and chord notation used in the text.

\begin{figure}[H]
    \begin{center} 
    \includegraphics[width=5.2in]{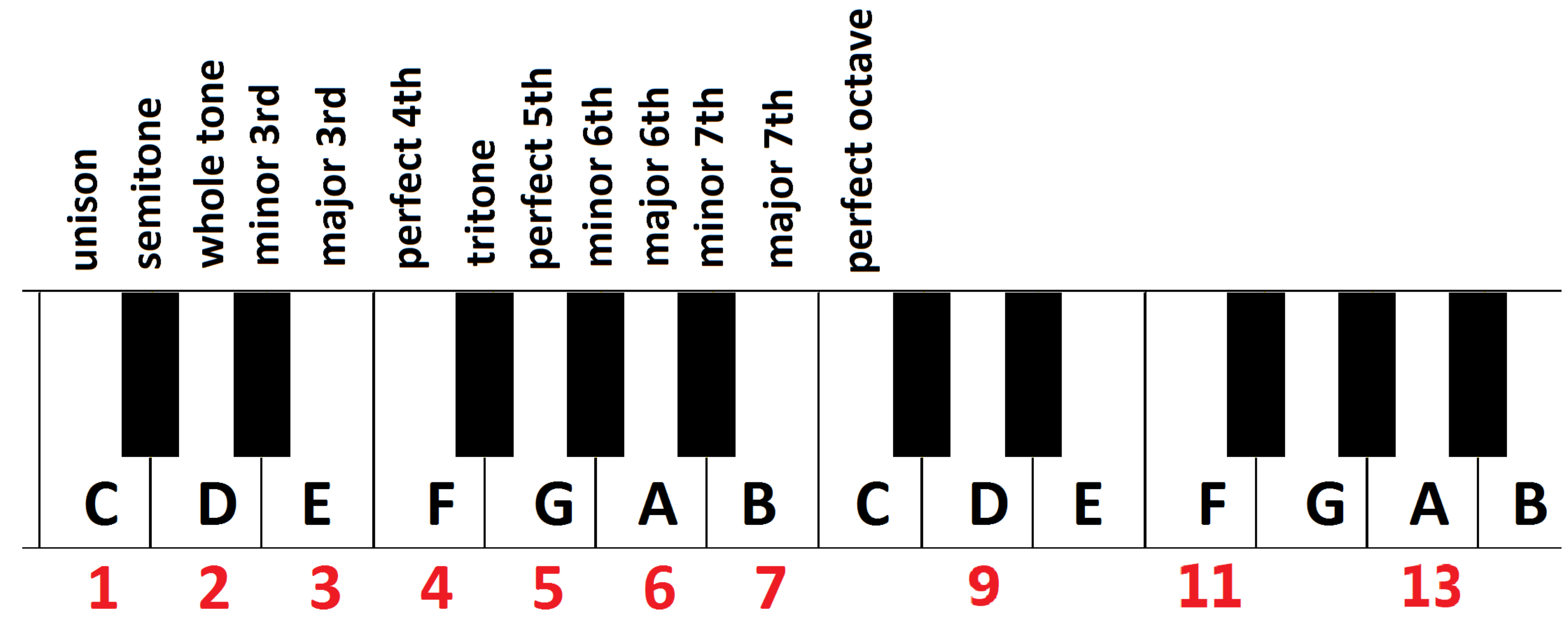}
    \caption{Piano intervals starting at C.}
    \label{fig:intervals}
    \end{center}
\end{figure}

Figure~\ref{fig:intervals} displays the piano intervals starting at C.
In jazz, we often use intervals extending beyond an octave.  Most commonly, these are the 9th, 11th and 13th.  Starting from C, these correspond to D, F, A that are marked by 9, 11, and 13 in the figure.

Table~\ref{tbl:triads} lists the standard triads (starting in C) with their chord symbols and common alternatives.
\begin{table}[H]
    \begin{center}
    \begin{tabular}{l|l|l}
    chord name & chord symbol & chord tones (starting in C) \\
    \hline
    Major Triad & C, CM, Cmaj& (C, E, G) \\
    Minor Triad  & Cm, Cmin, C- & (C, E${\flat}$, G) \\
    Augmented Triad  & Caug, C+ & (C, E, G${\sharp}$) \\
    Diminished Triad  & Cdim & (C, E${\flat}$, G${\flat}$) \\
    \end{tabular}
    \end{center}
    \caption{Basic triads.}
    \label{tbl:triads}
\end{table}
For each of these triads, we can form inversions by moving one of the tones up or down an octave. For example, an inversion of Cmaj is given by (E, G, C) where the last note is obtained by moving C up an octave.  For the purposes of harmony, we tend not to distinguish between a chord and its inversions.

Now for common 4-note chords.  The C$^6$ chord adds the 6th degree to a major triad (alternatively written Cmaj$^{(\text{add 6})}$).

Most 4-chords are 7th chords, common in jazz, built by adding a 3rd and 7th to the root and 5th.  Table~\ref{tbl:4chords} lists the standard ones; each extends a basic triad (e.g., Cmaj$^7$ adds a major 7th to a major triad).
\begin{table}[H]
    \begin{center}
    \begin{tabular}{l|l|l}
    chord name & chord symbol & chord tones (starting in C) \\
    \hline
    Major 7th & Cmaj$^7$, C$^{\Delta 7}$ & (C, E, G, B)\\
    Major 6th & C$^6$ & (C, E, G, A)\\
    Minor 7th  & Cmin$^7$,  C-$^7$ & (C, E${\flat}$, G, B${\flat}$) \\
    Minor 11th  & Cmin$^{11}$,  C-$^{11}$ & (C, E${\flat}$, F, B${\flat}$) \\
    Dominant 7th  & C$^7$ & (C, E, G, B${\flat}$) \\
    Suspended 7th & C$^{7(sus4)}$ & (C, F, G, B${\flat}$) \\
    Diminished 7th  & Cdim$^7$, C$^{\circ}$ & (C, E${\flat}$, G${\flat}$, A) \\
    Half-Diminished 7th & Cmin$^{7(\flat 5)}$, C$^{\text{\O}}$ &  (C, E${\flat}$, G${\flat}$, B${\flat}$) \\
    \end{tabular}
    \end{center}
    \caption{Common 4-chords.}
    \label{tbl:4chords}
\end{table}
For all these chords, we can also consider inversions that move one of the tones up or down an octave.  For example, if we move the A in C$^6$ down an octave, we obtain the Amin$^7$ chord.

Higher extensions involve the 9th, 11th, and 13th degrees.  For example, C$^9$, C$^{11}$, and C$^{13}$ extend C$^7$ by adding these tones, typically dropping the 5th to keep the chord at four notes.  There is inherent ambiguity in these extensions: they indicate which tones to make explicit, not which notes to play exactly.  This flexibility allows the performer to create their own arrangement.
\newpage 
\section{A Beginner's Guide to Improvisation}
\label{sec:appB}
This appendix provides a starting point for learning to improvise, aimed at absolute beginners.  Packings---exemplified by the pentatonic scale---offer a structured way to form melodic lines with minimal complexity.  Because they lack dissonance, almost any sequence of packing notes sounds interesting and any subset forms a usable chord.  This generalizes to all irreducible packings.

The missing ingredient is \textbf{when} to use each packing.  Jazz chord notation provides the answer: even without understanding what the symbols mean, we can associate each chord with a packing.  This gives the beginner a concrete set of notes over each chord, enabling improvisation from the very first lesson.

Table~\ref{tbl:chordpack} lists common chords (all in C) with a suggested packing.  The beginner need not understand what the chord symbol means---it simply triggers a packing choice.
\begin{table}[H]
\begin{center}
\begin{tabular}{l|l}
chord symbol (in C) & packing(s) \\
\hline
Cmaj (C) & C \textbf{penta} \\
Cmaj$^7$ (C$\Delta^7$,  CM$^7$) & G \textbf{penta} or C \textbf{penta} \\
Cmin$^7$ (C$-^7$,  Cm$^7$) & E$\flat$  \textbf{penta} or B$\flat$ \textbf{penta} \\
C$^7$ & C \textbf{dpenta} \\
C$^{7(sus4)}$ & B$\flat$ \textbf{penta}\\
C$^{7(\sharp 11)}$ & D \textbf{dpenta} \\
C$^{7(\flat 9, \flat 13)}$ &  B$\flat$ \textbf{dim} or  D$\flat$ \textbf{hmaj}\\
C$^{\text{7alt}}$ &  F$\sharp$ \textbf{dpenta}\\
Cmin$^{7(\flat 5)}$ (C$^{\text{\O}}$) & C \textbf{dim} or G$\flat$ \textbf{hmaj}\\
Cdim$^7$ (C$^{\circ}$) & C \textbf{dim} \\
Cmin$^{maj 7}$ & F \textbf{dpenta}\\
\end{tabular}
\end{center} 
\caption{Chord Symbols/Packings Translation.}
\label{tbl:chordpack}
\end{table}
The table gives packings for C in the bass; in practice, transpose to match the chord.  For example, over Dmin$^7$, the table says to take the pentatonic starting at the minor 3rd---which is F when the bass is D, giving F \textbf{penta}.

To practice: find the packing for a chord, play it with the right hand while holding the chord in the left.  We hope this translation provides a concrete entry point that motivates deeper study of the theory behind it.
\newpage
\section{Pythagorean Tuning}
\label{sec:appC}
We take a closer look at how the piano is tuned.  Start with a string tuned to a fundamental frequency $f_0$.  The simplest way to add related notes is to double the frequency (halve the string length), producing an octave; for the purposes of harmony, we do not distinguish tones an octave apart.  The next new overtone is $3f_0$, which we place in the octave between $f_0$ and $2f_0$ by halving to $\frac{3}{2}f_0$.  The ratio 3:2 is the \textbf{perfect fifth}, the simplest nontrivial consonant ratio.

Pythagorean tuning iterates this: take $\frac{3}{2}$ of $\frac{3}{2}f_0 = \frac{9}{4}f_0$, then divide by 2 to get $\frac{9}{8}f_0$ back in the octave.  Continuing indefinitely gives finer subdivisions.  The ratio is always a power of 3 over a power of 2, so we can never return exactly to $f_0$:
\begin{equation}
3^m \neq 2^n
\end{equation}
because the left side is odd and the right side even.  We must stop at some reasonable point.

\begin{figure}[H]
    \begin{center}
    \begin{tabular}{ccc}
        \includegraphics[width=1.7in]{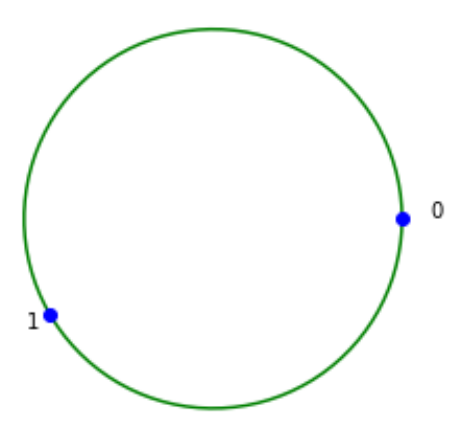}& 
        \includegraphics[width=1.8in]{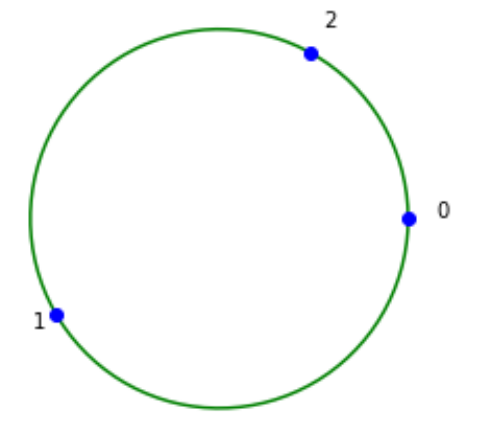} & 
        \includegraphics[width=1.7in]{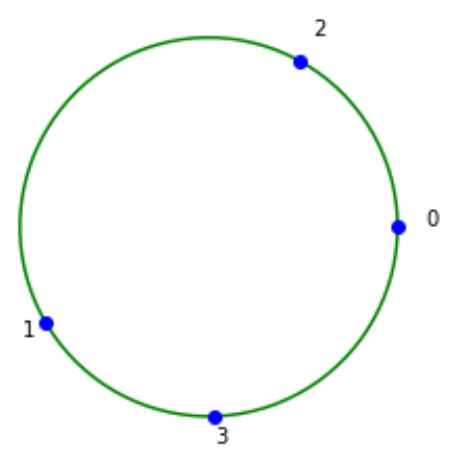}\\
        (a) & (b) & (c) \\
        \includegraphics[width=1.7in]{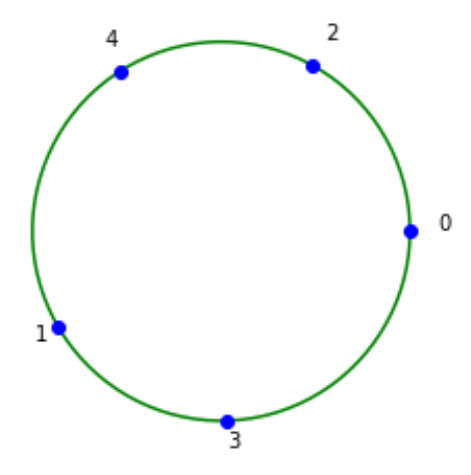}& 
        \includegraphics[width=1.8in]{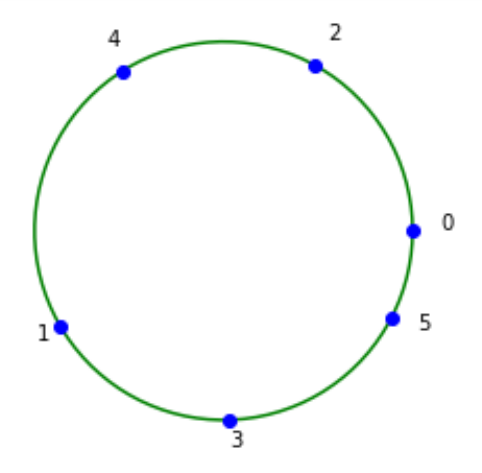} & 
        \includegraphics[width=1.7in]{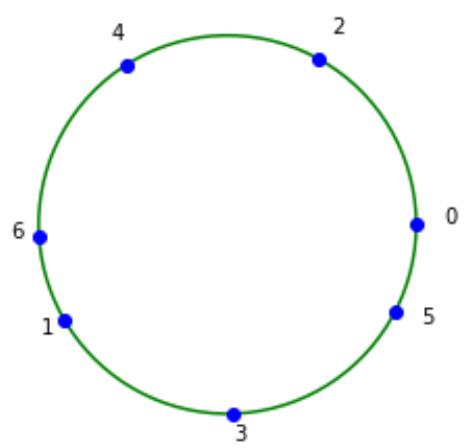}\\
        (d) & (e) & (f) \\

        \includegraphics[width=1.7in]{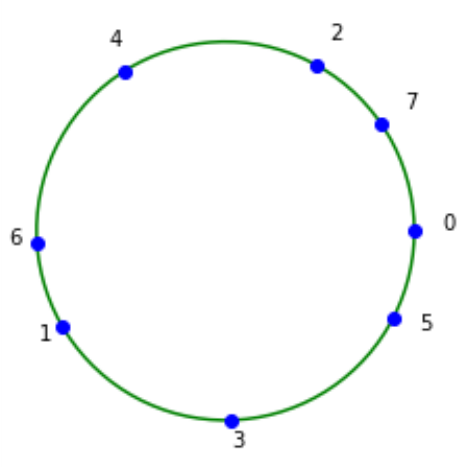}& 
        \includegraphics[width=1.8in]{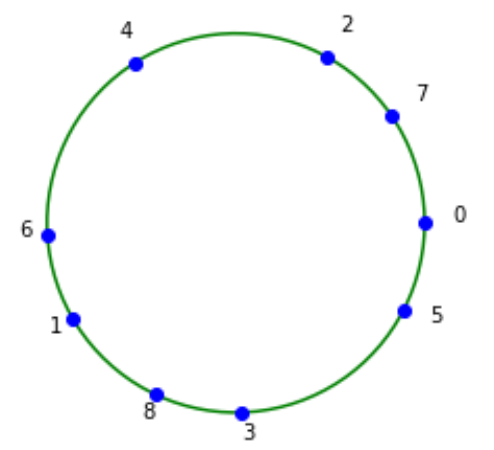} & 
        \includegraphics[width=1.9in]{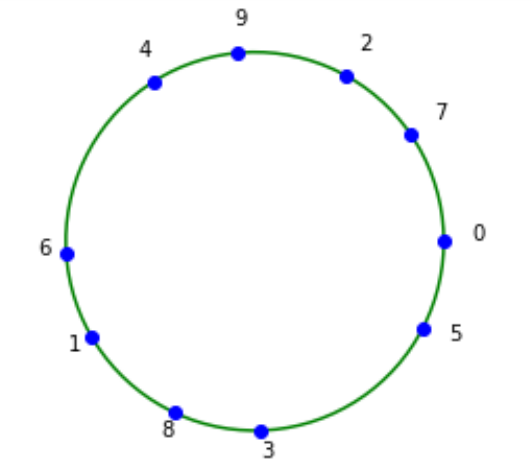}\\                 
        (g) & (h) & (i) \\

        \includegraphics[width=1.8in]{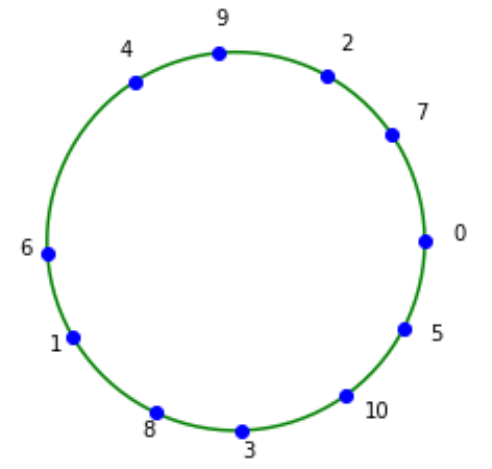}& 
        \includegraphics[width=1.8in]{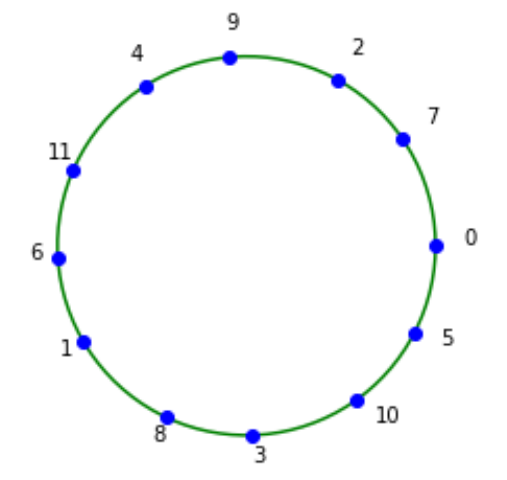} & 
        \includegraphics[width=1.7in]{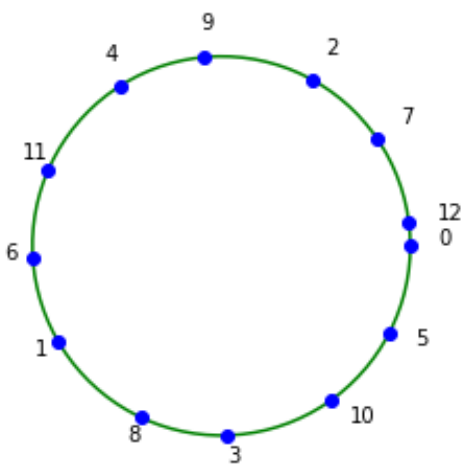}\\ 
        (j) & (k) & (l) \\

    \end{tabular}
    \caption{Pythagorean tuning.}
    \label{fig:pyth}
    \end{center}
\end{figure}

To visualize the procedure, we work with the base-2 logarithm of frequency multiples.  Recall that 
\begin{equation}
    log(2^n) = n
\end{equation}
and that 
\begin{equation}
    log(A\cdot B) = log(A) + log(B) 
\end{equation}
\begin{equation}
    log(\frac{A}{B}) = log(A) - log(B) 
\end{equation}
The logarithm turns ratios into differences, making the relationships linear and easy to visualize.  Since we identify frequencies up to octaves, it is natural to place the logarithmic values on the unit circle: one full rotation equals doubling the frequency, and $\frac{3}{2}f_0$ corresponds to an angle of $2\pi \cdot \log(3/2)\approx 1.58$ radians ($\approx 210°$).  The Pythagorean procedure amounts to repeatedly rotating by this angle.  See Figure~\ref{fig:pyth}.

We can fill the circle as densely as we like but never return exactly to the start.  Our stopping criterion is:

\textbf{Stopping Criterion:} The next iteration will bring us closer to the starting point than anything previously.  Concretely, the arc distance to the initial point will be less than half the minimal distance we have seen before.

In Figure~\ref{fig:pyth}, starting with the base frequency and the perfect fifth, the next note (labeled 2) lands much closer to the start than either the 5th or the 4th ((b)).  Taking F as the fundamental, we get F, C, G---with G only a tone from F.  So we could stop at two notes, but this is trivial.

The next place to stop is after five notes: F, C, G, D, A.  The sixth note, E ((e)), is a semitone from F---much closer than the whole-tone intervals seen so far.  The notes F, C, G, D, A form the F major pentatonic scale, subdividing the circle relatively uniformly (the largest interval is at most double the smallest).  This explains why pentatonic scales are so prevalent in world music: they are the simplest Pythagorean subdivision of the octave.

Continuing, the next natural stop is at 12 notes: note 13 lands much closer to the start than any previous interval ((l)).  The circle is now roughly equally subdivided into semitone-sized intervals; adding more notes would violate this uniformity.  Further iterations would introduce microtonal intervals---possible in principle, but requiring at least a doubling of notes, a high cost in complexity.

\begin{figure}[H]
    \begin{center}
    \includegraphics[width=1.3in]{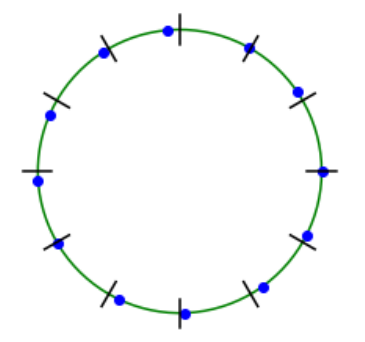}
    \caption{Pythagorean tuning vs equal temperament tuning for a 12-tone subdivision.}
    \label{fig:equalA}
    \end{center}
\end{figure}

Modern tuning systems use equal temperament, subdividing the octave into 12 equal intervals.  Figure~\ref{fig:equalA} compares the Pythagorean 12-tone tuning to equal temperament: the Pythagorean subdivision approximates equal temperament reasonably well.  Why bother with the Pythagorean approach at all?  One can always subdivide an octave into $n$ equal parts, but such divisions need not approximate the 3:2 ratio, potentially producing dissonant chords.  The Pythagorean approach guarantees consonance by construction.  See Appendix~\ref{sec:appD} for more on equal temperament.

As a final comment, there is an important dichotomy.  \textit{Melodically}, microtones are quite natural: the human voice continuously interpolates between pitches, and pitch bending emulates this in the 12-tone system.  \textit{Harmonically}, however, using microtones means building chords in an expanded system---typically at the edge of what the ear can distinguish.

\newpage
\newpage
\section{Equal Temperament Tuning}
\label{sec:appD}
As observed in Section~\ref{sec:piano}, the 3:2 ratio does not quite close up after 12 steps.  After 12 iterations, we have
\begin{equation}
    1 < \frac{3^{12}}{2^{19}} \approx 1.013\cdots <2
\end{equation}
very close to, but not exactly, the starting frequency.  The equal temperament solution replaces the perfect fifth with an approximation that closes exactly after 12 steps:
\begin{equation}
r = 2^\frac{7}{12} \approx 1.498
\end{equation}
which gives us
\begin{equation}
   r^{12}  = (2^\frac{7}{12})^{12} = 2^7
\end{equation}
exactly 7 octaves from the base frequency.  This subdivides the octave into 12 equal semitones, each with ratio $2^\frac{1}{12}$---hence the name equal temperament.

\begin{figure}[H]
    \begin{center} 
    \includegraphics[width=4.5in]{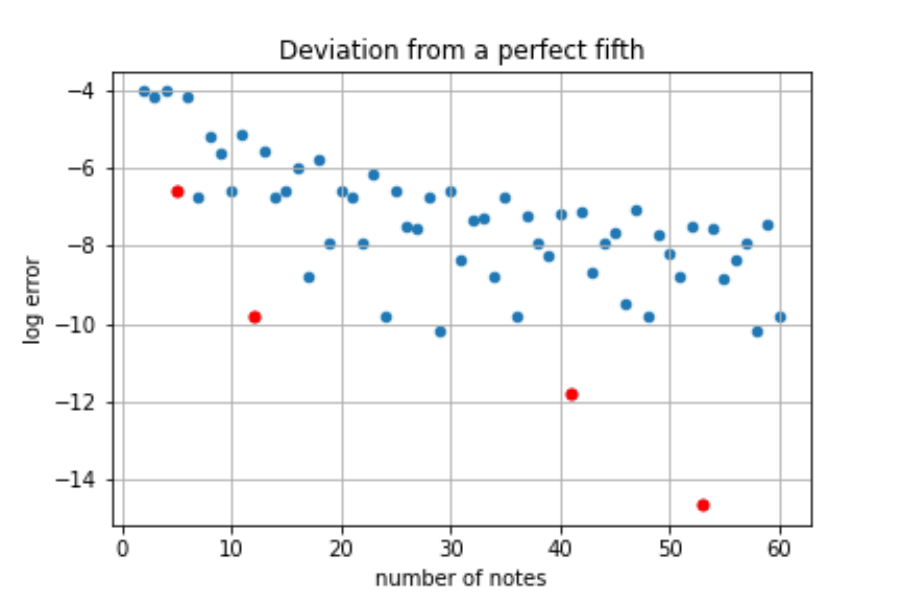}
    \caption{Error plot for equal temperament tuning.}
    \label{fig:equal}
    \end{center}
\end{figure}

More generally, we can subdivide the octave into $n$ equal intervals (ratio $2^{\frac{1}{n}}$), but such a tuning need not approximate the perfect fifth well.  Figure~\ref{fig:equal} plots the log error between a perfect fifth and the closest tone in an $n$-note equal temperament scale.
Interesting values are highlighted in red.  The first occurs at $n=5$ (the pentatonic scale), confirming it as the simplest useful subdivision.  A large drop occurs at $n=12$ (the standard system), with further drops at $n=41$ and $n=53$, suggesting larger tuning systems.

\textbf{Good tunings and continued fraction expansions:} These tuning sizes arise systematically from the \textit{continued fraction expansion} of $\log_2(\frac{3}{2})$, which provides progressively better rational approximations.
We seek fractions $\frac{a}{b}$ such that $2^{\frac{a}{b}}$ approximates $\frac{3}{2}$, or equivalently, $\frac{a}{b} \approx \log(\frac{3}{2}) \approx 0.58496250$.  The continued fraction gives:
\begin{equation}
 \frac{1}{1},   \frac{1}{2}, \frac{3}{5}, \frac{7}{12}, \frac{24}{41}, \frac{31}{53},  \frac{179}{306},  \frac{389}{665},  \frac{9126}{15601}, \dots
\end{equation}
The denominator gives the number of subdivisions; the numerator identifies the tone approximating the fifth.  For example, $\frac{7}{12}$ is the standard 12-tone scale with the fifth at tone 7.  After the pentatonic and 12-tone scales, 41 and 53 tones give good approximations; the next is 306, far too large for practical use.  The 53-tone system has been explored: in 1876, Robert Bosanquet built a harmonium with fifty-three notes per octave.  For more on continued fractions and music, see Benson~\cite{benson2006}.

\section{The Overtone Series and Tunings}
\label{sec:appE}
\begin{figure}[H]
    \begin{center}
    \begin{tabular}{ccc}
        \includegraphics[width=1.84in]{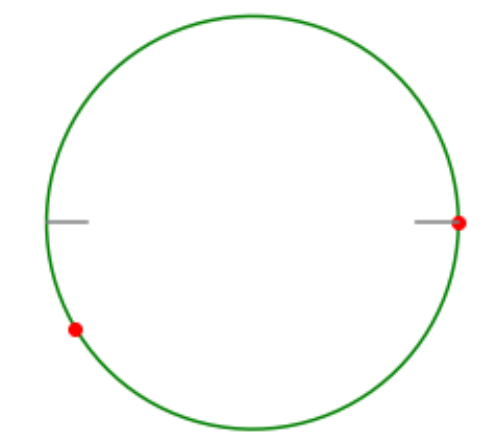}& 
        \includegraphics[width=1.8in]{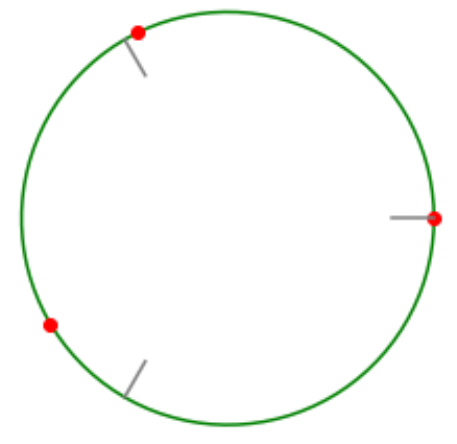} & 
        \includegraphics[width=1.8in]{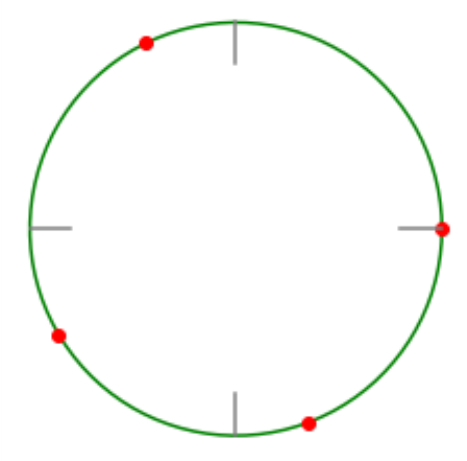}\\
        (a) & (b) & (c) \\
        \includegraphics[width=1.8in]{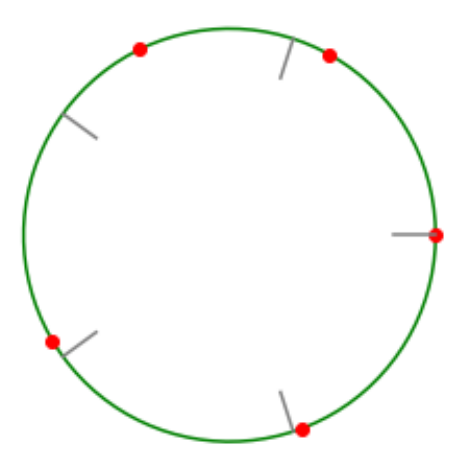}& 
        \includegraphics[width=1.8in]{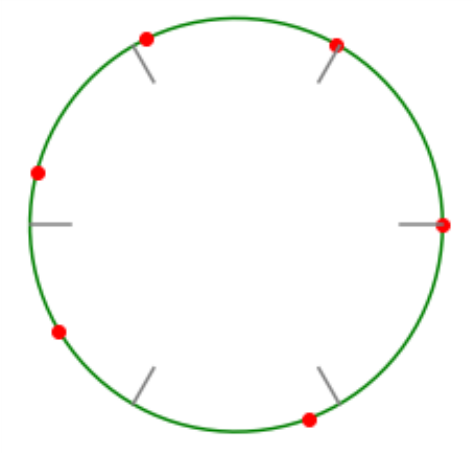} & 
        \includegraphics[width=1.8in]{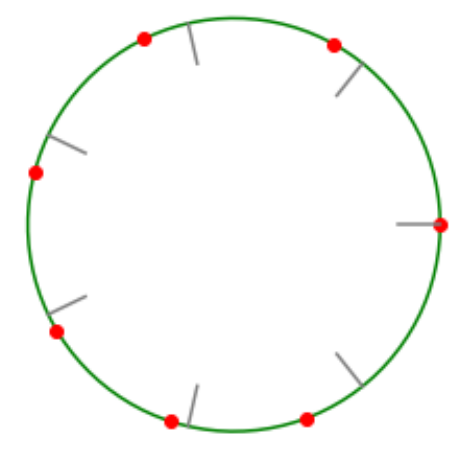}\\
        (d) & (e) & (f) \\
        \includegraphics[width=1.8in]{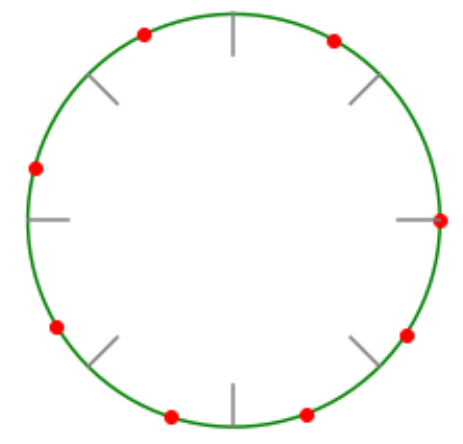}& 
        \includegraphics[width=1.8in]{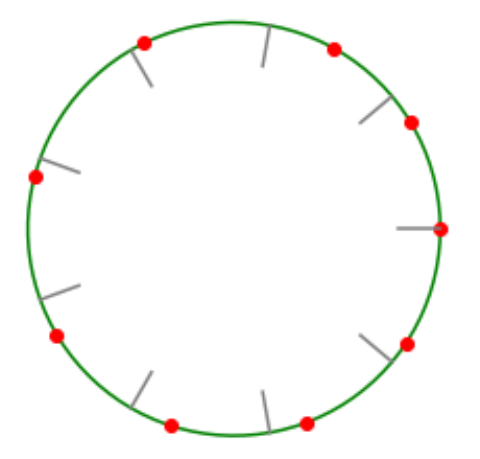} & 
        \includegraphics[width=1.8in]{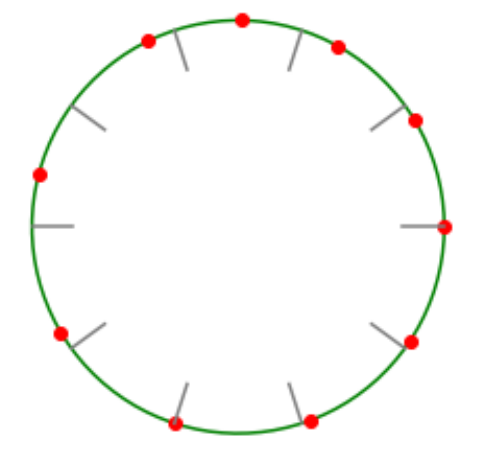}\\                 
        (g) & (h) & (i) \\
        \includegraphics[width=1.8in]{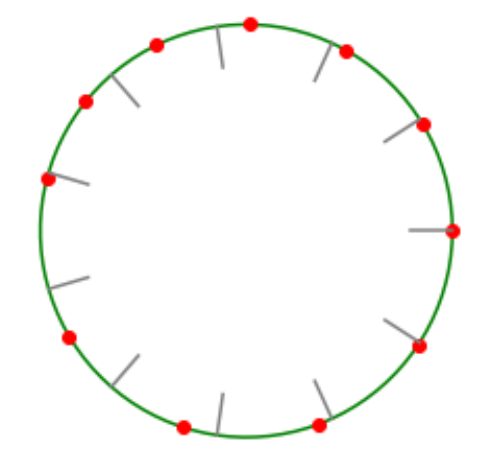}& 
        \includegraphics[width=1.8in]{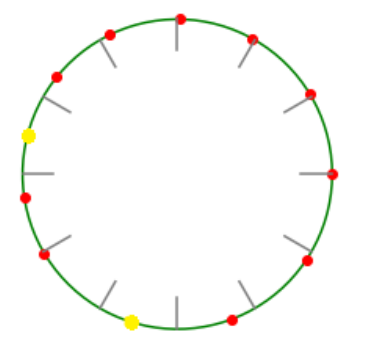} & 
        \includegraphics[width=1.8in]{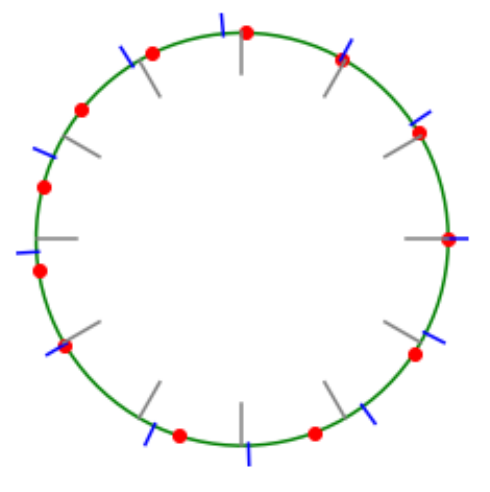}\\ 
        (j) & (k) & (l) \\

    \end{tabular}
    \caption{Overtone series tuning.}
    \label{fig:harmonic}
    \end{center}
\end{figure}

We briefly explore a tuning based on the full overtone series, rather than just the 3:1 ratio of Pythagorean tuning.  For a bass frequency $f_0$, the overtone (harmonic) series is:
\begin{equation}
    \text{$f_0$, $2f_0$, $3f_0$, $4f_0$, $5f_0$, $\dots$}
\end{equation}
The first overtone ($2f_0$) defines an octave; $3f_0$ gives the perfect fifth (ratio 3:2).  When an instrument plays a note, the string vibrates at the fundamental and its overtones, whose relative strengths determine the instrument's timbre.  Ignoring even multiples (which are octave copies), the odd overtones are:
\begin{equation}
    \text{$f_0$, $3f_0$, $5f_0$, $7f_0$, $9f_0$, $11f_0$, $13f_0$, $\dots$}
\end{equation}
Note that in the Pythagorean tuning, we only used overtones that are powers of 3:
\begin{equation}
    \text{$f_0$, $3f_0$, $9f_0$, $27f_0$, $81f_0$, $243f_0$, $\dots$}
\end{equation}
Pythagorean tuning uses only powers of 3 among these overtones.  Since higher overtones occur naturally, we ask whether the full series can serve as a tuning.  Figure~\ref{fig:harmonic} plots the logarithm of the first 12 odd overtones against equal temperament.  The fit is poor: in (k), the 7th and 10th tones (yellow) fall approximately halfway between equal temperament positions.  In (l), comparison with the Pythagorean 12-tone tuning (blue) confirms that powers of 3 approximate equal temperament much better.

We conclude that higher overtones are inconsistent with equal temperament.  Moreover, it is unclear how perceptually significant the 10th overtone is for timbre, so including all overtones in a tuning framework may be misguided.

\newpage
\newpage
\section{Triad Chords and Harmony}
\label{sec:appF}
We briefly examine triads---central to traditional harmony---and explain how they fit into our framework.  Two distinct tones form a \textbf{dissonant tone block} if they are a semitone or tone apart (up to octave shifts).  This is the ``block'' analogue of the tone cells from our voicing discussion.  The basic triads admit a clean characterization:

\textbf{Characterization of Triads:} The tone block complete sets of chords with zero blocks correspond exactly to the major, minor, augmented and diminished triads.

This parallels the classification of harmonies and packings.  One could also consider the irreducible case; the ``Sus chord'' (e.g., \{C,D,G\}) is a tone-block irreducible example.

From our perspective, triads are block-incomplete objects obtained from their completions by removing tones.  For example, the major triad is a major 6th chord minus the 6th degree.

\begin{figure}[H]
    \begin{center} 
    \begin{tabular}{ccc}
    \includegraphics[width=1.9in]{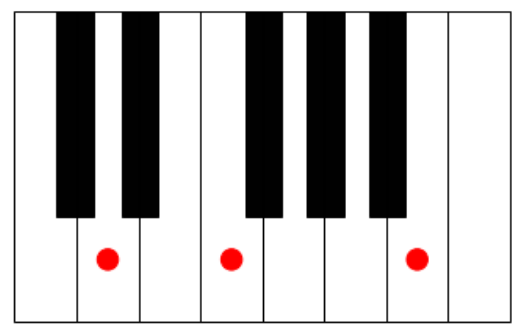} &  \includegraphics[width=1.9in]{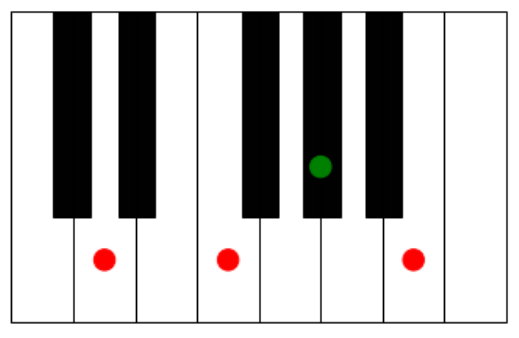} &  \includegraphics[width=1.9in]{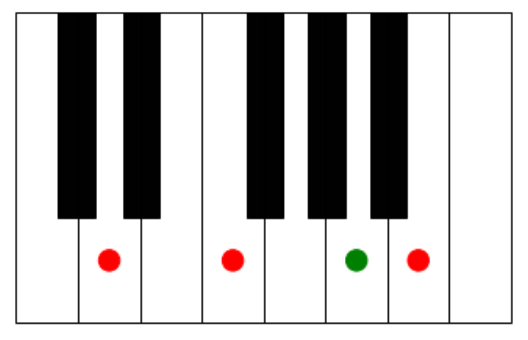} \\
    (a) & (b) & (c) \\   
     \end{tabular}
\caption{Possible completions of a diminished triad.}
\label{fig:triadComp}
\end{center}
\end{figure}

A subtler question is how triads complete within a given harmony.  Consider the Bdim triad in C major (Figure~\ref{fig:triadComp}(a)).  The natural completion to Bdim$^7$ ((b)) does not fit inside any major scale.  Indeed, accommodating such extended chords motivated our consideration of harmonies beyond the major.  However, there is a completion of Bdim \textit{inside} major harmony by an irreducible voicing ((c)).  This holds generally:

\textbf{Completion of Triads by Irreducible Voicings:}  For a given harmony and a basic triad inside this harmony, there is an irreducible voicing that completes the triad and is contained in the same harmony.

This is verified by direct checking and gives a satisfying result: all basic triads of a harmony are obtained from its irreducible voicings.  In this precise sense, our theory of irreducible voicings subsumes the classical theory of triads.
\newpage
\section{Harmony and Irreducible Scales}
\label{sec:appG}
This work has focused on complete scales with no cells (harmonies), of which there are seven.  Nothing prevents us from considering complete scales with cells---these correspond to more dissonant harmonies.  Casenave~\cite{casenave2014} has classified such scales and their modes; we refer the reader there for full details.  Here we discuss how the notion of irreducibility applies to this broader setting, paralleling our treatment of packings in Section~\ref{sec:packings}.

Table~\ref{tbl:comp} enumerates complete scales by number of cells.  At one extreme are the 7 harmonies (zero cells); at the other, the chromatic scale (12 cells).  There are 59 complete scales in total.
\begin{table}[H]
\begin{center}
\begin{tabular}{l|l|l|l|l|l|l|l|l|l|l|l|l|l}
number of cells & 0 & 1 & 2 & 3 & 4 & 5 & 6 & 7 & 8 & 9 & 10 & 11 & 12 \\
\hline
number of complete scales & 7  & 9  & 11  & 9  & 8  & 7  & 4  & 1  & 1  & 1  & 0  & 0  & 1 \\
\end{tabular}
\end{center}
\caption{Complete scales by number of cells.}
\label{tbl:comp}
\end{table}
This is a large collection.  Just as incomplete scales are derived from complete ones by removing notes, a complete scale can be enlarged by adding missing tones to obtain another complete scale; such scales are \textbf{reducible}.  The complementary notion is:

\textbf{Definition:}  A complete scale is called \textbf{scale irreducible} if no subset of it is complete.

\begin{figure}[H]
    \begin{center} 
    \begin{tabular}{cc}
    \includegraphics[width=2.0in]{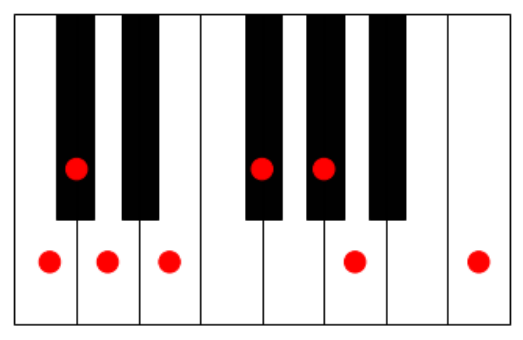} &  \includegraphics[width=2.0in]{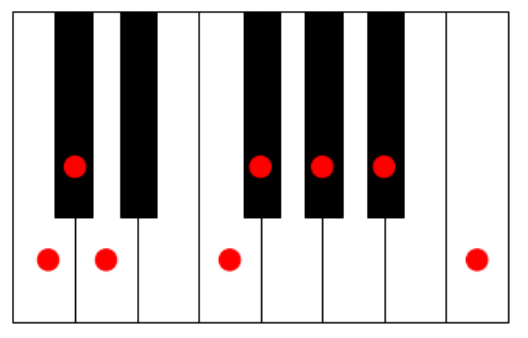} \\ \vspace{.4cm}
    (a)  & (b)  \\
    \includegraphics[width=2.0in]{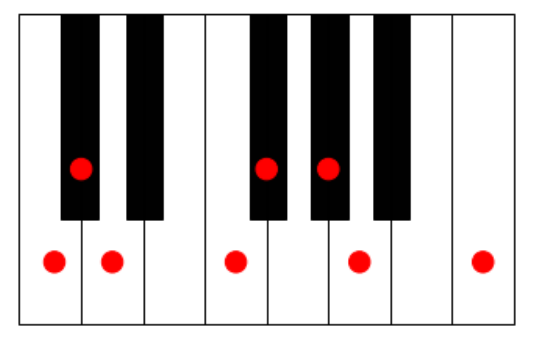} &  \includegraphics[width=2.0in]{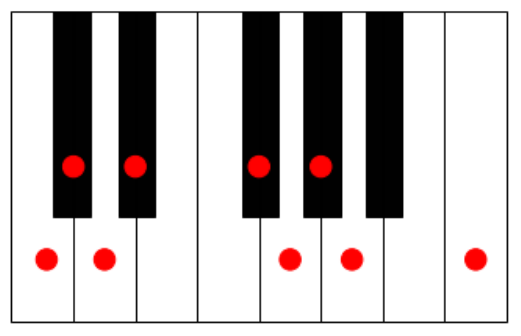} \\
    (c)  & (d) \\
     \end{tabular}
\caption{Irreducible scales with cells.}
\label{fig:irredscale}
\end{center}
\end{figure}

In other words, irreducible scales are not derived from smaller complete scales.  The classification simplifies considerably:

\textbf{Classification of Irreducible Scales:} There are 11 irreducible scales.  Seven of them correspond to complete scales with no cells (harmonies).  The remaining four are given by the examples in Figure~\ref{fig:irredscale}.
Scales (a), (b), and (c) each have one cell; (d) has four.  Scale (c) is the double harmonic scale (also known as Byzantine/Arabic/Gypsy Major/Bhairav Raga), used in ``Misirlou.''  All remaining 48 complete scales are obtained by adding tones to these 11 irreducible ones.

This illustrates the power of irreducibility for studying harmonic structure.  We note, however, that we do not have a duality between irreducible scales and some analogue of packings---the clean bijection of Section~\ref{sec:dual} does not extend to scales with cells.

\newpage 
\section{The 251 Chord Progression}
\label{sec:appH}
We briefly summarize common chord substitutions and progressions in jazz.  None of this is original; it is included for the reader's convenience.

The basic idea behind substitution is to replace one mode with another that shares enough harmonic content for the change to sound natural.  A basic example: in jazz, Lydian and Dorian are interchangeable since they use the same major scale.  For instance, F Lydian can be replaced by D Dorian---both use the C major scale, so only the bass and chord symbol change (Fmaj$^{7}$ to Dmin$^7$).

Dominant chords admit richer substitutions: any dominant chord sharing the same tritone can replace the original, provided the melody fits.  Starting from G Mixolydian, one can sharpen the 11th (Lydian Dominant, 4th mode of \textbf{MEL}) or switch to the 2nd mode of F \textbf{DIM} (G$^{7(\flat 9)}$).  All these share the tritone \{F,B\}.  A more dramatic option is the tritone substitution: replace G with D$\flat$ (a tritone away), which shares the same tritone \{F,B\}.  A particularly useful choice is D$\flat^{\text{7alt}}$, whose scale is identical to G Lydian Dominant.  From this perspective, G$^{7\sharp 11}$ $\to$ D$\flat^{\text{7alt}}$ fixes the harmony (D melodic minor) and only changes the bass.

The most common progression in music is the 5--1 (dominant to tonic): in F, C$^7$ $\to$ Fmaj (see the end of Figure~\ref{fig:agl1}).  The 4--1 progression (B$\flat$maj $\to$ Fmaj in F) is also common.  In jazz, the 4-chord is often replaced by its relative minor (a chord substitution): in F, Gmin$^7$ replaces B$\flat$maj$^7$.  Since G is the 2nd degree, this gives the ``2-chord.''

\begin{figure}[H]
    \begin{center} 
    \includegraphics[width=6.0in]{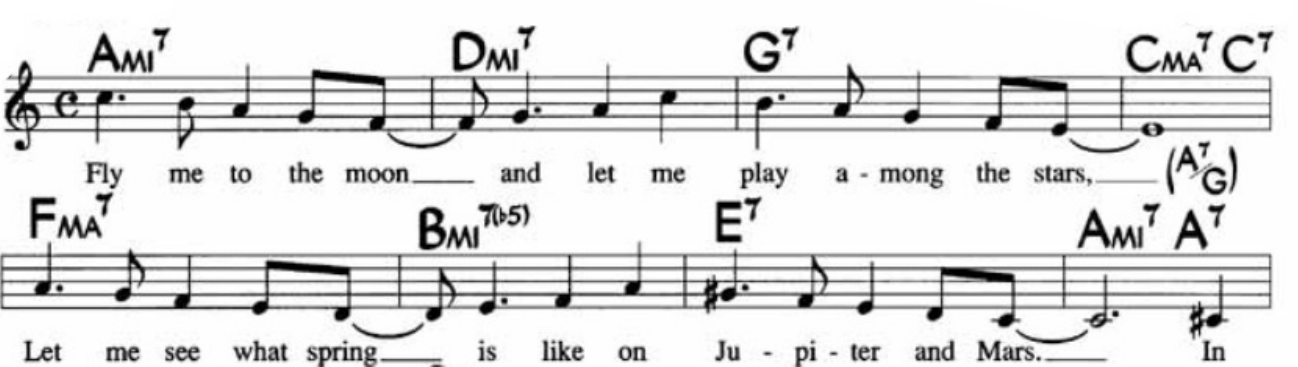} \\
    \caption{Lead Sheet for ``Fly me to the moon''.}
    \label{fig:fly}
    \end{center}
\end{figure}

The most common jazz progression is the ``251'':
\begin{equation}
    \text{Dmin}^7 \rightarrow \text{G}^7 \rightarrow \text{Cmaj}^7
\end{equation}
following the circle of fifths.  In terms of modes:
\begin{center}
 D Dorian (C major scale) $\rightarrow$ G Mixolydian (C major scale) $\rightarrow$ C Lydian (F major scale)
\end{center}
The first line of ``Fly me to the moon'' provides an example (Figure~\ref{fig:fly}).  A common minor variant is:
\begin{equation}
    \text{Dmin}^{7(\flat 5)} \rightarrow  \text{G}^{7(\flat 9,\flat 13)} \rightarrow \text{Cmin}^7  
\end{equation}
In terms of modes:
\begin{center}
 2nd mode of C $\textbf{HMIN}$ $\rightarrow$  5th mode of C $\textbf{HMIN}$  $\rightarrow$ C Dorian
\end{center}
The second line of Figure~\ref{fig:fly} gives an example:  $$\text{Bmin}^{7\flat 5} \rightarrow  \text{E}^{7(\flat 9,\flat 13)} \rightarrow \text{Amin}^7$$   Note that lead sheet notation is often vague here: $\text{E}^{7(\flat 9,\flat 13)}$ may appear as $\text{E}^{7(\flat 9)}$ or simply $\text{E}^{7}$.

\newpage 
\section{Related Work}
\label{sec:related}

The combinatorial approach to harmony developed in this work intersects with
several traditions in mathematical music theory. We briefly survey the most
relevant prior work and clarify what our framework adds.

\subsection{Pitch-Class Set Theory}

The systematic study of subsets of the 12-tone chromatic scale originates with
Forte~\cite{forte1973}, who classified all pitch-class sets up to transposition
and inversion and introduced the interval vector as an invariant. Lewin~\cite{lewin1987}
further developed the algebraic foundations using group actions on pitch-class
space. Within this tradition, the complement of a pitch-class set has been
extensively studied: Forte observed that a set and its complement share certain
interval-vector properties, a fact formalized by the hexachord theorem.

Our work can be situated within this tradition: we study subsets of
$\mathbb{Z}_{12}$ and their complements. However, the approaches differ
fundamentally. Pitch-class set theory classifies \emph{all} subsets and
observes that complements exist. Our approach \emph{derives} the musically
relevant subsets from dissonance constraints (blocks and cells) and
\emph{proves} that the two constraint families produce
complementary objects (Section~\ref{sec:dual}).

\subsection{Maximally Even Sets}

Clough and Douthett~\cite{clough1991} introduced maximally even (ME) sets:
subsets of the chromatic scale whose elements are distributed as uniformly as
possible. They showed that many musically important collections---the diatonic
scale (7-in-12), the pentatonic scale (5-in-12), the whole-tone scale
(6-in-12), and the diminished scale (8-in-12)---are maximally even. The
complement of a maximally even set is also maximally even, providing one
explanation for the pentatonic--diatonic complement relationship. This theory
has been extended by Amiot~\cite{amiot2016} using discrete Fourier analysis.

Our framework overlaps with ME theory for the most symmetric harmonies:
\textbf{MAJ}, \textbf{WTONE}, and \textbf{DIM} are all maximally even, and
their dual packings are the corresponding ME complements. However, the
block-cell framework captures harmonies that ME theory does not.
\textbf{The melodic minor scale (MEL), harmonic minor (HMIN), harmonic major
(HMAJ), and augmented scale (AUG) are not maximally even}---they have three
distinct step sizes rather than the two permitted by ME sets. These scales,
which account for the majority of non-major harmony in jazz and late-Romantic
practice, fall outside ME theory but arise naturally from the
cell-completeness condition. This is the most significant advantage of
our approach: all seven harmonies emerge from a single constraint, whereas ME
theory captures only a subset.

\subsection{Well-Formed Scales}

Carey and Clampitt~\cite{carey1989} introduced well-formed scales: scales
generated by iterated transposition of a single interval, where the ordering
of notes by generation matches their ordering by pitch. The diatonic and
pentatonic scales are well-formed (both generated by the perfect fifth), as
is the chromatic scale. This theory connects scale structure to continued
fractions and the Stern--Brocot tree, placing pentatonic (5), diatonic (7),
and chromatic (12) scales in a hierarchy of convergents of
$\log_2(3/2)$---an idea we touch on in Appendix~\ref{sec:appD}.

Well-formed scale theory provides an elegant explanation for scales generated
by a single interval but, like ME theory, does not naturally accommodate scales
with more irregular structure. Harmonic minor, harmonic major, melodic minor,
and augmented scales are not well-formed---they cannot be generated by
iterating any single interval. Our framework handles these cases uniformly
alongside the well-formed scales.

\subsection{Voice-Leading Geometry}

Tymoczko~\cite{tymoczko2006,tymoczko2011} developed a geometric approach in
which chords are represented as points in orbifolds (quotients of Euclidean
space by permutation symmetry). This framework is particularly powerful for
studying voice leading---the smooth motion between chords---and has yielded
deep results about the geometry of common chord progressions. Callender, Quinn,
and Tymoczko~\cite{callender2008} further unified several music-theoretic
equivalence relations within this geometric setting.

Our work is complementary rather than competitive. We address a different
question: which static harmonic objects (scales and chord shapes) satisfy
certain dissonance constraints, rather than how chords \emph{move} in
voice-leading space. Our framework says nothing about voice leading or
temporal structure; conversely, the geometric approach does not provide a
constraint-based \emph{derivation} of the scale vocabulary. Combining the
two---using block-cell constraints to identify harmonic objects and then
studying their voice-leading geometry---is a natural direction for future work.

\subsection{Balzano's Group-Theoretic Approach}

Balzano~\cite{balzano1980} used group theory to explain properties of the
diatonic scale, showing that the unique factorization
$\mathbb{Z}_{12} \cong \mathbb{Z}_3 \times \mathbb{Z}_4$ makes the
diatonic scale algebraically distinguished. This provides a complementary
answer to ``why is the diatonic scale special?''  Our framework does not use
group structure directly, but the connection deserves investigation: the
symmetries of diminished harmony (minor-third invariance) and augmented
harmony (major-third invariance) reflect the $\mathbb{Z}_3$ and
$\mathbb{Z}_4$ subgroups in Balzano's decomposition.

\subsection{Casenave's Harmonic Structure Levels}

The combinatorial approach to defining harmony via semitone cells originates
with the work of Casenave~\cite{casenave2014}. Specifically, Casenave
introduced the notion of a semitone cell (three consecutive semitones within
a scale) and used cell-completeness to classify all complete scales, organized
by their number of cells. The seven harmonies (complete scales with zero cells)
are a special case of this broader classification. Our work takes Casenave's
foundational insight in a different direction by: (a) introducing the parallel
notion of semitone blocks and block-complete packings; (b) proving the duality
between harmonies and irreducible packings (Section~7); and (c) combining
blocks and cells to classify irreducible chord shapes (Section~9). To our
knowledge, the packing concept, the duality theorem, and the chord
classification are new contributions of the present work.

\end{document}